\documentclass[prd,10pt,aps,longbibliography,onecolumn]{revtex4-1}  

\usepackage[top=3cm, bottom=3cm, left=3cm, right=3cm]{geometry}
\usepackage{amsmath}
\usepackage{amsfonts}
\usepackage{amssymb}
\usepackage{graphicx}%
\usepackage{hyperref}
\usepackage{aligned-overset}
\usepackage{natbib}
\usepackage{longtable}
\usepackage{xcolor}

\providecommand{\U}[1]{\protect\rule{.1in}{.1in}}
\begin{document}

\title{Role of bound states and resonances in scalar QFT at nonzero temperature}
\author{Subhasis Samanta}
\email{subhasis.samant@gmail.com} 
\affiliation{\textit{Institute of Physics, Jan-Kochanowski University, }\\\textit{ul. Uniwersytecka 7, 25-406 Kielce, Poland.}}
    
\author{Francesco Giacosa}
\email{fgiacosa@ujk.edu.pl}
   \affiliation{\textit{Institute of Physics, Jan-Kochanowski University, }\\\textit{ul. Uniwersytecka 7, 25-406 Kielce, Poland.}}
   \affiliation{\textit{Institute for Theoretical Physics, J. W. Goethe University, }\\\textit{ Max-von-Laue-Str. 1, 60438 Frankfurt, Germany.}}

%\author{Subhasis Samanta$^{1}$ and Francesco Giacosa$^{1,2},$\\$^{1}$\textit{Institute of Physics, Jan-Kochanowski University, }\\\textit{ul. Swietokrzyska 15, 25-406, Kielce, Poland.}\\$^{2}$\textit{Institute for Theoretical Physics, J. W. Goethe University, }\\\textit{ Max-von-Laue-Str. 1, 60438 Frankfurt, Germany.}\\
%}
%\date{}

\begin{abstract}
We study the thermal properties of quantum field theories (QFT) with three-leg interaction vertices $g\varphi^{3}$ and $gS\varphi^{2}$ ($\varphi$ and $S$ being scalar fields), which constitute the relativistic counterpart of the Yukawa potential. 
We follow a non-perturbative unitarized one-loop resummed technique for which the theory is unitary and well-defined for a large range of values of the coupling constant $g$.
Using the partial wave decomposition of two-body scattering we calculate the phase shifts, whose derivatives are used to infer the pressure of the system at nonzero temperature by using the so-called phase shift formalism. 
A $\varphi \varphi$ bound state is formed when the coupling $g$ is larger than a certain critical value.
As the main outcomes of this work, we estimate the influence of particle interaction on the pressure (both without and with the bound state), and we demonstrate that the latter is always continuous as a function of the coupling constant $g$ (no sudden jumps occurs when the bound state forms), and we show that the contribution of the bound state to the pressure does not count as \textit{one} state in the thermal gas, since a cancellation with the residual $\varphi \varphi$ interaction occurs. The amount of this cancellation depends on the details of the model and its parameters and a variety of possible scenarios is presented. 
%Moreover, even when no bound state occurs, we estimate the role of the interaction (including a resonance in the $gS\varphi^{2}$ theory), which is in general non-negligible. 
We also show how the overall effect of the interaction, including eventual resonances and bound states, can be formally described by a unique expression that makes use of the phase shift continued below the threshold.  

%In the  $\varphi^{3}$ case, we also investigate the role of an additional  $\varphi^{4}$ term.  In the $S\varphi^2$ case the resonance $S$ is present: the bound state (if it forms) and the resonance $S$ can be consistently taken into account by the $\varphi\varphi$ interaction. 
%At the critical value of the coupling, pressure due to $\varphi\varphi$ interaction in s-wave changes sign from positive to negative.  The decrease of this pressure is exactly compensated by the pressure of the bound state and makes the total pressure continuous as a function of $g$. 
%We have also investigated the formation of a bound state in presence of an intermediate state %$S$ of mass $M$ which have type interaction with $\varphi$. For $M > 2m$, we observe a %resonance like structure in the derivative of s-wave phase shift near the centre of mass energy $\sqrt{s} = M$.
%The bound state is also formed in this case above some critical value of the coupling $g$. 
%However critical value of $g$ in $S\varphi^2$ interaction is different from the $\varphi^{3}$ %interaction. Total pressure as a function of $g$ is also continuous here.

\end{abstract}

\maketitle

\section{Introduction}
The production of hadronic bound states such as deuteron ($^{2}\text{H}$), tritium ($^{3}\text{H}$), helium-3 ($^{3}$He), helium-4 ($^{4}\text{He}$), hypertritium
($_{\Lambda}^{3}$H) and their antiparticles in high energy collisions has created a lot of interest in the community \cite{Cocconi:1960zz,
Abelev:2010rv,Agakishiev:2011ib,Adam:2015vda,Adam:2019phl,Acharya:2019xmu,Acharya:2020sfy}. The temperature at the chemical freezeout is much larger than the binding energies of those bound states. Hence the natural question is how such weakly bound objects can form in such a hot environment? Moreover, the size of such bound states is usually large compared to the inter-particle spacing of the fireball. It is, therefore, important to understand the mechanism of the formation of those bound states in a thermal system. Further, a whole new class of $X, Y$ and $Z$ resonances are observed in the QCD spectrum which are not predicted by the quark model; see \cite{Esposito:2014rxa} references therein. Some of these resonances can be mesonic molecular bound states. Experimentally observed pentaquarks \cite{Aaij:2019vzc} can also be understood as molecular objects.

Two successful phenomenological models  describing the production of bound states in high energy collisions are discussed in the literature: the thermal model \cite{Siemens:1979dz,Andronic:2010qu,Andronic:2012dm,Cleymans:2011pe,Ortega:2017hpw,
Ortega:2019fme} and the coalescence model \cite{Butler:1963pp, Schwarzschild:1963zz, Gutbrod:1988gt,Sato:1981ez,Mrowczynski:1992gc,Csernai:1986qf,
Mrowczynski:2016xqm,Bazak:2018hgl,Dong:2018cye,Sun:2016rev,
Sun:2018jhg,Polleri:1997bp, Mrowczynski:2019yrr, Bazak:2020wjn}.
In the thermal model, bound states are formed directly from the source according to the corresponding probability at a given temperature $T$. On the other hand,  the coalescence model works in a two-step process: first, nucleons are produced from a fireball and, second, the bound states (such as nuclei) are formed long after the emission nucleons when the relative momenta of the nucleons become small. Along the same line, conventional mesons, such as pions and kaons, etc. are directly produced, while molecular states emerge as a secondary product.  
Till now, it is not clear which assumption is correct, although both models describe the production yields of bound states in high-energy collisions quite well. Transport \cite{Danielewicz:1991dh} and hybrid dynamical  \cite{Oliinychenko:2018ugs} models are also applied to describe bound states.

In Ref. \cite{Samanta:2020pez}, we investigated how to take into account the effect of the interaction in a thermal gas in the context of quantum field theory (QFT) by using a scalar $\lambda\varphi^4$ interaction ($\lambda$ being the dimensionless coupling), which corresponds to a delta-potential in the non-relativistic limit. For that (relatively simple) QFT, it was possible to consider only the $s$-channel Feynman diagrams. The temperature dependence was incorporated via the  so-called phase shift (or S-matrix) formalism \cite{Dashen:1969ep,Venugopalan:1992hy,Broniowski:2015oha,Lo:2017ldt,Lo:2017sde,Lo:2017lym,Dash:2018can,Dash:2018mep,Lo:2019who,Lo:2020phg} (see Appendix A for a brief QM recall of this approach). 
This approach allows calculating the effect of the interaction in the medium
by using the derivative of the scattering phase shifts calculated in the
vacuum. In particular, a positive (negative) derivative implies an increase
(decrease) of the pressure.
For $\lambda<0$ (for which attraction occurs) and $|\lambda|$ above a certain critical value, a bound state forms. It was shown that this bound state is relevant at nonzero temperature, but it counts less than  what a single state with the same mass would contribute since a partial cancellation with the residual $\varphi\varphi$ interaction takes place. Note, this is in partial agreement with the quantum mechanical (QM) approach of Ref. \cite{Ortega:2017hpw}, where a similar (but even more pronounced) cancellation was shown to occur.

A natural continuation of the work in Ref. \cite{Samanta:2020pez} is to investigate the role of bound states in more complex QFTs, that go beyond the simple contact interaction. The next step is then to consider scalar theories that correspond to the Yukawa interaction, in which, besides the $s$-channel, also the $t$-channel and $u$-channel Feynman diagrams must be included. 
The simplest of such theories contains the interaction $g\varphi^3$ ($g$ being the coupling constant), in which the exchanged particle is of the same type as the scattering ones. 
The interaction is always attractive (for any value of the coupling constant $g$) and, if strong enough, a bound state forms. The $t$ and the $u$ exchange channels are crucial for the attraction and thus for the formation of the bound state and generate also a left-hand cut in the complex plane, which must be properly taken into account in the unitarized version of the theory. 
 Thus, the understanding of such scalar QFTs is an important intermediate step toward the application to realistic cases (such as the deuteron or the $X(3872)$ or mentioned above) that involve particles with spin. Namely, some interesting and quite general issues can be addressed by the simple but nontrivial models presented in this work.

At nonzero $T$ the bound state (if it forms) must be included as an additional state in the thermal gas, yet its effect is typically partially canceled by the residual $\varphi\varphi$ interaction, in a way that resembles the results of  Ref. \cite{Samanta:2020pez}. 
%Yet, this cancellation gets smaller and smaller for increasing coupling.
%and in the strong coupling limit, one may also see an overall contribution that is even larger than the simple bound state thermal contribution. 
As expected, the pressure of the system turns out to be continuous as a function of $g$ at any given temperature, showing that the emergence of the bound state does not cause any discontinuity in the pressure, because the abrupt contribution of the bound state is compensated by a jump in the interaction part.   
%[Adding a $\varphi^4$-interaction modifies the numerical results but not their qualitative interpretation.] 

Moreover, in order to quantify the role of the interaction and of the
(eventually forming) bound state, we aim also to calculate the following quantities:

\begin{itemize}
\item By denoting with $P_{\varphi, free}$ the non-interacting pressure of a
gas of particles at a given temperature $T$, when no bound state occurs the
the pressure of the system can be expressed as $\eta P_{\varphi,free}$, where $\eta$
is a constant that quantifies how much the interaction modifies the simple
free gas contribution. As we shall see, $\eta\geq1,$ thus showing that the
interaction increases the pressure. The goal is to quantify $\eta$ in
dependence of the coupling. In other words, we can estimate the error that one
would do by neglecting the effect of the interaction.

\item When the attraction is strong enough to generate a bound state, the
total pressure can be written as $P_{\varphi,free}+\zeta P_{B},$ where $P_{B}$
is the pressure contribution of a free gas of bound state particles. In the
limit $\zeta=1$, one has the sum of two free gases, thus the
bound state counts as a normal state. Yet, we shall
show that $0\leq\zeta\leq1$: one may interpret this result as a
partial cancellation of the bound state contribution due to the $\varphi \varphi$
interactions above the threshold. In some cases, $\zeta$ can deviate, even sizably, from unity,
showing that the simple inclusion of the bound state to the pressure may not
be accurate.
\end{itemize}

As a next step, we repeat the study above for the QFT that contains two distinct fields $S$ and $\varphi$ which interact via a term of the type $gS\varphi^2$. The state $S$ with mass $M$ is exchanged by the two $\varphi$-fields. Assuming that $M>2m$, the field $S$ corresponds to resonance with a certain decay width into $\varphi\varphi$, see e.g. Ref. \cite{Giacosa:2007bn} for a detailed description. As it is well-known, the thermal properties of the resonance $S$ can be described via the $\varphi\varphi$ phase shift above the threshold, see e.g. Refs. \cite{Weinhold:1996ts,Weinhold:1997ig,Florkowski:2010zz,Broniowski:2015oha,Lo:2017lym,Lo:2019who,Lo:2015cca,Lo:2017ldt} and Refs. therein. 
Moreover, besides the resonance $S$, also a bound state can form if $g$ is large enough, making this system quite interesting. %(obviously, the critical value differs from the previous $\varphi^3$ theory). 
Also, in this case, we estimate the effect of the interaction and of the bound state on the pressure and we verify that the latter is always continuous as a function of the coupling $g$.

Finally, for the QFTs mentioned above and in agreement with Ref. \cite{Samanta:2020pez}, we show how to formally generalize the phase shift approach for the thermal description of the system  by extending it below the $\varphi\varphi$ threshold. In this case, (eventual) bound state(s) and resonance(s), if present, are automatically incorporated into the finite-$T$ properties of the thermal gas. 

The paper is organized as follows: in Sec. \ref{sec:phi3Theory} we briefly present the vacuum's properties of the $\varphi^3$- QFT. Here we discuss scattering phase shifts, the unitarization procedure, and the formation of a bound state. Then, in Sec. \ref{sec:FiniteTemp} we discuss the formalism and the results of the system at finite temperature.  Further, in Sec. \ref{sec:IntermediateState} we consider a second state $S$ with the three-leg interaction $S \varphi^2$. Finally, in Sec. \ref{sec:Conclusions} we summarize and conclude the paper.
 Additional topics (brief recall of the phase shift formula, study of causality via the Wigner condition and the speed of sound, and the addition of a $\varphi^4$-interaction term to the potential) are presented in four separate Appendices.

\section{Vacuum phenomenology of the scalar $\varphi^{3}$-QFT}\label{sec:phi3Theory}
In this Section, we describe (some) vacuum properties of the $\varphi^3$ theory, with special attention to elements of scattering and the employed unitarization approach. These properties will be used later on when presenting the results of the system at nonzero $T$. 

\subsection{Lagrangian and amplitudes}
\label{sec:scttering}
The Lagrangian under consideration reads
\begin{equation}
\mathcal{L}=\frac{1}{2}\left(  \partial_{\mu}\varphi\right)  ^{2}-\frac{1}%
{2}m^{2}\varphi^{2}-\frac{g}{3!}\varphi^{3}, \label{eq:L}%
\end{equation}
where the first two terms describe a free particle with mass $m$ and the last
term corresponds to the interaction. The coupling constant $g$ has dimension
[Energy], therefore the theory is renormalizable \cite{Peskin:1995ev}. 
Yet, as we shall comment later on, we shall introduce a non-perturbative unitarization
procedure on top of Eq. \ref{eq:L}, in such a way to make the theory finite and unitary at each energy.

The potential in Eq. \ref{eq:L}
\begin{equation}
V(\varphi)=\frac{m^{2}}{2}\varphi^{2}+\frac{g}{3!}\varphi^{3}\label{pot}%
\end{equation}
has a local minimum at $\varphi=0$, but is unbounded from below. Namely,
choosing $g>0,$ one has $V(\varphi\rightarrow-\infty)=-\infty$, meaning that
the underlying system is only metastable: instabilities are expected to occur for $g$ large enough.

One can study small oscillations around the minimum by applying the perturbation theory that holds when $g$ is sufficiently small (as we shall
see, small implies $g/m\lesssim1$). Yet, in this work, we are interested in the
emergence of a bound state, which is a nonperturbative phenomenon. Thus, one
must consider intermediate values of $g$ (as it will be clear later, of the order of $g/m\sim10$), for
which one needs to go beyond perturbation theory: to this end, a unitarization
approach shall be applied. An important discussion point in the following is
indeed the determination of the range of $g$ up to which the employed 
unitarization approach can be considered reliable.

We now turn to $\varphi\varphi$ scattering. In the centre of the mass frame, the differential cross-section is given by \cite{Peskin:1995ev}
\begin{equation}
\frac{d\sigma}{d\Omega}=\frac{|A(s,t,u)|^{2}}{64\pi^{2}s},
\end{equation}
where $A(s,t,u)$ is the scattering amplitude as evaluated through Feynman
diagrams, and $s,t$ and $u$ are Mandelstam variables:
\begin{align}
s  &  =(p_{1}+p_{2})^{2}\geqslant4m^{2}\text{ ,}\\
t  &  =(p_{1}-p_{3})^{2}=-\frac{1}{2}(s-4m^{2})(1-\cos\theta)\leq0\text{ ,}\\
u  &  =(p_{2}-p_{3})^{2}=-\frac{1}{2}(s-4m^{2})(1+\cos\theta)\leq0\text{ ,}%
\end{align}
where $p_{1},p_{2},p_{3}$ and $p_{4}$ are the four-momenta of the particles in the centre of the mass frame 
($p_{1},p_{2}$ ingoing and $p_{3},p_{4}$ outgoing), and $\theta$ is the
scattering angle. The sum of these three variables is $s+t+u=4m^{2}.$

In this example (as well as in the following ones in the paper) we limit our
study to two-body elastic scattering. The first inelastic channel opens at
$s=(3m)^{2}=9m^{2},$ the second at $s=(4m)^{2}=16m^{2},$ etc. The treatment of
the problem including inelastic channels is much more involved and is not
attempted here. Fortunately, inelastic channels are expected to deliver subleading
contributions to the pressure up to (at least) temperatures of the order of $T\sim3m$.

In the particular case of our Lagrangian of Eq. \ref{eq:L}, the tree-level
scattering amplitude $A(s,t,u)$ takes the form
\begin{equation}
A(s,t,u) =-\frac{g^{2}}{s-m^{2}+i\epsilon}-\frac{g^{2}}{t-m^{2}+i\epsilon
}-\frac{g^{2}}{u-m^{2}+i\epsilon} \text{ .}
\end{equation}
The amplitude on-shell reads:
\begin{align}
A(s =4m^{2},0,0) & =-\frac{g^{2}}{4m^{2}-m^{2}}-\frac{g^{2}}{-m^{2}}%
-\frac{g^{2}}{-m^{2}}  =\frac{g^{2}}{m^{2}}\left(  -\frac{1}{3}+1+1\right)  =\frac{5g^{2}}{3m^{2}%
}>0 \text{ ,}
\end{align}
thus attraction wins at the threshold because the attractive $t$- and $u$-channels overcome the $s$-channel repulsion. 

%The s-wave ($l=0$, where $l$ is the orbital angular momentum) scattering length at tree-level is:
%\begin{equation}
%ma_{0}^{\text{SL}}=\frac{1}{32\pi}A(s=4m^{2},0,0)=\frac{1}%{32\pi}\frac{5g^{2}%
%}{3m^{2}}\Rightarrow a_{0}^{\text{SL}}=\frac{5g^{2}}{96\pi %m^{3}}\text{ .}%
%\end{equation}
Next, we turn to partial wave expansion \cite{Messiah}:%
\begin{equation}
A(s,t,u)=A(s,\theta)=\sum_{l=0}^{\infty}(2l+1)A_{l}(s)P_{l}(\cos
\theta)\text{{ ,}}%
\end{equation}
where $P_{l}(\xi)$ with $\xi=\cos\theta$ are the Legendre polynomials with
\begin{equation}
\int_{-1}^{+1}d\xi P_{l}(\xi)P_{l^{\prime}}(\xi)=\frac{2}{2l+1}\delta
_{ll^{\prime}}\text{ .}%
\end{equation}
In general, the $l$-th wave contribution to the amplitude is given by
\begin{equation}
A_{l}(s)=\frac{1}{2}\int_{-1}^{+1}d\xi A(s,\theta)P_{l}(\xi) \text{ .}
\end{equation}
At tree-level the s-wave amplitude's contribution takes the form:
\begin{equation}
A_{0}(s)=\frac{1}{2}\int_{-1}^{+1}d\xi A(s,\theta)=-\frac{g^{2}}{s-m^{2}%
}+2g^{2}\frac{\ln\left[  1+\frac{s-4m^{2}}{m^{2}}\right]  }{s-4m^{2}}\text{
,}%
\end{equation}
which at threshold reduces to $A_{0}(s=4m^{2})=\frac{5g^{2}}{3m^{2}}$. 
%\begin{equation}
%A_{0}(s=4m^{2})=\frac{5g^{2}}{3m^{2}}\text{ }.
%\end{equation}
The s-wave scattering length (at tree-level) $a_{0}^{\text{SL}}$can
be reobtained as\footnote{Note, in order to avoid confusion, we denote with the Latin alphabet s-, d-, g- ...as the partial waves with increasing $l$. On the other hand, we employ calligraphic $s$, $t$, and $u$ when referring to the Mandelstam variables. Finally, the calligraphic $g$ refers to the three-leg coupling constant entering into the various Lagrangian(s).}:
\begin{equation}
a_{0}^{\text{SL}}=\frac{1}{2}\frac{A_{0}(s=4m^{2})}{8\pi\sqrt{4m^{2}}}%
=\frac{1}{32\pi m}\frac{5g^{2}}{3m^{2}}=\frac{5g^{2}}{96\pi m^{3}}\text{ .}%
\end{equation}

There are two important properties of $A_{0}(s)$ that need to be discussed since they will be useful later on:

(i) $A_{0}(s)$ contains a pole for $s=m^{2}$: this is the pole of the
single  particle in the $s$-channel, 
$\left\vert A_{0}(s=m^2)\right\vert =\infty$; this simple pole (with nonzero
residuum) should be preserved when unitarizing the theory. We  assume
that the position of the pole is not shifted by loop corrections: this can be achieved by a suitable subtraction.

(ii) $A_{0}(s)$ diverges for $s=3m^{2}$. This is due to the left-hand-cut
induced by the $t$- and the $u$-channels onto the s-wave. (The
residuum is zero, thus no particle corresponds to this divergence).
We stress that the existence of the branch cut for
$s\leq3m^{2}$ is also a consequence of the single-particle pole
described in point (i) \cite{Frazer:1969euo,Taylor}. 
As for (i), we will impose this feature when unitarizing the theory.

We now turn to higher waves. Clearly, $A_{2n+1}(s)=0$, since each odd wave vanishes. 
For the d-wave, the corresponding amplitude reads:  
\begin{equation}
A_{2}(s)=\frac{-2g^{2}}{(s-4m^{2})^{3}}\left(  3\left(  8m^{4}-6m^{2}%
s+s^{2}\right)  +\left(  2m^{4}+2m^{2}s-s^{2}\right)  \ln\left[
1+\frac{s-4m^{2}}{m^{2}}\right]  \right) \text{ ,}
\end{equation}
which for $s$ close to the threshold is approximated by:%
\begin{equation}
A_{2}(s) \simeq \frac{1}{15}\frac{g^{2}}{m^{6}}(s-4m^{2})^{2}\text{ .}%
\end{equation}
For the g-wave we have:%
\begin{align}
A_{4}(s)  &  =\frac{-g^{2}}{3(s-4m^{2})^{5}}\Biggl(-5\left(  46m^{4}%
-2m^{2}s-5s^{2}\right)  \left(  s-2m^{2}\right)  \left(  s-4m^{2}\right) \nonumber \\
&  +6\left(  74m^{8}-124m^{6}s+54m^{4}s^{2}-4m^{2}s^{3}-s^{4}\right)
\ln\left[  1+\frac{s-4m^{2}}{m^{2}}\right]  \Biggr) \text{ ,}
\end{align}
which close to the threshold reads:
\begin{equation}
A_{4}(s)\simeq \frac{g^{2}}{315m^{8}}(s-4m^{2})^{4} \text{ .}
\end{equation}
Both  $A_{2}$ and $A_{4}$ vanish at the threshold, like any other wave with $l>0$. 
The total cross-section reads
\begin{equation}
\sigma(s)=\frac{1}{2}2\pi\frac{1}{64\pi^{2}s}\int_{0}^{\pi}\left\vert
A(s,\theta)\right\vert ^{2}\sin\theta d\theta\text{ =}\frac{1}{2}2\pi\frac
{1}{64\pi^{2}s}\sum_{l=0}^{\infty}2(2l+1)\left\vert A_{l}(s)\right\vert
^{2} \text{ ,}
\label{eq:sigma}%
\end{equation}
which at the threshold reduces to:
\begin{equation}
\sigma(s_{th}=4m^{2})=\frac{1}{2}2\pi\frac{1}{64\pi^{2}s}2\left\vert
A_{0}(s_{th})\right\vert ^{2}=8\pi\left\vert a_{0}^{\text{SL}}\right\vert
^{2} \text{ .}
\end{equation}

Next, let us introduce an important quantity of this work, the phase shifts.  The $l$-th wave phase shifts $\delta_{l}(s)$ is defined as:
\begin{equation}
\frac{e^{2i\delta_{l}(s)}-1}{2i}=\frac{1}{2}\cdot\frac{k}{8\pi
\sqrt{s}}A_{l}(s) \text{ ,} \label{defdeltal}%
\end{equation}
where $k=\sqrt{\frac{s}{4}-m^{2}}$. 
%and where the `running' length $a_{0}(s)$
% by construction such that $a_{0}(s=4m^{2})=a_{0}^{\text{SL}}.$
Note, for $s$ just above threshold we have $\frac{e^{2i\delta_{0}(s)}-1}{2i}%
\simeq\delta_{0}(s)\simeq ka_{0}^{\text{SL}}$.
In general, the phase shift $\delta_{l}(s)$ can be calculated as:
\begin{equation}
\delta_{l}(s)=\frac{1}{2}\arg\left[  1+i\frac{k}{8\pi\sqrt{s}}A_{l}(s) \right]  \text{ .}
\label{eq:ps_tree}%
\end{equation}

\begin{figure}[ptb]
\centering
\includegraphics[width=0.32\textwidth]{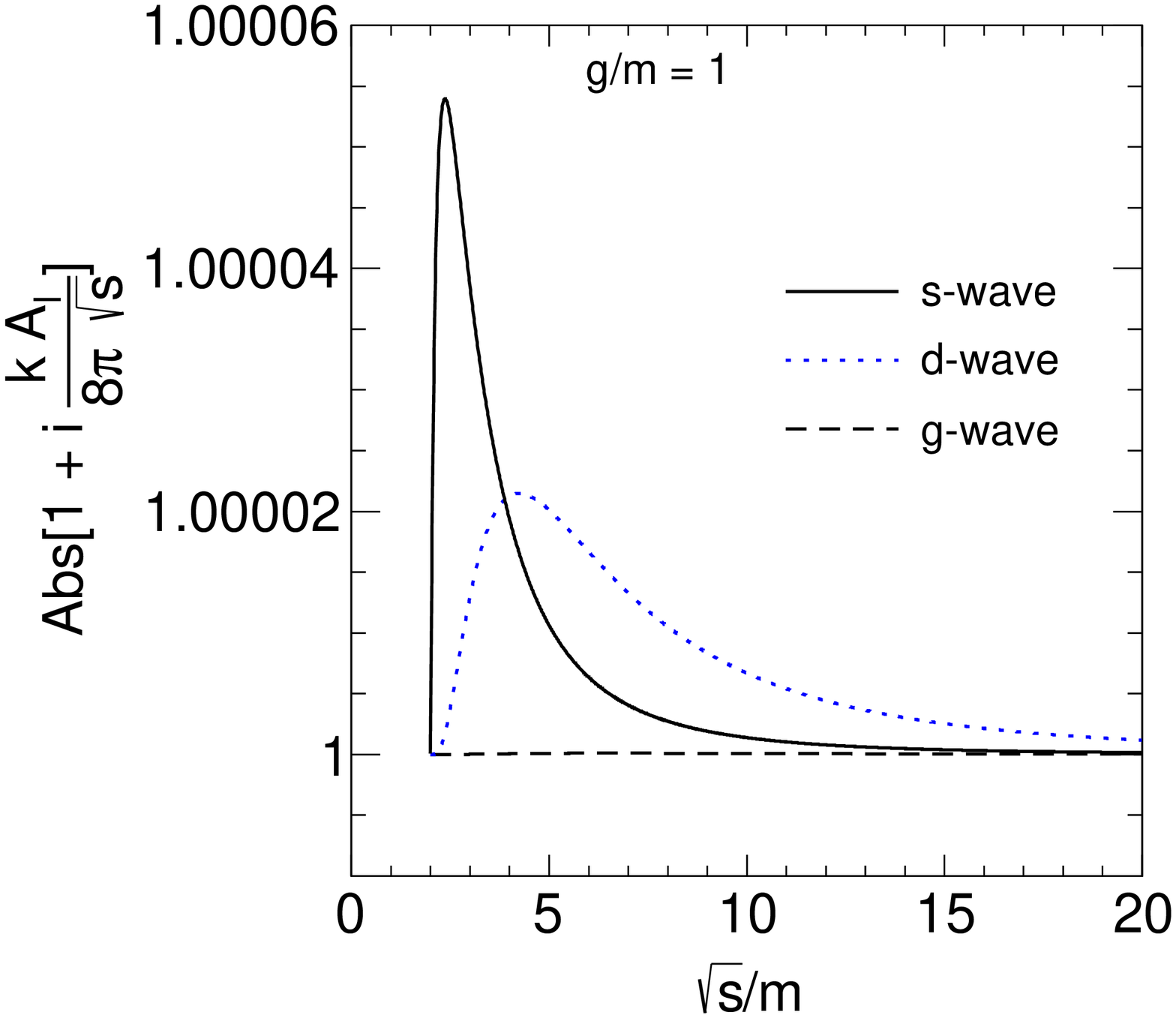}
\includegraphics[width=0.32\textwidth]{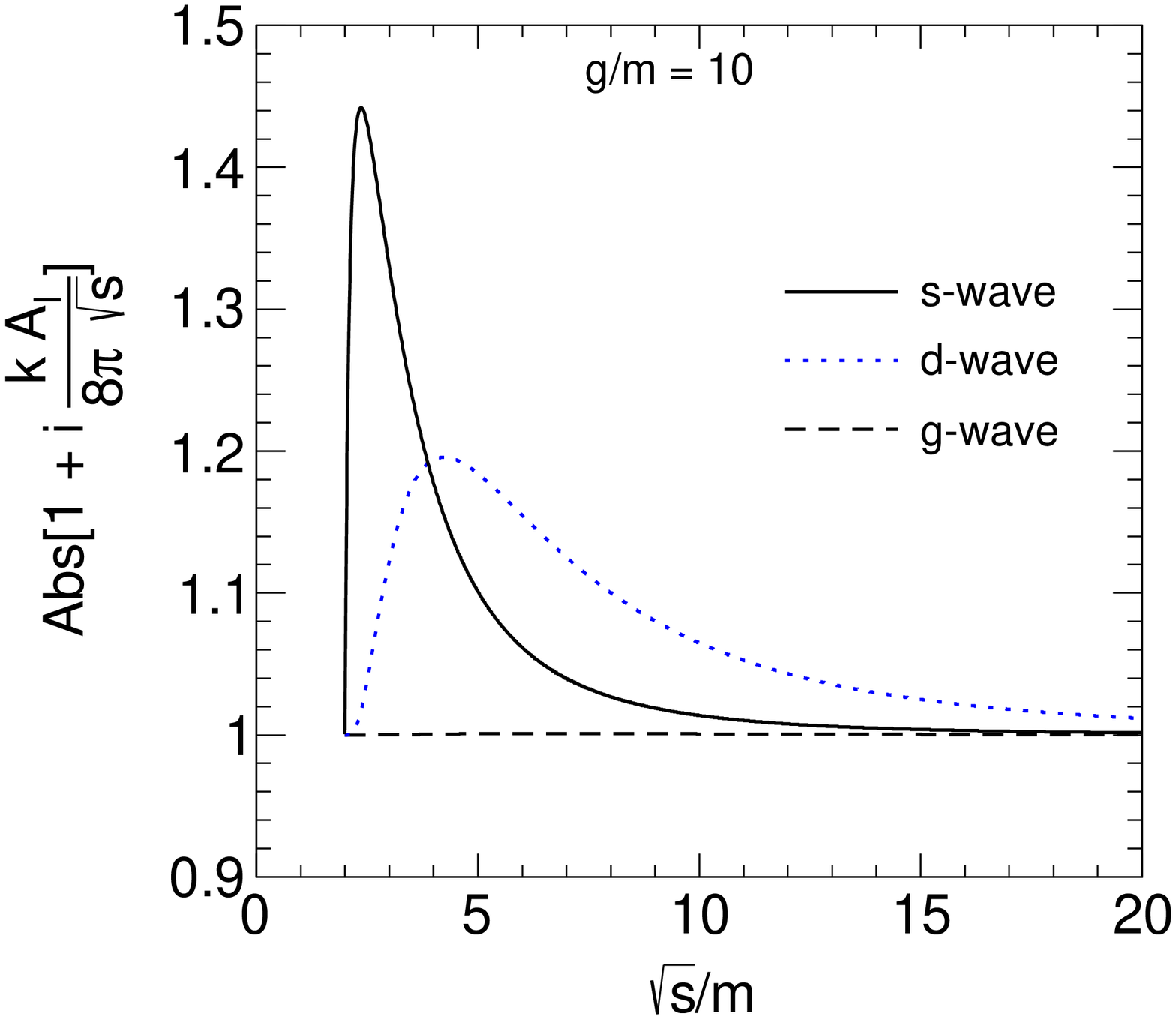}
\includegraphics[width=0.32\textwidth]{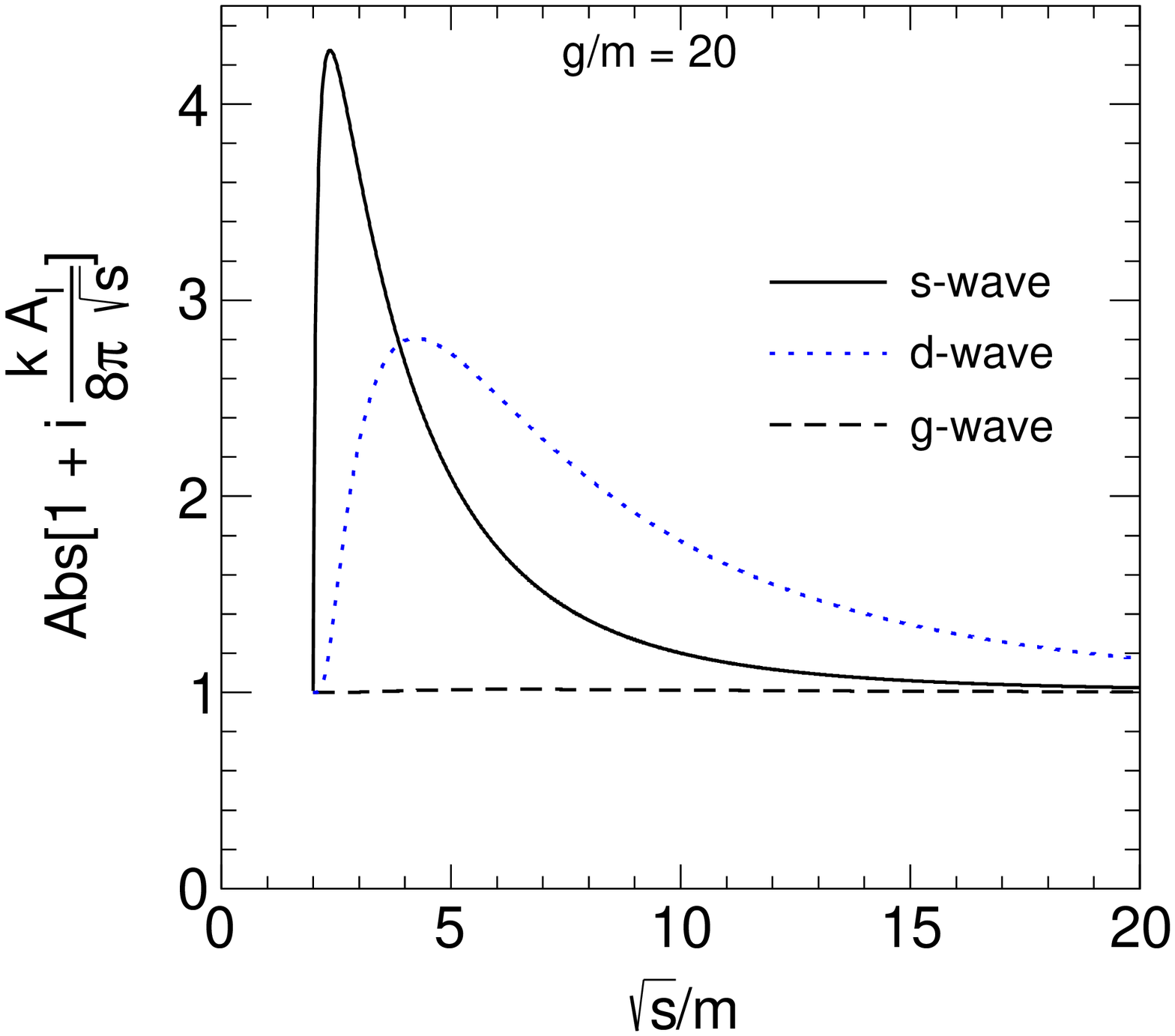}
\caption{Energy dependence of the absolute value of $1 + i \frac{k A_{l}}{8\pi \sqrt{s}}$ for $l = 0, 2, 4$ at $g=1m$ (Left), $g = 10m$ (Middle) and $20m$ (Right) at the tree-level. }
\label{Fig:violation_unitarization}
\end{figure}

 We can rewrite Eq. \ref{defdeltal} as
\begin{equation}
  e^{2i\delta_{l}(s)}=1+ i\frac{k}{8\pi\sqrt{s}}A_{l}(s) 
  \text{ .}
  \end{equation}
In Fig. \ref{Fig:violation_unitarization} we show the energy dependence of
$\left\vert 1+i\frac{kA_{l}}{8\pi\sqrt{s}}\right\vert $ for $l=0,2,4$ and for
$g/m=1,$ $10$, $20$. Clearly, if unitarity is strictly fulfilled, this
quantity is one for any value of $s.$ As it is known, perturbation theory
fulfills unitarity only perturbatively (up to the considered order). For $g/m=1$ violations of
unitarity are rather small, showing that $g/m=1$ can be considered as
a small coupling. On the other hand, this is not the case for $g/m=10$ or
$20,$ for which these violations are large for the s-wave and non-negligible
for the d-wave, (yet, they are still rather small for the g-wave). 
Since, as we shall
see in the next two subsections, the range $10\lesssim g/m\lesssim20$ is relevant in this paper, an unitarization approach is required.

\subsection{Unitarization}
\label{sec:Unitarization}
In this subsection, we describe the so-called on-shell unitarization \cite{Dobado:1992ha,Oller:1997ng}. 
First, we need to introduce the $\varphi \varphi$ loop $\Sigma(s)$. Its imaginary part above the threshold is the usual phase-space kinematic factor (see e.g. Ref. \cite{Giacosa:2007bn}):
\begin{equation}
I(s)=\operatorname{Im}\Sigma(s)=\frac{1}{2}\frac{\sqrt{\frac{s}{4}-m^{2}}%
}{8\pi\sqrt{s}}\text{ for }\sqrt{s}>2m \text{ .}
\label{imsigma0}
\end{equation}
We use no cutoff, hence the above equation is considered valid up to arbitrary
values of the variable $s$. The imaginary part alone does not fix the form of $\Sigma(s)$ completely. Here, the loop function $\Sigma(s)$ for a complex $s$
is chosen by considering two subtractions \cite{Giacosa:2021brl}:
\begin{equation}
\Sigma(s)=-\dfrac{(s-m^{2})(s-3m^{2})}{\pi}\int_{4m^{2}}^{\infty}\frac
{\frac{1}{2}\frac{\sqrt{\frac{s^{\prime}}{4}-m^{2}}}{8\pi\sqrt{s^{\prime}%
}}}{(s-s^{\prime}+i\varepsilon)(s^{\prime}-m^{2})(s^{\prime}-3m^{2}%
)}ds^{\prime} \text{ .}\label{loop}%
\end{equation}
The subtractions guarantee that $\Sigma(s=m^{2})=0$ and $\Sigma
(s=3m^{2})=0.$
 In this way, the choice of $\Sigma(s)$ fulfills the following requirements: (i) it preserves the pole corresponding to $s=m^{2}$ (in other words, the tree-level mass is also preserved at the unitarized level); (ii) it assures that the unitarized amplitude diverges at the branch point $s=3m^{2}$ generated by the single-particle pole for $m^{2}$ along the $t$ and $u$ channels. 

Note, that the convergence of the integral is guaranteed by a single subtraction.  Yet, it turns out that in the present approach a single subtraction is not appropriate to study our system, since,  whenever a bound state forms, also a ghost state appears \cite{Donoghue:2019fcb}. This problem does not take place when two subtractions as in Eq. \ref{loop} are implemented. A single subtraction is possible if a different unitarization loop is implemented, see Appendix B for details.  
Interestingly, the same unitarization procedure has been used in the recent work of Ref. \cite{Giacosa:2021brl} dealing with glueball-glueball scattering in an effective dilaton model of Yang-Mills theory.

\begin{figure}[ptb]
\centering
\includegraphics[width=0.48\textwidth]{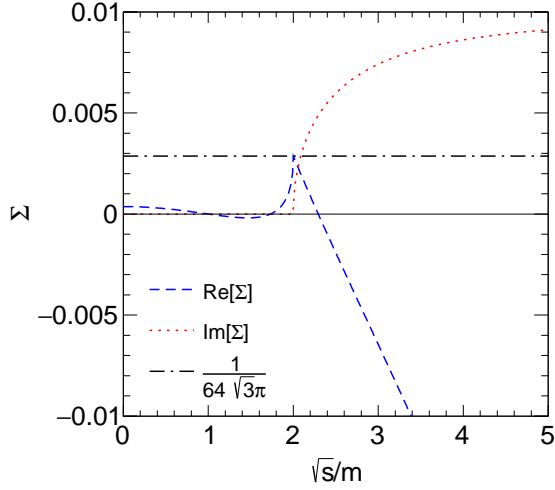}\caption{Energy dependence of the real and imaginary parts of the loop function $\Sigma$ (see Eq. \ref{loop}). The value of the $\Sigma$ at the threshold ($= 1/(64\sqrt{3}\pi)$) is also indicated in this plot.}
\label{Fig:ReImloop}
\end{figure}

The real and the imaginary parts of $\Sigma$ as a function of $\sqrt{s}/m$ are shown in Fig. \ref{Fig:ReImloop}. [Note: throughout this paper, we consider $m = 1$ in arbitrary unit (a.u.) and all the variables are normalized w.r.t. $m$.]
The function $\operatorname{Re}\Sigma(s)$ vanishes at $\sqrt{s}/m = 1$ and $\sqrt{3}$ as a consequence of the subtractions.
In particular, it is positive between $\sqrt{s}/m = \sqrt{3}$ and the threshold  $\sqrt{s}/m = 2$, where it reaches the  value
\begin{equation}
\Sigma(s=4m^{2})=\frac{3\pi+i\ln\left(  \frac{1728(3-i\sqrt{3})^{3}%
}{(-36-12i\sqrt{3})^{3}}\right)  }{64\sqrt{3}\pi^{2}}=\frac{1}{64\sqrt{3}\pi} \approx 0.0028715\text{ .}%
\end{equation}
In this way a bound state, if existent, has a mass within  $(\sqrt{3}m,2m)$. Above threshold, $\operatorname{Re}\Sigma(s)$  decreases and becomes negative at large $\sqrt{s}/m$. Conversely, the imaginary part $\operatorname{Im}\Sigma(s)$  is zero (or better infinitesimally small) below the threshold, while above the threshold it increases according to Eq. (\ref{imsigma0}). It is also useful to re-express the imaginary part $\operatorname{Im}\Sigma(s)$ as:
\begin{equation}
\operatorname{Im}\Sigma(s)=\left\{
\begin{array}
[c]{c}%
\frac{1}{2}\frac{\sqrt{\frac{s}{4}-m^{2}}}{8\pi\sqrt{s}} + \epsilon\text{  for }s>\left(
2m\right)  ^{2}\\
\epsilon\text{ for }s<\left(  2m\right)  ^{2}%
\end{array}
\right.  \text{ ,} 
\label{imsigma}%
\end{equation}
where $\epsilon$ is an infinitesimal positive quantity, 
 see Appendix B for proper treatment of this issue.
As discussed later on, this formal point is relevant when extending the phase shift below the threshold upon considering a small but finite $\varepsilon$.

%The real part rises below the threshold, has a cusp at it, and then decreases
%monotonically and becomes negative at large $\sqrt{s}/m$. 
%The imaginary part is zero (infinitesimally small) below the threshold, then
%it rises above it and saturates at larger $\sqrt{s}/m$. The derivative at
%the threshold is infinite.
%Below $\sqrt{s}/m = 1$, $\operatorname{Re}\Sigma(s)$  is slightly positive. Where as it is negative in between $\sqrt{s}/m = 1$ and $\sqrt{3}$. 
 
 The unitarized amplitudes are obtained by a resummation of the  tree-level amplitudes $A_{k}(s)$ by assuming that the loop function of Eq. (\ref{loop}) can be factorized out:
\begin{equation}
A_{k}^{U}(s)=A_{k}(s)+A_{k}(s)\Sigma(s)A_{k}(s)+...=A_{k}(s)+A_{k}%
(s)\Sigma(s)A_{k}^{U}(s) \text{ ,}
\end{equation}
thus a Bethe-Salpeter-type equation (e.g. Ref. \cite{Nieves:1998hp} ) is obtained. Hence, the final expression reads:
\begin{equation}
A_{k}^{U}(s)=\left[  A_{k}^{-1}(s)-\Sigma(s)\right]  ^{-1}\text{ .}\label{eq:Au}
\end{equation} 

It is interesting to notice that, if we would keep only the $s$ channel for the case $l=0$, the tree-level amplitude reduces to $A_{l=0}(s)\simeq\frac{-g^{2}}{s-m^{2}},$ then
the corresponding unitarized version is $A_{l=0}^{\text{U}}(s)\simeq\frac{-g^{2}}%
{s-m^{2}+g^{2}\Sigma(s)}$ with $\Sigma(s)$ is the loop function given in\ Eq.
(\ref{loop}). Thus, the resummed propagator in the $s$-channel emerges. Yet,
this is not a good approximation here, since the $t$ and $u$ channels are
relevant, especially close to the threshold and for the eventual emergence of a
bound state. The situation is different when two scalar fields are considered, in which one of them represents a resonance: in that case, the $s$-channel might be a good approximation, see Sec. IV.

Once the unitarized amplitudes are determined, the unitarized phase shift is calculated as:
\begin{equation}
\frac{e^{2i\delta_{l}^{U}(s)}-1}{2i}=\frac{1}{2}\cdot\frac
{k}{8\pi\sqrt{s}}A_{l}^{U}(s)\text{ ,} \label{deltalU}%
\end{equation}
hence:
\begin{equation}
\delta_{l}^{U}(s)=\frac{1}{2}\arg\left[  1+i\frac
{k}{8\pi\sqrt{s}}A_{l}^{U}(s)\right]  . \label{eq:ps_unitarized}%
\end{equation}

Note, one can calculate $\delta_{l}^{U}(s)$ by using the equivalent
expressions%
\begin{align}
\delta_{l}^{U}(s)  &  =\frac{1}{2}\arcsin\left[  \frac{k}{8\pi\sqrt{s}%
}\operatorname{Re}\left[  A_{l}^{U}(s)\right]  \right]  =\frac{1}{2}\arccos\left[  1-\frac{k}{8\pi\sqrt{s}%
}\operatorname{Im}\left[  A_{l}^{U}(s)\right]  \right]  \text{ .} \label{delta0Uarccos}%
\end{align}
In fact, when unitarization is preserved, the expressions in Eq. (\ref{delta0Uarccos}) and (\ref{eq:ps_unitarized}) give rise to the same result for the phase shift.
This is also a practical useful check of the validity of unitarity.
The unitarization of amplitudes that contains a sizable contribution
of $t$ and $u$ channel exchanges is discussed in various works, especially in
the domain of pion-pion (or other hadronic) scattering phenomena
\cite{Black:2000qq,Dobado:1992ha,Delgado:2015kxa,Oller:1997ng,Guo:2006br,Gulmez:2016scm,Oller:2020guq,Giacosa:2021brl}. 
In
practice, any unitarization simplifies in some form the underlying
Lippmann-Schwinger equations. The on-shell approximation used here
means that the $t$ and the $u$ channels are evaluated on-shell, implying that our
results for the formation of a bound state are acceptable when the coupling constant $g$ is not too large, see next Section.
Namely, this unitarization does not describe properly the left-hand cut that
starts at $s=3m^{2}$.

The unitarized scattering length reads: 
%\begin{equation}
%a_{0}^{U,\text{SL}}=\frac{1}{2}\frac{1}{16\pi m}\frac
%{1}{A_{0}^{-1}(4m^{2})-\Sigma(4m^{2})}=\frac{1}{32\pi %m}\frac{1}{\frac{3m^{2}%
%}{5g^{2}}-0.0028715\text{.}}\text{.}%
%\end{equation}
\begin{equation}
a_{0}^{U,\text{SL}}=\frac{1}{2}\frac{1}{16\pi m}\frac
{1}{A_{0}^{-1}(4m^{2})-\Sigma(4m^{2})}=\frac{1}{32\pi m}\frac{1}{\frac{3m^{2}%
}{5g^{2}}-\frac{1}{64\sqrt{3}\pi}\text{.}}\text{.}%
\end{equation}

The critical value of $g$ is given by:
%\begin{equation}
%\frac{3m^{2}}{5g^{2}}-0.0028715=0\rightarrow\frac{g_{c}^{2}}{m^{2}}=\frac
%{3}{5\cdot0.0028715}=208.95\text{ .}%
%\end{equation}
\begin{equation}
\frac{3m^{2}}{5g^{2}}-\frac{1}{64\sqrt{3}\pi}=0\rightarrow\frac{g_{c}^{2}}{m^{2}}= \frac{192 \sqrt{3}\pi}{5} \approx 208.95\text{ ,}%
\end{equation}
thus
\begin{equation}
\frac{g_{c}}{m} \approx 14.4551 \text{ .}
\end{equation}
For $g=g_{c}$ the divergence of the scattering length signalizes the emergence of a bound state just at the threshold, see next subsection.

 When $g$ approaches $g_{c}$ the tree-level scattering length becomes less and
less accurate. While for $g/m=1$ the ratio of the tree-level and the unitarized scattering length is $a_{0}^{\text{T,SL}}/a_{0}%
^{\text{U,SL}}=0.995,$ already for $g/m=5$ one gets $a_{0}^{\text{T,SL}%
}/a_{0}^{\text{U,SL}}=0.88.$ By further increasing the coupling to $g/m=10$
implies $a_{0}^{\text{T,SL}}/a_{0}^{\text{U,SL}}=0.52,$ which corresponds to
a sizable underestimation w.r.t. the unitarized result. For values
exceeding $g_{c}$ the tree-level results are not meaningful any longer, since no bound state forms. These
numbers, together with the previously presented Fig. 1, show that a unitarization is required if $g/m$ of the order of 10-20 needs to be investigated.

\begin{figure}[ptb]
\centering
\includegraphics[width=0.48\textwidth]{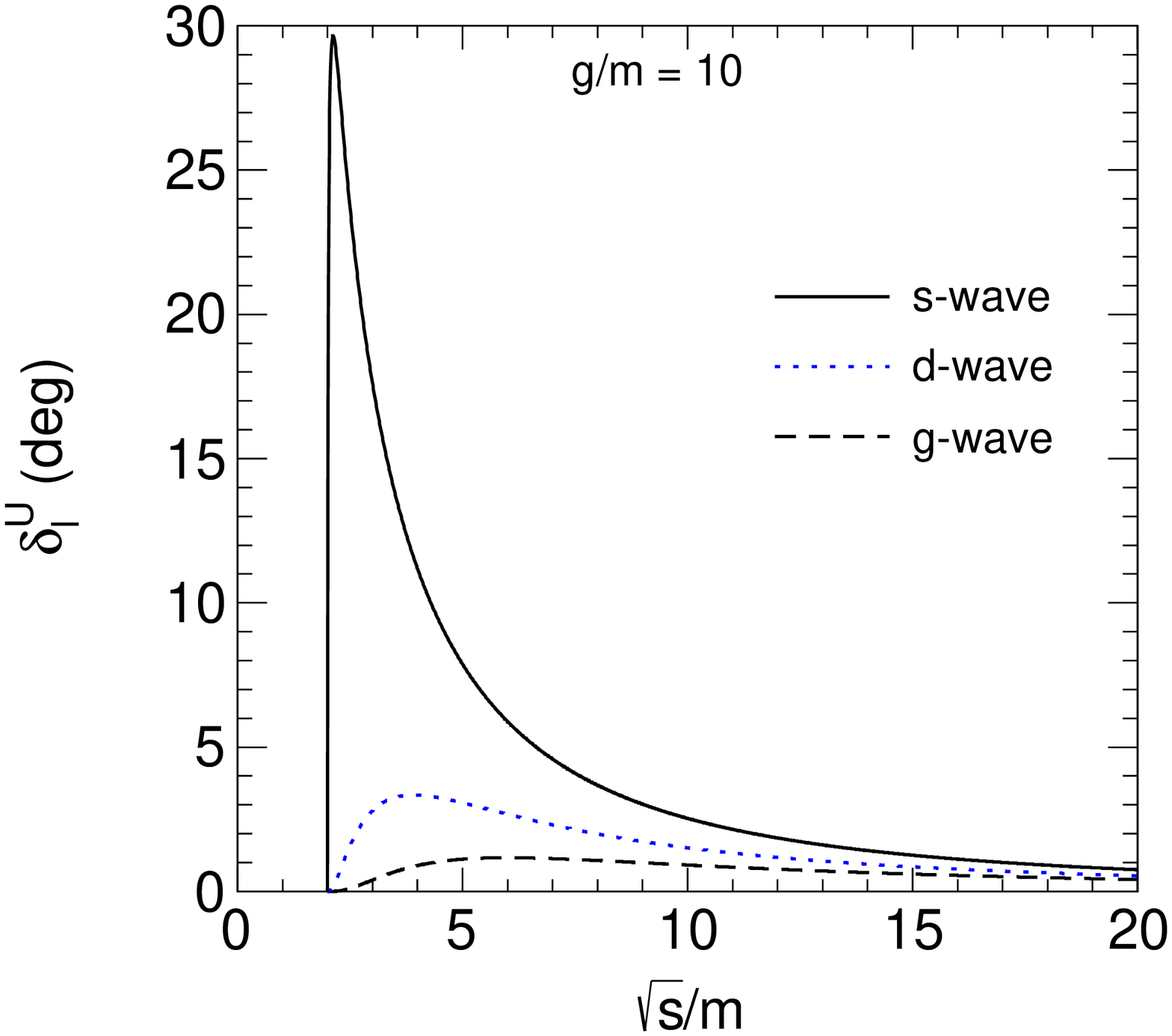}
\includegraphics[width=0.48\textwidth]{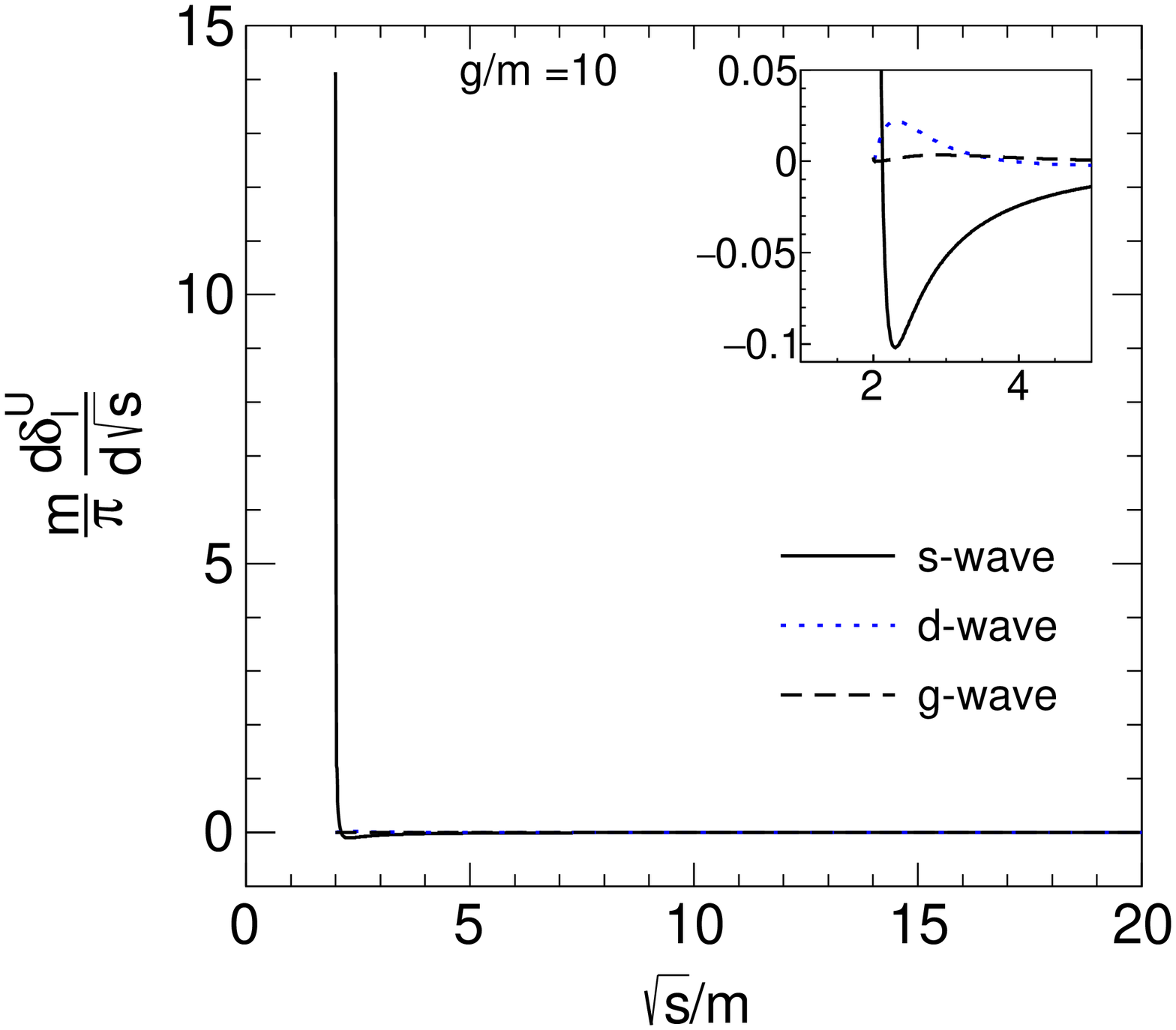}
\caption{(Left) Energy dependence of the phase shifts (using Eqs. \ref{eq:Au} and \ref{eq:ps_unitarized}) for $l = 0, 2, 4$ at $g = 10m<g_c$. (Right) Energy dependence of derivatives of the phase shifts.}
\label{Fig:ps_der_ps_g_10}
\end{figure}
 
Before going into the details of the numerical results of the phase shifts, let us mention the convention adopted in this work. We impose that the phase shifts for any partial waves vanish at the threshold:
\begin{equation}
    \delta_{l}^{U}(s = 4m^2) = 0 \text{ .}
\end{equation}
Sometimes a different convention is used, according to which the phase space at the threshold equals $n_{B}\pi$, where $n_{B}$ is the number of bound states below the threshold \cite{Taylor}. Since the physical quantities are related to the difference and derivative of the phase shifts, the results are unaffected by the choice of the convention of the phase shift at the threshold.

The left panel of Fig. \ref{Fig:ps_der_ps_g_10} shows the energy dependence of the unitarized phase shifts for s-wave ($l = 0$), d-wave ($l = 2$), and g-wave ($l = 4$) at the coupling $g = 10m<g_c\approx 14.45m$. 
The s-wave phase shift increases rapidly just above the threshold $\sqrt{s} = 2m$ and then decreases approaching zero at large energies. 
The d-wave and g-wave show similar behavior, yet their magnitudes are smaller than that of the s-wave. (Note, the degeneracy factor ($2l+1$) is not displayed in Fig. 2). 
The corresponding  phase shift derivatives are depicted in the right panel of Fig. \ref{Fig:ps_der_ps_g_10} .
%of Fig. $\sqrt{s}/m$ derivative of s-wave phase shift takes a large positive value. It then decreases sharply and becomes slightly negative as $\sqrt{s}/m$ increases and reached a minimum afterwards. It starts increasing again becomes zero at large $\sqrt{s}/m$. With the increase of $\sqrt{s}/m$, derivatives of $d$ and $g$- waves first increase and then start decreasing, even become negative and then start increasing again, eventually go towards zero at large $\sqrt{s}/m$. Magnitude wise $d$- wave is smaller than that of s-wave and $g$- wave is even smaller.

\begin{figure}[ptb]
\centering
\includegraphics[width=0.48\textwidth]{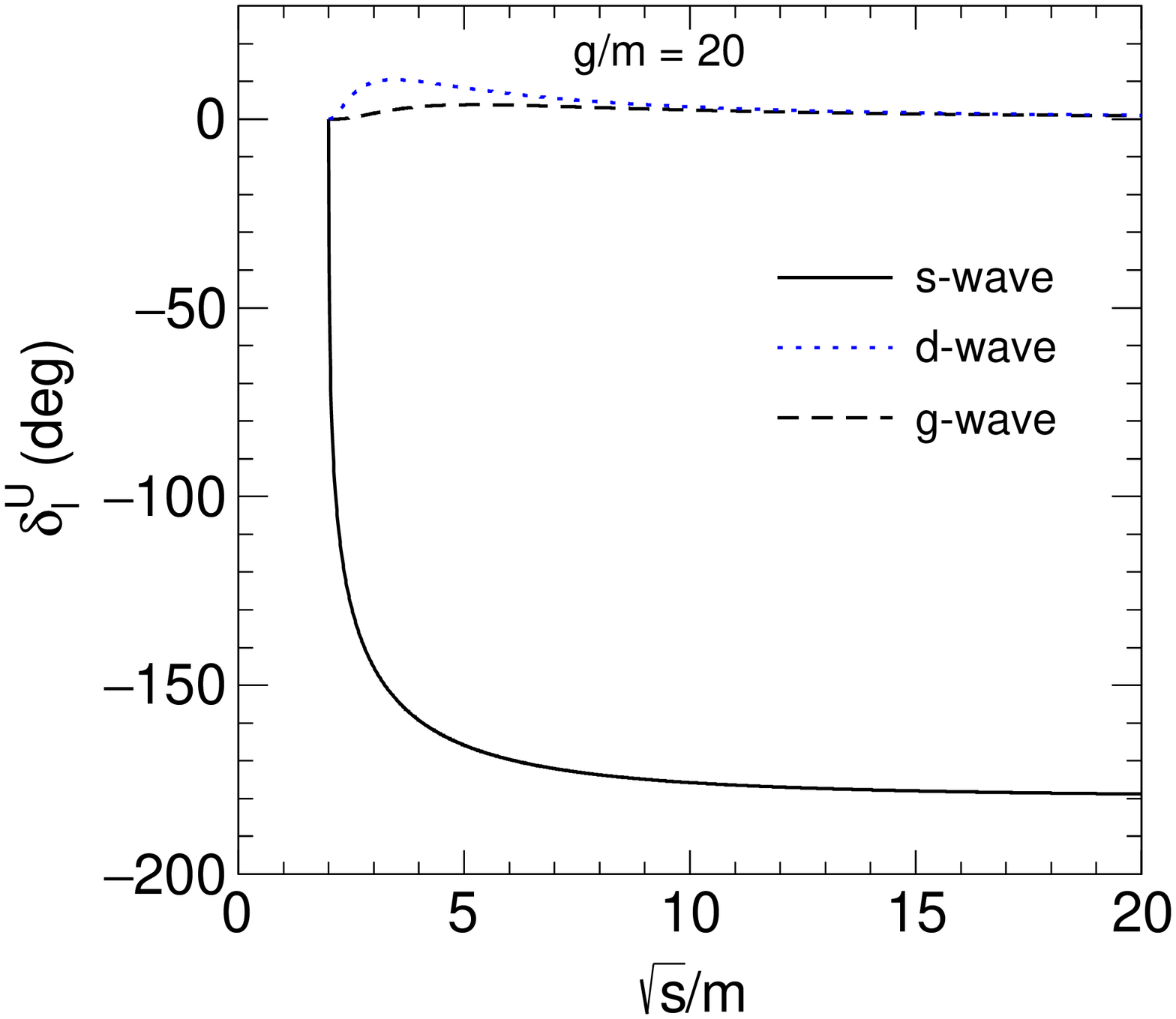}
\includegraphics[width=0.48\textwidth]{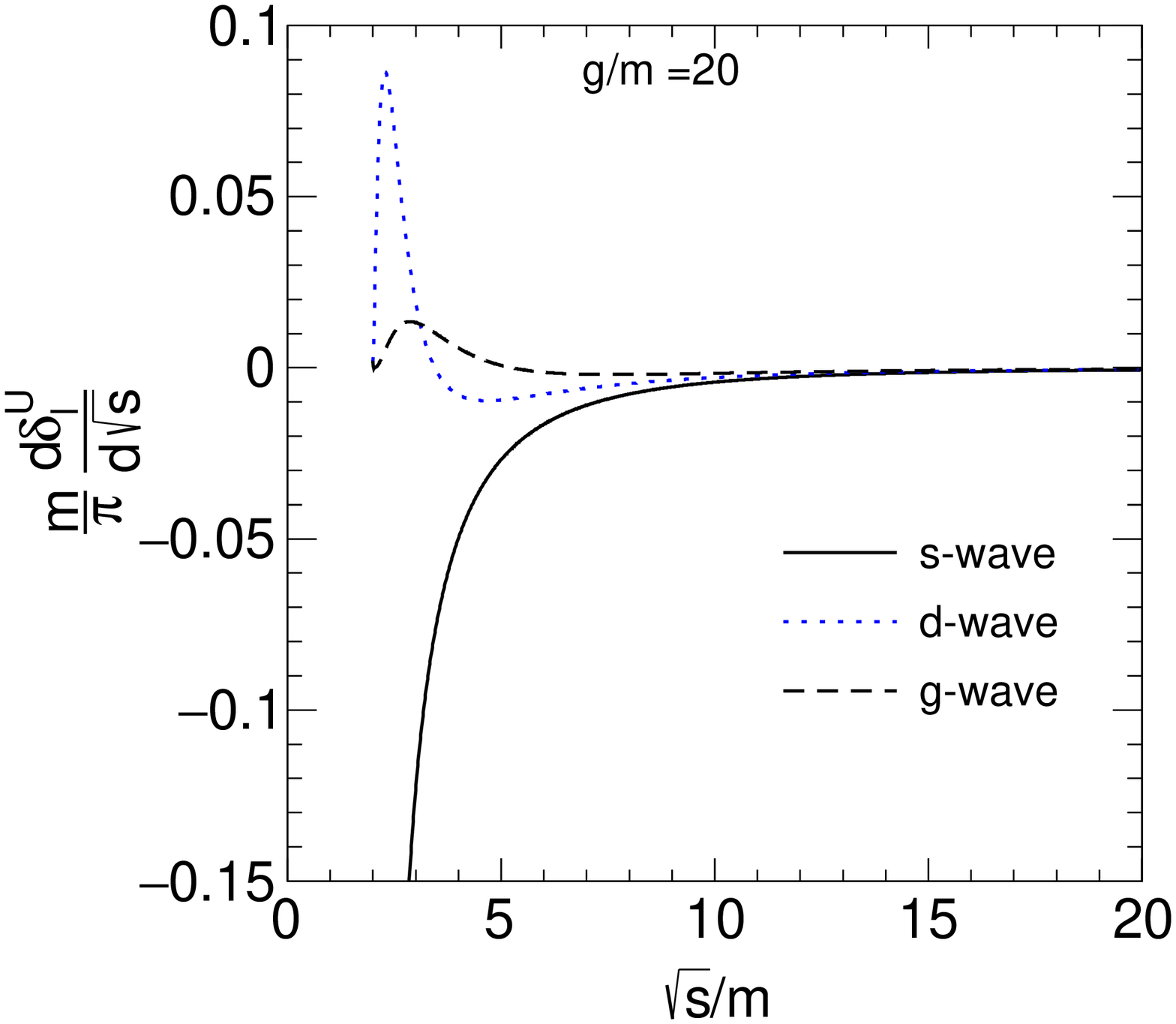}
\caption{Energy dependence of the phase shifts (using Eqs. \ref{eq:Au} and \ref{eq:ps_unitarized}) for $l = 0, 2, 4$ at $g = 20m>g_c$. (Right) Energy dependence of the derivatives of the phase shifts.}
\label{Fig:ps_der_ps_g_20}
\end{figure}

Figure \ref{Fig:ps_der_ps_g_20} shows the same quantities as Fig. \ref{Fig:ps_der_ps_g_10} but for $g = 20m>g_c$. The s-wave phase shift, shown in the left panel, decreases rapidly above the threshold and saturates at $-\pi$ at large $\sqrt{s}/m$. 
%Note that s-wave phase shift was positive in the previous case which we showed in Fig.  \ref{Fig:ps_der_ps_g_10}. 
This behavior indicates the presence of a bound state below the threshold, see the next subsection. 
%In fact, we will see later that one bound state is formed when $g \ge g_c$. 
The phase shifts of d- and g-waves are similar to the previous case. 
%waves first increase with increase of $\sqrt{s}/m$ and then slowly decrease to zero at high $\sqrt{s}/m$. Magnitudes are also smaller than that of s-wave. Derivatives of the phase shifts are shown in the right panel of Fig. \ref{Fig:ps_der_ps_g_20}. 
The right panel shows the derivatives of the phase shifts, where it is visible that the s-wave contribution is sizable.

 The phase shift in Eq. (\ref{deltalU}) is defined only for $s\geq4m^2.$ Yet,
following the discussion of Ref. \cite{Samanta:2020pez}, upon using Eq. (\ref{imsigma}) one can
extend the phase shift also below the threshold by considering the expression upon considering an arbitrarily small but nonzero $\varepsilon$:%
\begin{equation}
\delta_{l}^{U}=\frac{1}{2}\arg\left[  1+2iA_{l}^{U}\operatorname{Im}%
\Sigma\right]  \text{ }.\label{deltabt0}%
\end{equation}
Above the threshold, we recover Eq. \ref{eq:ps_unitarized}. Below the threshold, the equation
$\operatorname{Im}\Sigma=\epsilon$ implies that $\delta_{l}^{U}=n\pi=const$
for any finite $A_{l}^{U}$ (the choice $n=0$
guarantees that the phase shift is continuous at the threshold). Yet, the situation is different if pole(s) of
$A_{l}^{U}$ appear(s) (indeed, the single-particle pole below the threshold
for $s=m^{2}$ is always present in the $\varphi^3$-theory), see the next Section.

 Here, we briefly discuss the intuitive meaning of the extension above. The $\varepsilon$ prescription makes the
particle $\varphi$ `slightly unstable', because the pole is realized for $%
s=m^{2}-i\varepsilon /2$, see Appendix B. Since the particle $\varphi$ is unstable and its
width is proportional to $\varepsilon$, its mass distribution is not exactly a
Dirac delta (it becomes such for $\varepsilon \rightarrow 0$). Then, the
scattering below the threshold $2m$ is possible (even though very small and
vanishing for $\varepsilon \rightarrow 0$ as expected). The bound state, if
existent, is itself slightly unstable, the width is also proportional to $%
\epsilon \propto \varepsilon$ (see the next subsection), and thus can be seen as a resonance produced in a scattering process. In this way, as discussed in Sec. III.A an alternative view is possible to understand the inclusion of the bound state and one may show that the pressure as a function of the coupling is continued when the latter forms. Yet, the result is not dependent on $\varepsilon$ in the limit $\varepsilon \rightarrow 0$.

\bigskip

 We conclude this subsection with some general considerations. As discussed above, Eq. (\ref{imsigma}) does not fix the real part of the loop. The two
subtractions used for the determination of the loop function $\Sigma(s)$ of Eq. (\ref{loop}) guarantee convergence and the
absence of unphysical states, but any number of subtractions would be
consistent with Eq. (\ref{imsigma}), even though it would be hard to physically
justify a choice with many subtractions. Alternatively, one may also include
a form factor by modifying the imaginary part itself, $\operatorname{Im}%
\Sigma(s)=\frac{1}{2}\frac{k}{8\pi\sqrt{s}}e^{-k^{2}/\Lambda^{2}}$ with
$k=\sqrt{\frac{s}{4}-m^{2}}.$ Such form factors are sometimes used in hadronic
theories (e.g. \cite{Amsler:1995td,Faessler:2003yf,Giacosa:2007bn}) since they model the finite extension of
hadrons. In this case, the real part is finite even without applying a subtraction (for technical details, see e.g. Ref. \cite{Coito:2019cts}). 

Indeed, various unitarization approaches have been explored in the literature, e.g. Refs. 
\cite{Oller:2020guq,Mai:2022eur};
some of them go beyond the on-shell approximation used here. One interesting unitarization scheme that was widely used in the past
is the so-called $N/D$ approach \cite{Frazer:1969euo,Hayashi:1967bjx,Gulmez:2016scm,Oller:2020guq,Cahn:1983vi,Mai:2022eur,Giacosa:2021brl}, 
where $N$ stays for the numerator and $D$ for the
denominator. The basic idea is that the unitarized amplitude can be written as
a ratio of functions, where the numerator contains the left-hand cut and the
denominator the right-hand cut. In Appendix B we present this unitarization
scheme (for simplicity, at its lowest order) for the
$\varphi^{3}$-case. 
In this way, we can compare the results of this
alternative unitarization scheme (phase shifts, the critical value of the coupling for the
emergence of a bound state and the mass of the latter as a function of $g$, as well as the
behavior of the pressure) with the results presented in the main text. As the outcomes show, the overall qualitative picture is
left unchanged, which makes us confident that the features that we present are not just inherent to the employed unitarization approach.

\subsection{Bound state}
\begin{figure}[ptb]
\centering
\includegraphics[width=0.48\textwidth]{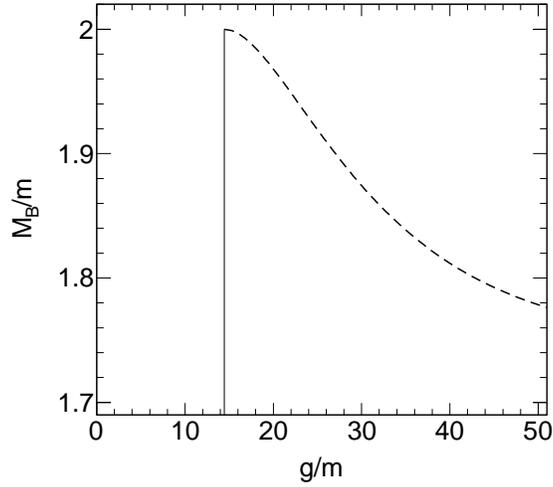}
\caption{Dependence of 
%Variation of 
the bound state mass (Eq. \ref{bs}) with the coupling $g$. The vertical line indicates the critical coupling $g_c$, for which $M_B=2m$.}
\label{Fig:mbs}
\end{figure}
We describe here the emergence of the bound state that takes place when the attraction is sufficiently strong.
The bound state equation ($s$-channel, for the Mandelstam variable $s$ continued below the threshold) corresponds to a pole of the amplitude in Eq. \ref{eq:Au}:
\begin{equation}
\left[  A_{0}^{-1}(s)-\Sigma(s)\right]  =0 \text{ .}
\label{bs}
\end{equation}
In Fig. \ref{Fig:mbs} the mass of the bound state $M_B/m$ is plotted as a function of the coupling $g$. For the critical value $g=g_c+\epsilon \approx 14.45 m$ the bound state forms exactly at threshold, and for $g>g_c$ it ranges between $\sqrt{3} < \sqrt{s}/m < 2$, in which $\operatorname{Re}\Sigma(s)>0$. Note that when a bound state is close to threshold forms, it can be also understood via non-relativistic approaches, see e.g. Ref. \cite{Baru:2003qq,Hayashi:1967bjx,Dong:2008mt}.
%Note, no bound state is observed in d- and g-waves, the reason being that the tree-level amplitude is non-zero at threshold only in case of the s-wave.

In particular, the limit $M_{B}(g\rightarrow\infty)=\sqrt{3}m$ holds. Thus,
the results are finite also for arbitrarily large values of $g$. However, this
property is a consequence of the employed unitarization but is not
physical. 
Namely, one could choose other unitarization approaches that deliver different results for the limit
$g\rightarrow\infty$. In the already mentioned $N/D$ method (see Appendix B), the bound state forms for a similar critical value of the coupling constant ($g_c=13.02m$) and the behavior of the bound state as a function of $g$ well compares to Fig. 5 for $g$ up to (and even slightly above) $20m$, but then the results deviate from each other. In particular, in the $N/D$ approach $M_{B}(g\rightarrow\infty)=m$. This is also an artifact of that particular method with the constraints described in Appendix B.

In general, different unitarization approaches usually come up with similar values for the critical coupling for which the bound state emerges, but depart from each other when the attraction is too strong. Actually, for $g$ very large one should encounter a
solution of the type $M_{B}^{2}<0$ that signalizes an instability, as the
potential of Eq. \ref{pot} suggests, but this feature is not described by
the twice-subtracted on-shell unitarization approach used in this work.

Summarizing our discussion, the plateau $M_{B}(g \rightarrow \infty) = \sqrt{3}m$, see also Fig. \ref{Fig:mbs}, is an artifact of our
unitarization. For this reason, we consider a value of $g/m$ of about $40$
(before the plateau sets in) as an upper limit of our approach
(in our numerical examples, we shall actually limit our
studies to $g/m\lesssim20$). Thus, the introduction of unitarization
allows us to go further than what the tree-level results do, yet not to
arbitrarily large coupling constants. More advanced unitarization approaches could
extend the range of $g$, but this is left for the future. 
%For our purposes that intend to study the role of the bound state close to threshold, the given interval is sufficient.

\begin{figure}[ptb]
\centering
\includegraphics[width=0.48\textwidth]{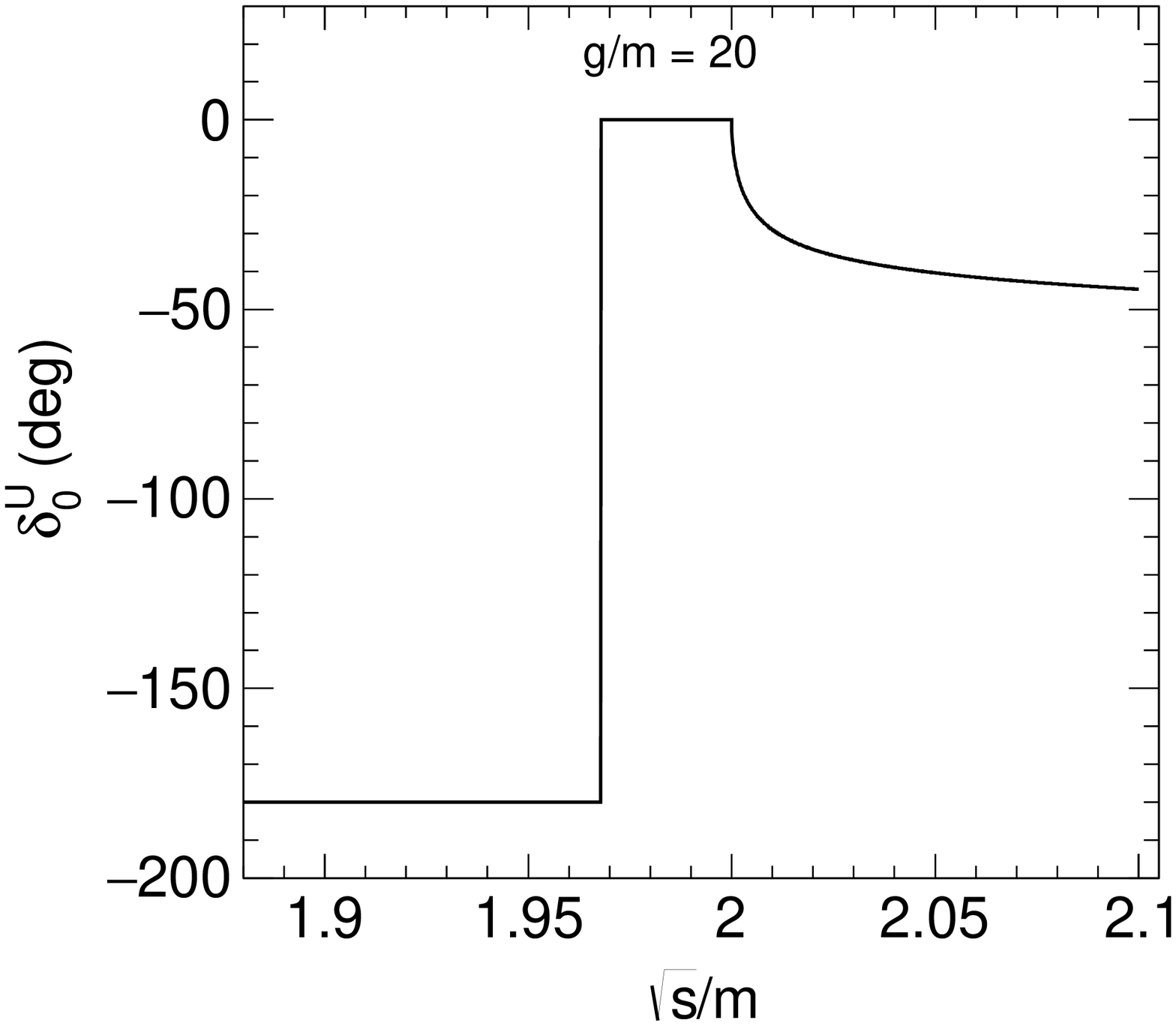}
\includegraphics[width=0.48\textwidth]{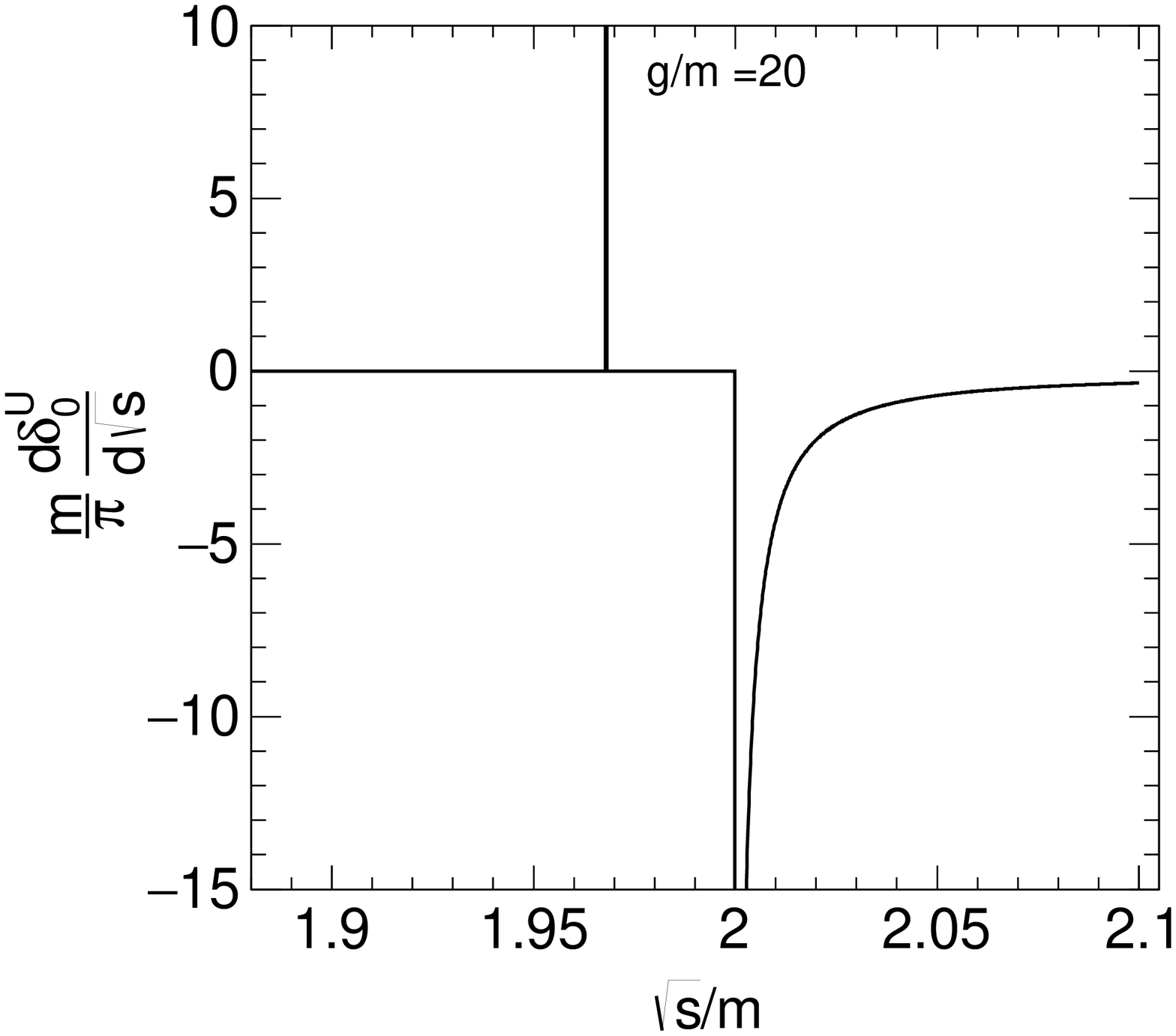}
\caption{(Left) Energy dependence of the s-wave phase shift at $g = 20m>g_c$ below and above the threshold. One may note the jump from $-\pi/2$ at the bound state mass $M_B$. (Right) Corresponding energy dependence of the derivative of the phase shift. At $\sqrt{s} = M_B$ a Dirac-delta function appears. }
\label{Fig:ps_der_ps_bs_g_20}
\end{figure}

 In\ Fig. \ref{Fig:ps_der_ps_bs_g_20} we present the s-wave phase shift and its derivative by using the
extended version of Eq. (\ref{deltabt}) that contains also the continuation below the threshold. 
This is indeed an important extension since it allows to accommodate the bound state in the phase-shift formula. In order to discuss these aspects,
we recall that for $s<4m^2$:
\begin{equation}
A_{l}^{U}\operatorname{Im}\Sigma=\frac{\epsilon}{A_{0}^{-1}-\Sigma}%
=\frac{\epsilon}{A_{0}^{-1}-\operatorname{Re}\Sigma-i\operatorname{Im}\Sigma
}=\frac{\epsilon}{A_{0}^{-1}-\operatorname{Re}\Sigma-i\epsilon} \text{ .}%
\end{equation}
For $g<g_{c}$ no bound state occurs, thus $\delta_{l}^{U}(x=\sqrt{s})=0$ for
$\sqrt{3}m\leq x\leq2m.$
For $g>g_{c}$ a bound state occurs for a certain $s=M_{B}^{2}$ belonging to the interval $(3m^2,4m^2)$, thus
\begin{equation}
\left(  A_{l}^{U}\operatorname{Im}\Sigma\right)  _{s=M_{B}^{2}}=\frac
{\epsilon}{-i\epsilon}=i\text{ ,}\label{eqatbs}%
\end{equation}
which in turn shows why it is important to keep track of the $i\epsilon$ factors in
the energy region below the threshold. It then follows that $\delta_{l}%
^{U}(x=M_{B})=n\frac{\pi}{2}.$ Upon requiring (for $\epsilon$ small but finite) a continuous phase shift, $\delta_{l}^{U}(x)$ for $\epsilon
\rightarrow0$ reads%
\begin{equation}
\delta_{l}^{U}(x=\sqrt{s})=-\pi+\pi\theta(x-M_{B})\text{ for }\sqrt{3}m \leq
x\leq 2m 
\text{ and }
g>g_c
\text{ ,}
\label{deltabt}
\end{equation}
whose derivative is a Dirac delta: $\pi\delta(x-M_{B}).$ 
In a certain sense, this result is quite intuitive: even though the physical
range is realized for $s\geq4m^{2},$ when the continuation to the region below
the threshold is performed the bound state corresponds to a very narrow object
(width proportional to $\epsilon$) that is encountered along the $s$-axis and
implies an increase of the phase shift of $\pi$ (see also Sec. 4 for the
conceptually analogous case of a resonance).

Moreover, the amplitude $A_{l}^{U}$ has a second pole for $s=m^{2}$ (the
single-particle pole that is also exchanged in the $s$-channel). Then, also for
this value Eq. \ref{eqatbs} applies. The phase shift
can be further extended below the threshold as:%

\begin{equation}
\delta_{l}^{U}(x=\sqrt{s})=-2\pi+\pi\theta(x-m)+\pi\theta(x-M_{B})\text{ for
}0\leq x\leq2m\
\text{ and }
g>g_c
\text{ ,}%
\end{equation}
whose derivative is $\pi\delta(x-m)+\pi\delta(x-M_{B})$, hence including both
the particle $\varphi$ and the bound state $B.$

%In the left panel of Fig. \ref{Fig:ps_der_ps_bs_g_20} we show the energy dependence of the s-wave phase shift for $g = 20 m > g_c$, in which, in addition to Fig. \ref{Fig:ps_der_ps_g_20}, also the sub-threshold range is displayed. 
%For $\sqrt{s} \approx 1.968 = M_B$ there is a sudden jump of the phase shift from -180 degree to zero (set as the conventional value at threshold). 
%As a consequence, there is a delta function in the derivative of the phase shift shown in the right panel of Fig. \ref{Fig:ps_der_ps_bs_g_20} .

%Note, Fig. \ref{Fig:ps_der_ps_bs_g_20} is consistent with the Levinson theorem \cite{Hartle:1965nj,levinson},  according to which the number of poles
%below threshold equals to the difference of the phase shift at infinity and at threshold:
%\begin{equation}
%n_{\text{poles-below-threshold}}=\frac{1}{\pi}\left(  \delta_{0}%
%^{U}(s\rightarrow\infty)-\delta_{0}^{U}(s=4m^{2})\right)  \text{ .}%
%\end{equation}

In the end, we have verified that the phase shifts studied above fulfill the so-called causality Wigner condition in the form discussed in Ref. \cite{Boglione:2002vv}: \begin{equation}
\frac{d\delta_{l}}{dx}\geq-\frac{x^{2}}{8\left(  \frac{x^{2}}{4}-m^{2}\right)
}\text{  ,}\label{wc}%
\end{equation}
see Appendix C for more details and related plots.

\section{Thermodynamic properties of the $\varphi^3$-QFT}\label{sec:FiniteTemp}
In this section, we study the $\varphi^3$-theory at nonzero temperature. In particular, we shall concentrate on the evaluation of the different contributions to the pressure. 
 To this end, we use the phase shift (or S-matrix) formalism \cite{Dashen:1969ep,Venugopalan:1992hy,Broniowski:2015oha,Lo:2017ldt,Lo:2017sde,Lo:2017lym,Dash:2018can,Dash:2018mep,Lo:2019who,Lo:2020phg} (for a brief recall see Appendix A), which allows calculating the pressure (or any other thermodynamic quantity) from the vacuum's phase shifts. Intuitively,  the idea behind this approach is that the energy levels $E_n$ entering into the partition function 
$Z=\sum_{n}e^{-E_{n}/T}$ are determined in the vacuum. Basically, the derivatives of phase shifts are proportional to the density of states that enter in $Z$. 

In this respect, this approach is different from thermal field theory \cite{Kapusta:2006pm} (see for instance works involving the CJT formalism as well as other thermal approaches in Refs. \cite{Rischke:2003mt,Brambilla:2014jmp,Lenaghan:1999si,Papazoglou:1996hf,Amelino-Camelia:1996sfy,Tolos:2008di,GomezNicola:2002tn,Carrington:1999bw} and refs. therein), since one needs only vacuum results to obtain thermodynamic quantities. 
The knowledge of the latter is relatively simple if one considers only elastic two-body scattering, but becomes more and more complicated when inelastic processes and  multiple channels become relevant \cite{Lo:2020phg}. 
Thus, the S-matrix approach is expected to be valid for (relatively) low temperatures. 

In practice, we shall evaluate the (various contributions to the) pressure of the system as a function of $g$ for fixed $T$ and as a function of $T$ for selected $g$. Then, we shall discuss deviations from the simple free gas results.

\subsection{Nonzero-$T$ formalism}
The non-interacting part of the pressure for gas of interacting particles
with mass $m$ reads:%
\begin{equation}
P_{\varphi \text{,free}}=-T\int_{k}\ln \left[ 1-e^{-\beta \sqrt{k^{2}+m^{2}}}%
\right] \text{ ,}
\end{equation}%
where $\int_{k}\equiv \int d^{3}k/(2\pi )^{3}$ and $\beta =1/T$. Let us then
include interaction and, at first, assume that no bound state forms. In the
scattering-matrix or the $S$-matrix formalism \cite%
{Dashen:1969ep,Venugopalan:1992hy,Broniowski:2015oha,Lo:2017ldt,Lo:2017sde,Lo:2017lym,Dash:2018can,Dash:2018mep,Lo:2019who}%
, the interacting part of the pressure is related to the derivative of the
phase shift with respect to the energy by the following relation: 
\begin{equation}
P_{\varphi \varphi \text{-int}}^{U}=-T\int_{2m}^{\infty }dx\frac{2l+1}{\pi }%
\sum_{l=0}^{\infty }\frac{d\delta _{l}^{U}(s=x^{2})}{dx}\int_{k}\ln \left[
1-e^{-\beta \sqrt{k^{2}+x^{2}}}\right] =\sum_{l=0}^{\infty }P_{\varphi
\varphi \text{-int},l}^{U}\text{ ,}  \label{eq:p_int}
\end{equation}%
where $x=\sqrt{s}$. In the previous equation, the usual thermal integral's
contribution for gas of particles with running mass $x$ is weighted by the
vacuum's phase shift derivatives, see Appendix A. Eq. \ref{eq:p_int} shows
that the contribution to the pressure of a certain wave is positive if $%
\frac{d\delta _{l}}{dx}>0$ and negative if $\frac{d\delta _{l}}{dx}<0.$ For $%
x$ close to the threshold, these two cases correspond to attraction and
repulsion, respectively (but this is not true in general). If, for instance, 
$\delta _{l}$ increases from the threshold up to a certain maximum and then
decreases to zero for large $x,$ the overall contribution is expected to be
positive, since low values of $\sqrt{s}=x$ are the dominant ones in the
thermal integrals.

It may be at first sight puzzling that an attraction generates an increase
of pressure, since one is rather used to thinking in terms of a classic
Van-der-Walls gas of particles in which (for a fixed number of particles) an
attraction implies a smaller pressure (and vice-versa, a repulsion means a
larger pressure). Yet, our case here is different: an attraction implies
that the density of states increases (roughly speaking, more states are
present), thus the pressure becomes larger (the opposite applies to a
repulsion).

Summarizing, the total pressure reads%
\begin{equation}
P_{\text{tot}}^{U}=P_{\varphi \text{,free}}+P_{\varphi \varphi \text{-int}%
}^{U}\text{ .}  \label{ptot1}
\end{equation}%
Above, it should be noted that the free part is calculated at the physical
mass $m$ which, due to suitable resummations, is left unchanged by loop
corrections. In this way, in the interacting case, $P_{\varphi \text{,free}}$
may be interpreted as the pressure of the asymptotic states of the system. 

\bigskip 

Next, let us move to the case in which a bound state exists, that is the
coupling $g$ is larger than the critical value $g=g_{c}$. The following
considerations are in order:

\begin{itemize}
\item The pressure of Eq. (\ref{ptot1}) has a jump as a function of $g$ at $%
g_{c}$ at $a$ given temperature $T$.\ This is due to the different behavior
of the s-wave phase-shift if the bound state is present: the corresponding
contribution changes sign. This abrupt jump in the pressure signalizes that
`something is missing' in Eq. (\ref{ptot1}). It is rather natural to think
that the missing element is indeed the emerging bound state. 

\item The first approach is to consider that a bound state is an additional
asymptotic state of the system. In this way, the bound state with mass $M_{B}
$ is expected to correspond to an additional state that needs to be added to
the expression of Eq. (\ref{eq:p_int}), thus: 
\begin{equation}
P_{B}\overset{\text{ }}{=}-\theta (g_{c}-g)T\int_{k}\ln \left[ 1-e^{-\beta 
\sqrt{k^{2}+M_{B}^{2}}}\right] \text{ ,}  \label{eq:pb}
\end{equation}%
where the theta function takes into account that for $g<g_{c}$ there is no
bound state $B.$ Indeed, this simple and intuitive expression turns out to
be consistent with the regulating procedure adopted in this work, see
below. The total pressure of the system is written as 
\begin{equation}
P_{\text{tot}}^{U}=P_{\varphi \text{,free}}+P_{B}+P_{\varphi \varphi \text{%
-int}}^{U}\text{ .}  \label{eq:ptot}
\end{equation}%
On top of the $\varphi \varphi $ interactions, there are also $B\varphi $
and $BB$ interactions, which however can be neglected in the energy and
temperatures of interest.

\item Another way to justify and obtain the result above goes via the $%
\varepsilon $-driven extension of the phase-shift formula below the
threshold presented in Sec. II.C (see also Appendix B and Ref. \cite%
{Samanta:2020pez}). Namely, the bound state (if it forms) can be seen a
manifestation of the interaction among the $\varphi $ particles and is
interpreted as a very narrow resonance below the threshold. Thus, upon extending
the interaction range in Eq. (\ref{eq:p_int}), the whole interaction
contribution (including the bound state if existent) reads: 
\begin{equation}
P_{\varphi \varphi \text{-int}}^{U}+P_{B}=-T\int_{\sqrt{3}m}^{\infty }dx%
\frac{2l+1}{\pi }\sum_{l=0}^{\infty }\frac{d\delta _{l}^{U}(s=x^{2})}{dx}%
\int_{k}\ln \left[ 1-e^{-\beta \sqrt{k^{2}+x^{2}}}\right] \text{ ,}
\label{pint}
\end{equation}%
where the lowest range of the integral is set to $\sqrt{3}m$, since the
bound state mass belongs to the interval $(\sqrt{3}m,2m)$. The bound state
contribution (if the bound state forms) corresponds to the integral range
between $(\sqrt{3}m,2m)$: 
\begin{align}
P_{B}=-T\int_{\sqrt{3}m}^{2m}dx\frac{2l+1}{\pi }\sum_{l=0}^{\infty }\frac{%
d\delta _{l}^{U}(s=x^{2})}{dx}\int_{k}\ln \left[ 1-e^{-\beta \sqrt{%
k^{2}+x^{2}}}\right]
\nonumber \\
=-\theta (g_{c}-g)T\int_{k}\ln \left[ 1-e^{-\beta \sqrt{%
k^{2}+M_{B}^{2}}}\right] \text{ ,}
\end{align}%
where the r.h.s. is obtained by using Eq. (\ref{deltabt}). The properties
discussed above are actually applicable to any QFT that displays bound
states: if present, they appear as infinitely narrow states that can be
(formally) created by particle scattering extended below the threshold.

\item The two interpretations above (including the bound state as an
additional asymptotic state or as part of the interaction as a narrow
resonance) lead to the same result of Eq. (\ref{eq:ptot}). Yet, in the latter
case, one still has the same set of asymptotic states that coincide with
the states of the theory realized for $g=0$, because the bound state is seen as part of the interaction.  (A subtle difference between
the $g>0$ and the $g=0$ case is present in the model described in\ Sec. IV).

\item As anticipated in the introduction and shown later on in various
examples, the expression in Eq. (\ref{pint}) is continuous as a function of $%
g$ for any fixed value of $T$. This is an additional confirmation of the
consistency of the proposed expression, since a non-analytic point of the pressure as a
function of the coupling constant would not be a physical feature. 
The jump
due to the emergence of the bound state is compensated by a jump in the
contribution to the pressure arising from the particle interaction above the
threshold. Indeed, upon considering a small quantity $\alpha$, one has that%
\begin{equation}
\frac{1}{\pi }\left[ \left( \frac{d\delta _{0}^{U}(x)}{dx}\right)
_{g=g_{c}-\alpha }-\left( \frac{d\delta _{0}^{U}(x)}{dx}\right)
_{g=g_{c}+\alpha }\right] \simeq \delta (x-2m)\text{ ,}
\end{equation}%
thus the jump of the interaction pressure due to the change in the phase
space behavior matches a state just a t threshold:%
\begin{equation}
\left( P_{\varphi \varphi \text{-int}}^{U}\right) _{g=g_{c}-\alpha }-\left(
P_{\varphi \varphi \text{-int}}^{U}\right) _{g=g_{c}+\alpha }=\left(
P_{B}\right) _{M_{B}=2m}\text{ ,}
\end{equation}%
thus showing that the pressure is continuous as a function of $g$.

\item Interestingly, in the particular case of the $\varphi ^{3}$-theory,
one may go even further and describe the total pressure via the phase shift
formula as 
\begin{equation}
P_{\text{tot}}^{U}=-T\int_{0}^{\infty }dx\frac{2l+1}{\pi }\sum_{l=0}^{\infty
}\frac{d\delta _{l}^{U}(s=x^{2})}{dx}\int_{k}\ln \left[ 1-e^{-\beta \sqrt{%
k^{2}+x^{2}}}\right] \text{ ,}
\end{equation}%
in which the lowest range of the integral is set to zero. This is not
possible in general but holds here because of the nature of the $\varphi ^{3}
$ self-interaction: the very same particle $\varphi $ is also exchanged in
the $s$-channel of $\varphi \varphi $ scattering.
\end{itemize}

In the end, we stress that above we have presented a series of plausible
arguments to properly include a bound state in the phase-shift formalism.
Surely a formally more rigorous approach would be needed in the future to fully
clarify the procedure and the issues raised here.

\bigskip

%As a final comment, we stress that the equations presented above shall be valid for all the QFT examples that we are going to investigate. 

\subsection{Numerical results}
Next, we turn to numerical examples and plots that allow showing the properties of the system at nonzero temperature.

\begin{figure}[ptb]
\centering
\includegraphics[width=0.48\textwidth]{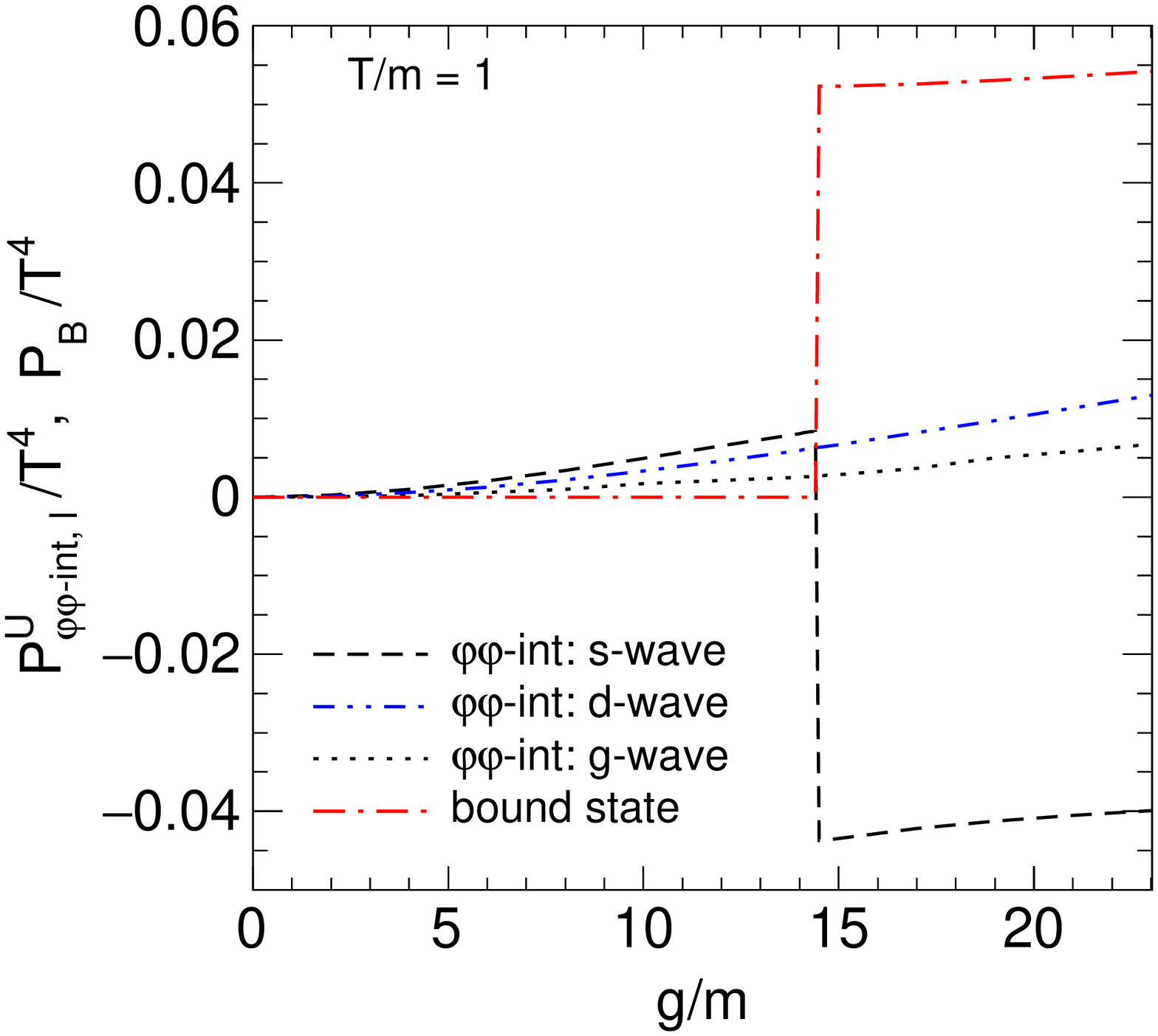}
\includegraphics[width=0.48\textwidth]{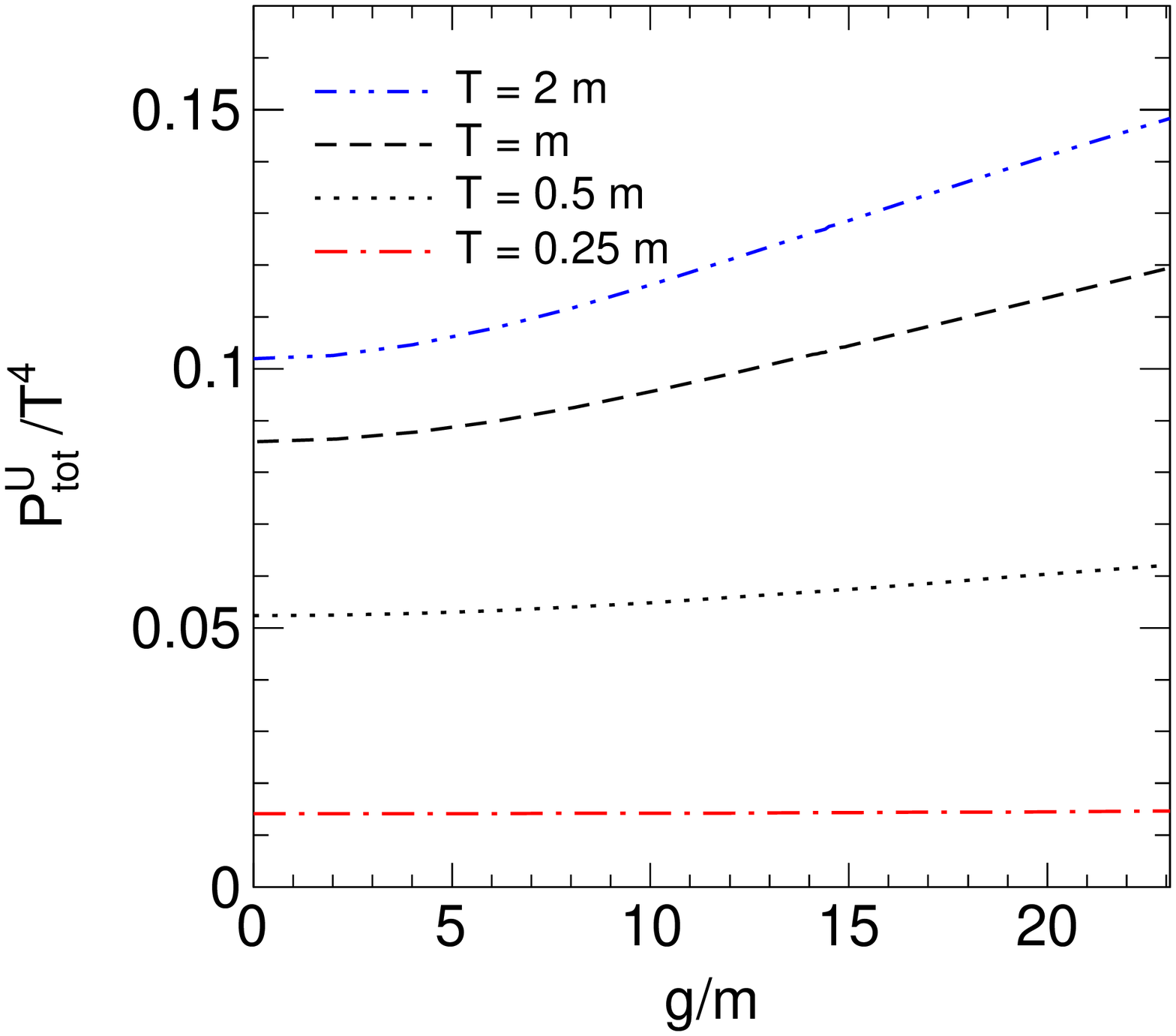}
\caption{Left panel shows $P^{U}_{\varphi\varphi\text{-int}, l}/T^4$ of Eq. \ref{eq:p_int}) as function of $g/m$ for three different partial waves corresponding to $l = 0$ (s-wave), $l = 2$ (d-wave) and $l = 4$ (g-wave) at $T/m =1$. The bound state contribution ($P_B/T^4$) (Eq. \ref{eq:pb}) is also shown. The right panel shows the normalized total pressure (Eq. \ref{eq:ptot}) at four different temperatures.}
%Normalized interacting pressure ($P^{U}_{\varphi\varphi\text{-int}, l}/T^4$ of Eq. \ref{eq:p_int}) as function of $g/m$ for three different partial waves corresponding to $l = 0$ (s-wave), $l = 2$ (d-wave) and $l = 4$ (g-wave) at $T/m =1$. The normalized pressure for bound state (Eq. \ref{eq:pb}) is also shown. }
\label{Fig:pvsg}
\end{figure}

The unitarized pressure  ($P^{U}_{\varphi\varphi\text{-int}, l}/T^4$) is shown in Fig. \ref{Fig:pvsg} as a function of $g/m$ at $T/m = 1$ for the partial waves corresponding to  $l = 0,2,4$.   
%The pressure is normalized with the temperature to make it dimensionless. 
%The interacting pressure is zero at $g = 0$ for all three waves since at this point there is no interaction. 
The s-wave is interesting: up to $g = g_c \approx 14.45 m$,  increases, then it abruptly jumps to negative values. The reason is that above $g_c$ a bound state exists. 
The normalized pressure ($P_B/T^4$) for the bound state (see Eq. \ref{eq:pb}) is also shown: it is zero below $g_c$ and nonzero (and positive) above this value. The jump has the same magnitude but the opposite sign of the s-wave interacting channel. Moreover, the normalized total pressure $P_{tot}^{U}/T^4$ with $g/m$ as evaluated via Eq. \ref{eq:ptot}, shown in the right panel of Fig. \ref{Fig:pvsg}, 
 varies continuously with $g$ in all four temperatures shown in this figure. This fact confirms one of the main outcomes of the paper: when interactions are taken into account, the formation of a new state does not correspond to any sudden jump in the pressure or energy density of the system.
%With a further increase of $g/m$, $P_B/T^4$ increases because of the decrease of mass of the bound state. $P^{U}_{\varphi\varphi\text{-int}, l}/T^4$ for $d$ and g-waves increase monotonically with the increase of $g/m$ within the range shown in this figure.

%We show results for four different $T/m$ ratio. 
%We observe that the discontinuity in the s-wave pressure (for $T/m = 1$ see Fig. \ref{Fig:pvsg}) is exactly compensated by the bound state pressure above $g = g_c$. As a result the total pressure as a function $g$ is continuous even at the critical value $g = g_c$. 
%Further, as expected, with increase of $T/m$, normalized total pressure increases.

%\begin{figure}[ptb]
%\centering
%\includegraphics[width=0.48\textwidth]{p_tot_vs_g_diff_T.eps}
%\caption{Normalized total pressure (Eq. %\ref{eq:ptot}) as function of $g/m$ at fixed $T/m$.}
%\label{Fig:ptotvsg}
%\end{figure}

\begin{figure}[ptb]
\centering
\includegraphics[width=0.48\textwidth]{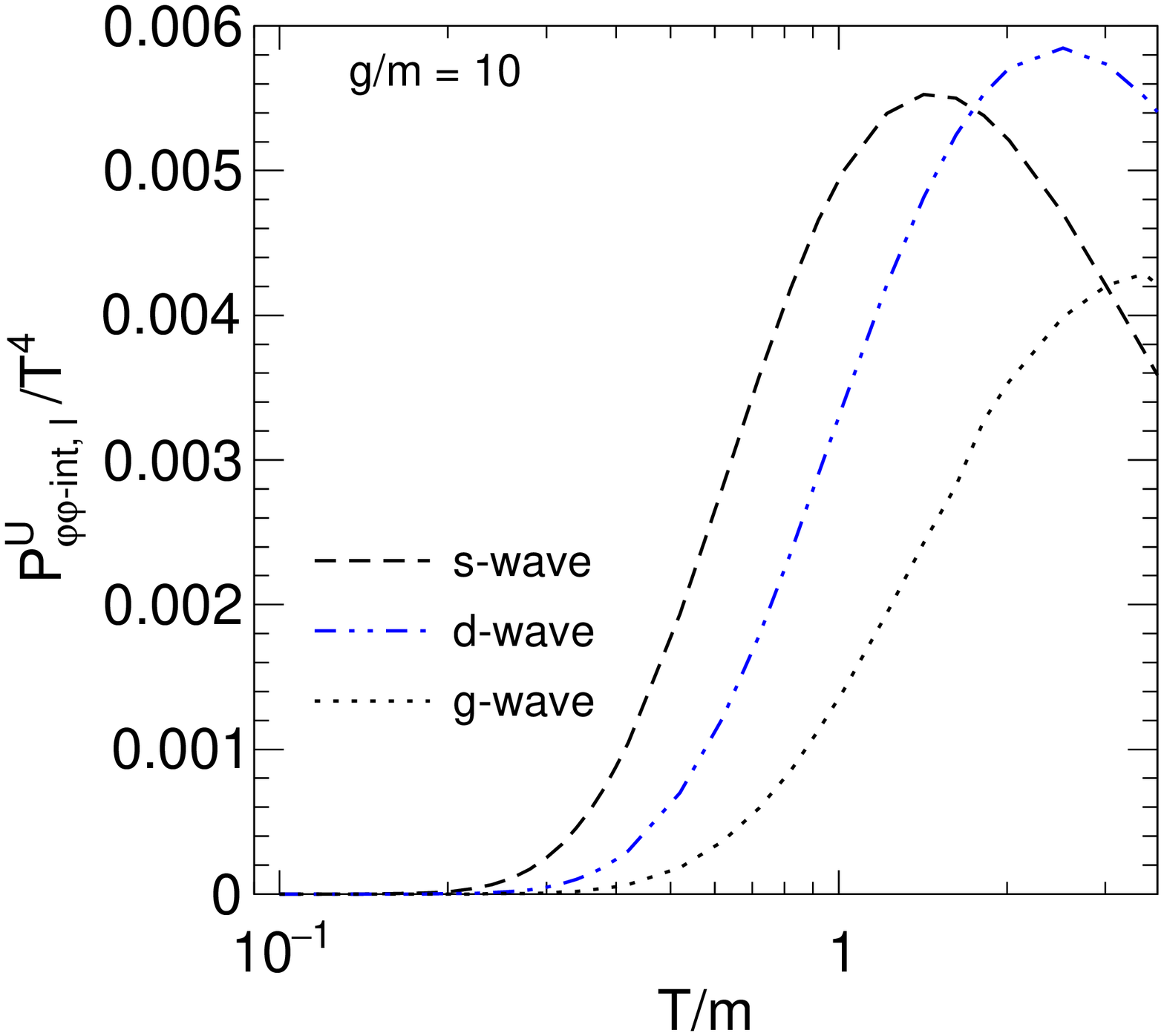}
\includegraphics[width=0.48\textwidth]{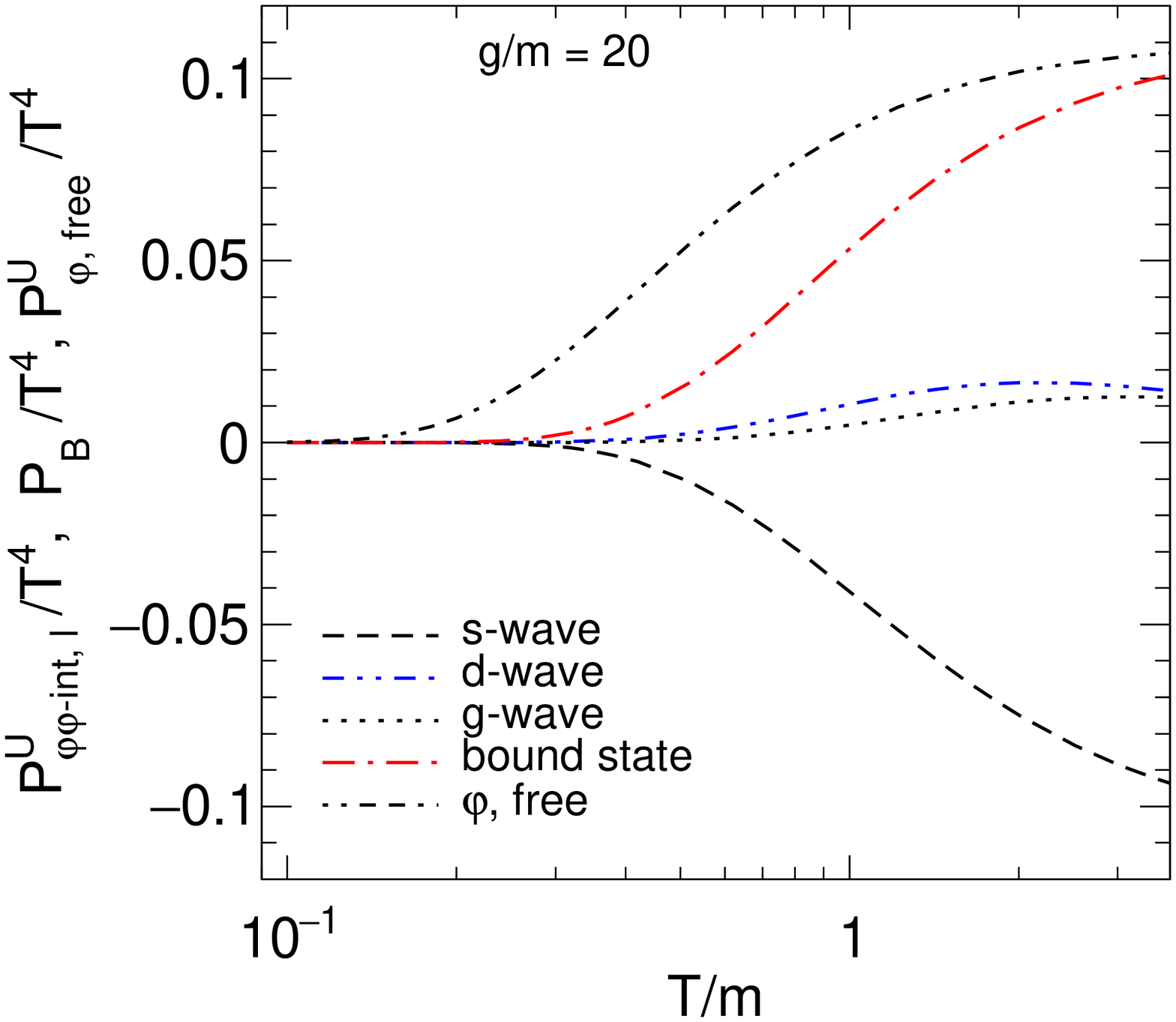}
\caption{The left panel shows the temperature dependence of the interacting normalized pressure (see Eq. \ref{eq:p_int}) for s-, d- and g-waves at $g = 10 m < g_c$. Right panel shows a similar plot but for $g = 20 m > g_c$ when bound state is produced. Besides the interactions of s-, d- and g-waves, the normalized pressure of the bound state and of free particles are also shown. }
\label{Fig:pintvsT_g_10}
\end{figure}

%\begin{figure}[ptb]
%\centering
%\includegraphics[width=0.48\textwidth]{p_int_s_d_g_bs_vs_T_uniterized_g_20.eps}
%\caption{Temperature dependence of the interacting normalized pressure (see Eq. \ref{eq:p_int}) for s-, d- and g-waves at $g = 20 m > g_c$ (bound state forms). Besides the interactions of s-, d- and g-waves, the normalized pressure of the bound state and of free particles are also shown.}
%\label{Fig:pintvsT_g_20}
%\end{figure}

Next, Fig. \ref{Fig:pintvsT_g_10} shows the interacting parts of the normalized pressure ($P^{U}_{\varphi\varphi\text{-int}, l}/T^4$) for the s-, d- and g-waves as a function of $T/m$ for $g = 10 m<g_c$ (no bound state). 
%All the contributions are positive. 
%Pressure for all three waves is positive implying attractive interaction. $P^{U}_{\varphi\varphi\text{-int}, l}$ for increases with the increase of temperature for all the three waves. However, normalized pressure $P^{U}_{\varphi\varphi\text{-int}, l}/T^4$, first increase with increase of $T/m$ and then starts decreasing at high $T/m$ and saturates at very high $T/m$. 
%We observe that at low $T$ the s-wave dominates over the d- and g-waves. However, with the increase of $T$, the d- and g-waves become larger due to their larger degeneracy factors (5 and 9, respectively).  
 This figure offers also an additional indication of
the range of temperatures for which we may trust our results. For $T/m \sim 1$ the s-wave dominates and the higher waves represent a small correction. 
For $T/m\sim2$, the d-wave contribution becomes as large as the
s-wave one, thus caution is required. Yet, the g-wave contribution is still safely small and we may argue that the further contributions are still negligible. In addition, we recall that at $T/m \sim 3$ inelastic channels become relevant. We thus consider  $T/m \sim 3$ as an upper limit of our study.
%(in some cases, we show also larger $T$-values since they might still be interesting as a guess of the behavior at higher $T$, but they should be regarded with care).
%We consider our results as useful, even if their accuracy is definitely smaller than in the low-$T$ domain.
%(Note, at about $T/m\sim 10$ the g-wave increases and approaches (and eventually overcomes) the d-wave.) 
For these reasons, in this figure (as well as in all other figures presenting the pressure contributions as a function of $T/m$) we use a logarithmic plot in order to underline the low-T part of our results up to $T/m \sim 3$. Whenever a fixed $T/m$ value is required, the maximal one used in this work is $T/m=2$, see e.g. Fig.\ref{Fig:pvsg}.

%In general, each further $l$-wave contribution is expected to follow the same qualitative
%behavior of Fig. 8: the maximum for a given wave is realized for a certain
%point $(T_{l},P_{\varphi\varphi\text{-int,}l}^{(\max)})$ with $T_{l}>T_{l+1}$
%and $P_{\varphi\varphi\text{-int,}l}^{(\max)}<$ $P_{\varphi\varphi
%\text{-int,}l+1}^{(\max)}$ However, at $T/m\sim10$ 

%Figure \ref{Fig:pintvsT_g_20} is similar to Fig. \ref{Fig:pintvsT_g_10} but for $g = 20 m>g_c$ (for which the bound state forms). 
%The right panel of Fig. \ref{Fig:pintvsT_g_10} is similar to the figure at the left panel but for $g = 20 m>g_c$ (for which the bound state is formed). The contribution of bound state and of free particles are also shown here.  The s-wave contribution is negative and decreasing, signalizing a repulsion. Here, also the bound state contribution is plotted. 
%Normalized pressure for both bound state and the free particle increase with increase of $T/m$. When $T/m$ is small, bound state contribution is smaller than the free particle because of its mass $M_B$ is almost double compared to that of free particle. However, at $T/m \gtrapprox 10$ both components saturate to the massless limit: $(P_{\varphi, \text{free}}/T^4)_{m=0} = (P_{\text{B}}/T^4)_{M_B = 0} = \pi^2/90 \approx 0.1096$.
%The behaviour of the pressure for the other two waves, $d$ and $g$, are similar to Fig. \ref{Fig:pintvsT_g_10}. 

\begin{figure}[ptb]
\centering
\includegraphics[width=0.48\textwidth]{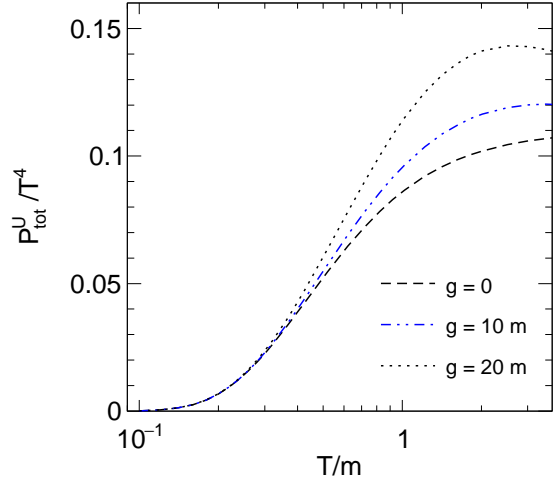}
\caption{Temperature dependence of the normalized total pressure (Eq. \ref{eq:ptot}).}
\label{Fig:ptotvsT} 
\end{figure}

The total normalized pressure as a function of $T/m$ is shown in Fig. \ref{Fig:ptotvsT} for three different values of $g$ (the free case $g = 0$, $g/m = 10$ and $g/m =20$). At high $T$ the massless limit $(P_{\varphi, \text{free}}/T^4)_{m=0} = \pi^2/90 \approx 0.1096$ is reached. 
%Apart from interactions of s, d and g-waves shown in Fig. the\ref{Fig:pintvsT_g_10}, result for $g = 10 m$ includes free part as well. No bound state is there since $g<g_c$. In this case up to $T/ \approx 2$ normalized pressure first increases with the increase of $T/m$ then it starts decreasing and eventually saturates at high $T/m$. Pressure for $g = 10 m$ is slightly greater than that of free particles because of the attractive nature of interactions of three partial waves. For $g = 20 m$ bound state contribution is also there. Different components are shown in Fig. \ref{Fig:pintvsT_g_20}. Behaviour is almost similar to that of $g = 10 m$. Although, $P/T^4$ is larger compared to the other two lines. 

 For completeness, we have studied the speed of sound
\begin{equation}
c_{s}^{2}=\frac{\partial P}{\partial\varepsilon}=\frac{P^{\prime}%
(T)}{\varepsilon^{\prime}(T)} = \frac{P^{\prime}(T)}{TP^{\prime \prime}(T)}
\label{speedsound}%
\end{equation}
which is safely smaller than one for all investigated temperatures, see Appendix C for details (in the r.h.s. above, the thermodynamical self-consistency $\varepsilon = T P^{\prime} -P$ has been used.)

An important point concerns the quantification of the contribution of the interaction to the pressure. We discuss the scenarios without and with the bound state separately. 
When no bound state forms ($g<g_c$),  it is useful to define the quantity:
\begin{equation}\label{eq:eta}
 \eta = \frac{P^{U}_{\text{tot}} }{P_{\varphi\text{,free}}} = 1 + \frac{P^{U}_{\varphi\varphi-\text{int}} }{P_{\varphi\text{,free}}} \text{ .}
\end{equation}
The absence of interactions ($g \rightarrow 0$) corresponds to $\eta = 1$ and departures from this value quantify the naive result that one obtains by neglecting them.  
%Since for $g<g_c$ the $\varphi^3$-interaction is attractive, $\eta\ge 1$.

\begin{figure}[ptb]
\centering
\includegraphics[width=0.48\textwidth]{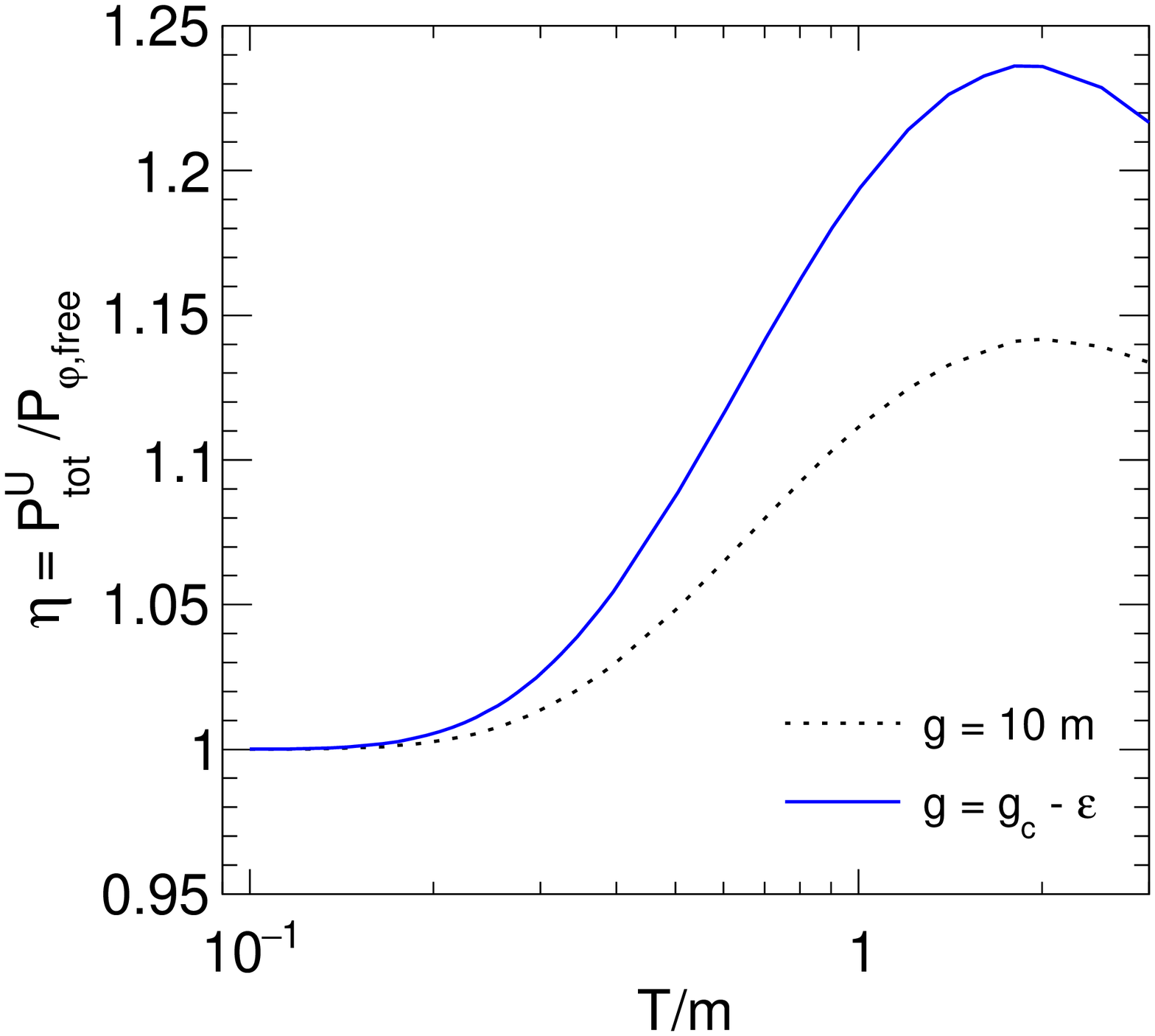}
\includegraphics[width=0.48\textwidth]{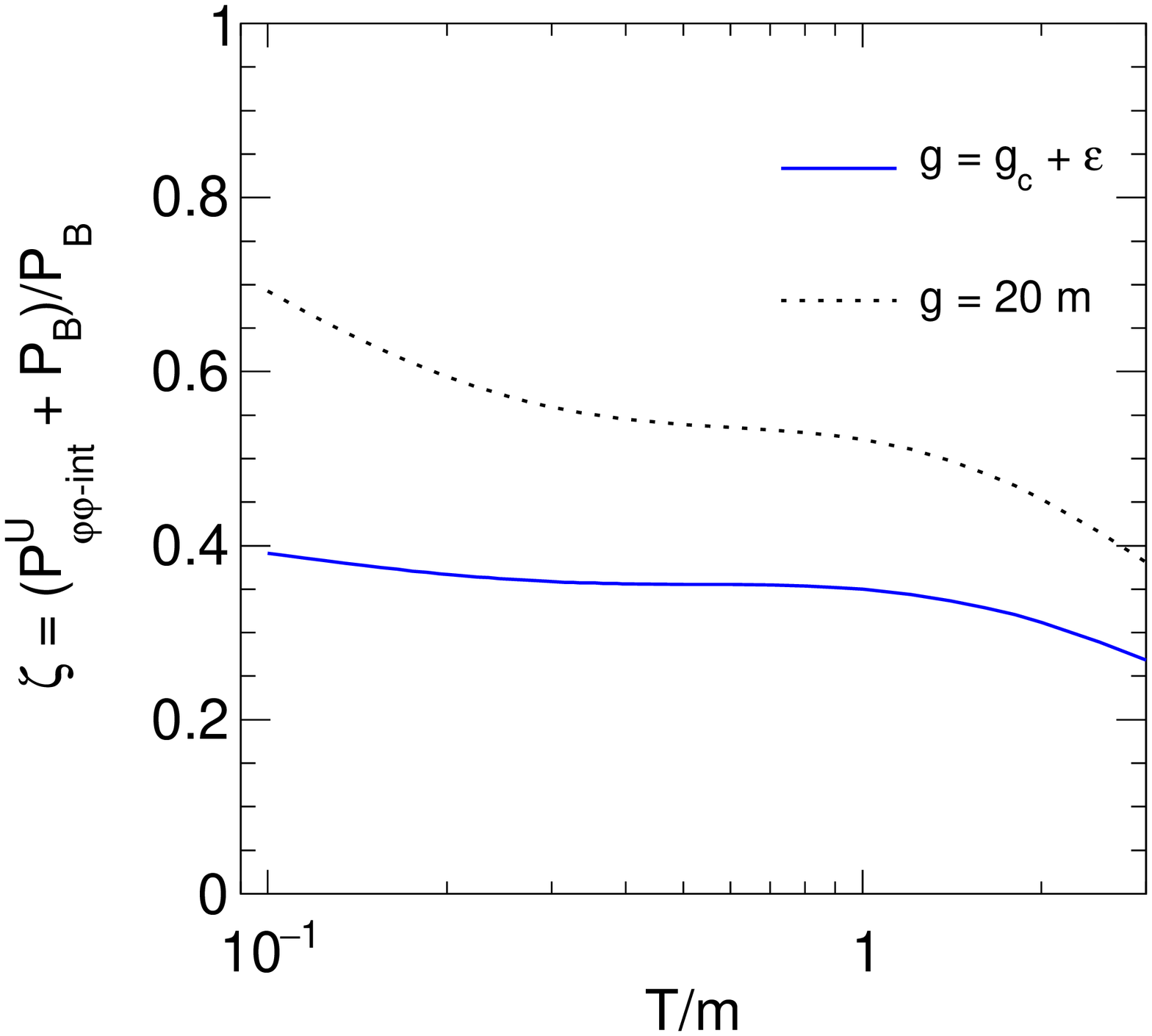}
\caption{Temperature dependence of the ratio $\eta$ (Eq. \ref{eq:eta}) (Left) and $\zeta$ (Eq. \ref{eq:zeta}) (Right).}
\label{Fig:etavsT}
\end{figure}
   
 The left panel of Fig. \ref{Fig:etavsT} shows the temperature dependence of $\eta$ for two different values of $g<g_c$. For both cases $\eta\approx 1$ when $T$ is low. This indicates that the role of the interaction is negligible at low temperatures. With the increase of $T/m$, $\eta$ increases and becomes maximal at around $T/m \approx 2$.  Moreover, we notice that for $g=g_c -\epsilon$ the maximum is reached for $\eta\approx 1.23$, which indicates that the effect of interaction is non-negligible at certain intermediate temperatures.
%The height at the maximum decreases -as expected- with the decrease of $g$, since the interaction is less intense. 

Let us now define an analogous quantity to be used when a bound state forms ($g>g_c$):
\begin{equation}\label{eq:zeta}
 \zeta = \frac{P^{U}_{\varphi\varphi-\text{int}} + P_{B} }{P_{B}} ,
\end{equation}
out of which the total pressure of the system is given by 
\begin{equation}
P^U_{\text{tot}} = P_{\varphi\text{,free}} + \zeta P_{B} \text{ ,}
\end{equation}
which is simply the sum of two free gases, one for the particles of the type $\varphi$ with mass $m$ and one for the bound state $B$ with mass $M_B$. Yet, the latter is rescaled by the factor $\zeta$, whose departure from unity quantifies how much of the bound state contribution remains after the partial cancellation induced by the interaction has been taken into account. 
%the departure from the naive result obtained by including only a simple thermal contribution of free particles with mass $M_B$. Note, the naive case corresponds to $\zeta =1$. In particular, the ratio $\zeta$ 

%\begin{figure}[ptb]
%\centering
%\includegraphics[width=0.48\textwidth]{p_int_p_bs_vs_T.eps}
%\caption{Temperature dependence of the ratio $\zeta$ (Eq. %\ref{eq:zeta}).}
%\label{Fig:zetavsT}
%\end{figure}
 
  In the right panel of Fig. \ref{Fig:etavsT} we show the temperature dependence of $\zeta$ for the values $g=g_c+\varepsilon$ and $g=20 m$.
%{\bf In Fig. \ref{Fig:zetavsT} we show the temperature dependence of $\zeta$ for the values $g=g_c+\varepsilon$, $g=20 m$.}
%, and $g=100 m$. 
For $g=g_c+\varepsilon$, the quantity $\zeta \approx 0. 4$ at $T/m = 0.1$, which then decreases for increasing $T/m$ and saturates to $\approx 0.08$. 
For $g=20 m$,  $\zeta \approx 0.7$ at low $T/m$ and decreases to $\approx 0.08$ at high $T/m$.
Thus, in both cases, the joint role of the bound state and interaction can be summarized by a contribution of free gas of bound state particles which is sizably reduced by $\zeta$. The amount of reduction depends both on the value of $g$ as the value of temperature, being typically larger at small $T$ and smaller at large $T$. 
%In general, there is not a unique simple answer to the question about the amount of cancellation between the bound state gas and the interaction contributions.

\begin{figure}[ptb]
\centering
\includegraphics[width=0.48\textwidth]{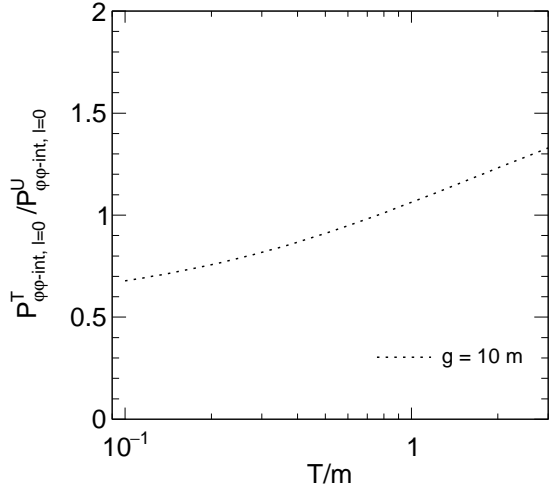}
\caption{Dependence of 
%Variation of 
the ratio of tree-level pressure and unitarized pressure for s-wave with $T/m$. }
\label{Fig:effect_uni}
\end{figure}
%The effect of unitarization on s-wave pressure is shown in Fig. \ref{Fig:effect_uni}.  The  ratio of tree-level to  unitarized pressure for s-wave ($P^{T}_{\varphi\varphi\text{-int}, l=0}/P^{U}_{\varphi\varphi\text{-int}, l=0}$) at $g = 10m$ is plotted as a function of $T/m$. 

%Up to the critical value of $g =g_c \approx 14.45 m$, $P^{T}_{\varphi\varphi\text{-int}, l=0}/P^{U}_{\varphi\varphi\text{-int}, l=0}$ is close to the one. However, above $g_c$, this ratio becomes negative since  $P^{U}_{\varphi\varphi\text{-int}, l=0}$ becomes negative while $P^{T}_{\varphi\varphi\text{-int}, l=0}$ is positive. We will see later that a bound state is formed at $g \ge g_c$ only after the unitarization is done. This is not possible at the tree level. This is the main motivation for the unitarization.

 We conclude this section with two additional remarks:

\begin{itemize}
\item The introduced unitarization is a suitable tool for the study of bound
states. It goes beyond perturbative results, which are not capable to generate
poles below the threshold. 
%In this respect, the need of unitarization is clear in this work, since the emergence of bound states is one of our main goals. 
Yet, even when no bound state forms but the coupling is not small, the role of
unitarization is also non-negligible. This point has been already mentioned in
Sec. II.B in connection with the scattering length.
We show this feature also at nonzero $T$ in Fig. \ref{Fig:effect_uni}, where
the ratio of the s-wave tree-level pressure and the unitarized one is
plotted as a function of $T$ for $g/m=10.$ It is visible that the effect of the
unitarization is in general non-negligible both at small and at large $T$.

\item The potential in Eq. \ref{eq:L} is unbounded from below. A simple improvement is
to add to the potential a term $\frac{\lambda}{4!}\varphi^{4}$ with
$\lambda>0.$ This case is presented in Appendix D: in general, there are
quantitative but not qualitative changes w.r.t. the results presented in this
section, but also some additional problems related to this extension appear.
The Wigner condition is not always fulfilled and the speed of sound exceeds 1
at high $T$, meaning that additional studies are required in the future. 
\end{itemize}

%Finally, we refer to Appendix C for the discussion of the case in which an additional $\varphi^4$ interaction term is added to the Lagrangian of Eq. \ref{eq:L}. There are quantitative but not qualitative changes w.r.t. the results presented in this section.

%Finally, the case $g=100 m$ can be regarded as a strong-coupling limit. Here, $\zeta>1$, shows that one has even a larger result than a simple free gas of  $B$ particles. A closer inspection indicates that for large $g$ the role of the d- and g-waves is sizable and the additional pressure is due to the higher waves.

%In order to understand this point, we show in Fig. \textbf{finish here!!!!!!}
%the behavior of an analogous ratio that however includes the s-wave only. Here, also for $g=100$ a cancellation does take place. It then follows that . 

\section{An intermediate state $S$ }\label{sec:IntermediateState}
In this section, we study the case in which two distinct particles, $\varphi$ with mass $m$ and $S$ with mass $M$, are  considered. Their interaction is a three-leg $gS\varphi^2$ vertex. Thus, the decay of $S$ into $\varphi\varphi$, if kinematically allowed, takes place. Then, the state $S$ is a resonance with a certain decay width.  
Since two $\varphi$ particles interact via an exchange of $S$, an attractive Yukawa interaction between them is induced: if it is strong enough, a bound state forms. 
The main question here is how this system behaves at nonzero $T$.
\subsection{Vacuum's formalism}
The Lagrangian under study takes the form:
\begin{equation}
\mathcal{L}=\frac{1}{2}\left(  \partial_{\mu}\varphi\right)  ^{2}-\frac{1}%
{2}m^{2}\varphi^{2}+\frac{1}{2}\left(  \partial_{\mu}S\right)  ^{2}-\frac
{1}{2}M^{2}S^{2}+\frac{g}{2!}S\varphi^{2},
\end{equation}
where $g$ is the coupling constant. 
The tree-level decay width $S \rightarrow \varphi\varphi$ (allowed for $M>2m$) reads (e.g. Ref. \cite{Giacosa:2007bn}):
\begin{equation}
\Gamma_S=\frac{g^2}{2}\frac{\sqrt{\frac{M^{2}}{4}-m^{2}}}{8\pi M^2} \text{ ,}
\end{equation}
and the tree-level scattering is:
\begin{equation}
A(s,t,u)=\frac{-g^{2}}{s-M^{2}+i\epsilon}+\frac{-g^{2}}{t-M^{2}%
+i\epsilon}+\frac{-g^{2}}{u-M^{2}+i\epsilon} \text{ .}
\end{equation}
The first three tree-level partial wave amplitudes are evaluated as: 
\begin{equation}
A_{0}(s)=\frac{1}{2}\int_{-1}^{+1}d\xi A(s,\theta)=-\frac{g^{2}}%
{s-M^{2}+i\epsilon}+2g^{2}\frac{\ln\left[  1+\frac{s-4m^{2}}{M^{2}}\right]
}{s-4m^{2}+i\epsilon}\text{ ;}%
\end{equation}
\begin{equation}
\begin{split}
    A_{2}(s)= & \frac{-2g^{2}}{(s-4m^{2})^{3}}(  3\left(  4m^{2}-s\right)
(4m^{2}-2M^{2}-s) \\ & + \left(  16m^{4}+6M^{4}+6M^{2}s+s^{2}-8m^{2}(3m^{2}%
+s)\right)  \ln\left[  1+\frac{s-4m^{2}}{M^{2}}\right]  ) \text{ ;}
\end{split}
\end{equation}
\begin{equation}
\begin{split}
   A_{4}(s) =& \frac{-2g^{2}}{3(s-4m^{2})^{3}} [ 5 (4 m^2-s) (4 m^2-2 M^2-s) (80 m^4-8 m^2 (21 M^2+5 s)+42 M^4  \\ & +42 M^2 s+5 s^2 ) - 6 (256 m^8-256
   m^6 (5 M^2+s)+96 m^4 (15 M^4+10 M^2 s+s^2) \\ & -16 m^2 (35 M^6+45 M^4 s+15 M^2 s^2+s^3)+70 M^8+140 M^6 s+90 M^4
   s^2 \\ & +20 M^2 s^3 +s^4) \ln (1+\frac{s-4 m^2}{M^2}) ] \text{ .}
\end{split}
\end{equation}

 Also, in this case, we limit our study to two-body $\varphi \varphi$ scattering processes. Within 
our framework, the state $S$ is a resonance, therefore it is not an asymptotic
state of the theory. 
Namely, the scattering of the type $S\varphi\rightarrow S\varphi$ should be understood as part of the more general process $\varphi
\varphi\varphi\rightarrow\varphi\varphi\varphi.$ They are omitted here since
they are not expected to contribute much to the energies and temperatures of interest.

The unitarization is carried out by repeating analogous steps as in Sec. \ref{sec:Unitarization}.
As previously, the loop function for $M<2m$ (when the state $S$ is stable) has therefore two subtractions, one at the mass $M$ and one at the branch point $4m^2-M^2$: 
\begin{equation}
\Sigma(s)=-\dfrac{(s-M^{2})(s-(4m^{2}-M^{2}))}{\pi}\int_{4m^{2}}^{\infty}%
\frac{\frac{1}{2}\frac{\sqrt{\frac{s^{\prime}}{4}-m^{2}}}{8\pi\sqrt{s^{\prime
}}}}{(s-s^{\prime}+i\varepsilon)(s^{\prime}-M^{2})(s^{\prime}-(4m^{2}-M^{2}%
))}ds^{\prime} \text{ .}
\end{equation}
Yet, for $M>2m$ (that is above the threshold) the state $S$ is a resonance, therefore we should only require that the real part of the loop vanishes at $M^2$, thus  $\operatorname{Re}[\Sigma(s=M^{2})]=0$
(the whole loop does not, since the imaginary part, proportional to the decay width of $S$, is nonzero).  Then, upon considering a single subtraction:
\begin{equation}
\Sigma(s)=-\dfrac{(s-(4m^{2}-M^{2}))}{\pi}\int_{4m^{2}}^{\infty
}\frac{\frac{1}{2}\frac{\sqrt{\frac{s^{\prime}}{4}-m^{2}}}{8\pi\sqrt
{s^{\prime}}}}{(s-s^{\prime}+i\varepsilon)(s^{\prime}-(4m^{2}-M^{2}%
))}ds^{\prime}+C\text{ ,}%
\label{Sonesub}
\end{equation}
where the subtraction $C$ is such that $\operatorname{Re}%
[\Sigma(s=M^{2})]=0.$

\begin{figure}[ptb]
\centering
\includegraphics[width=0.48\textwidth]{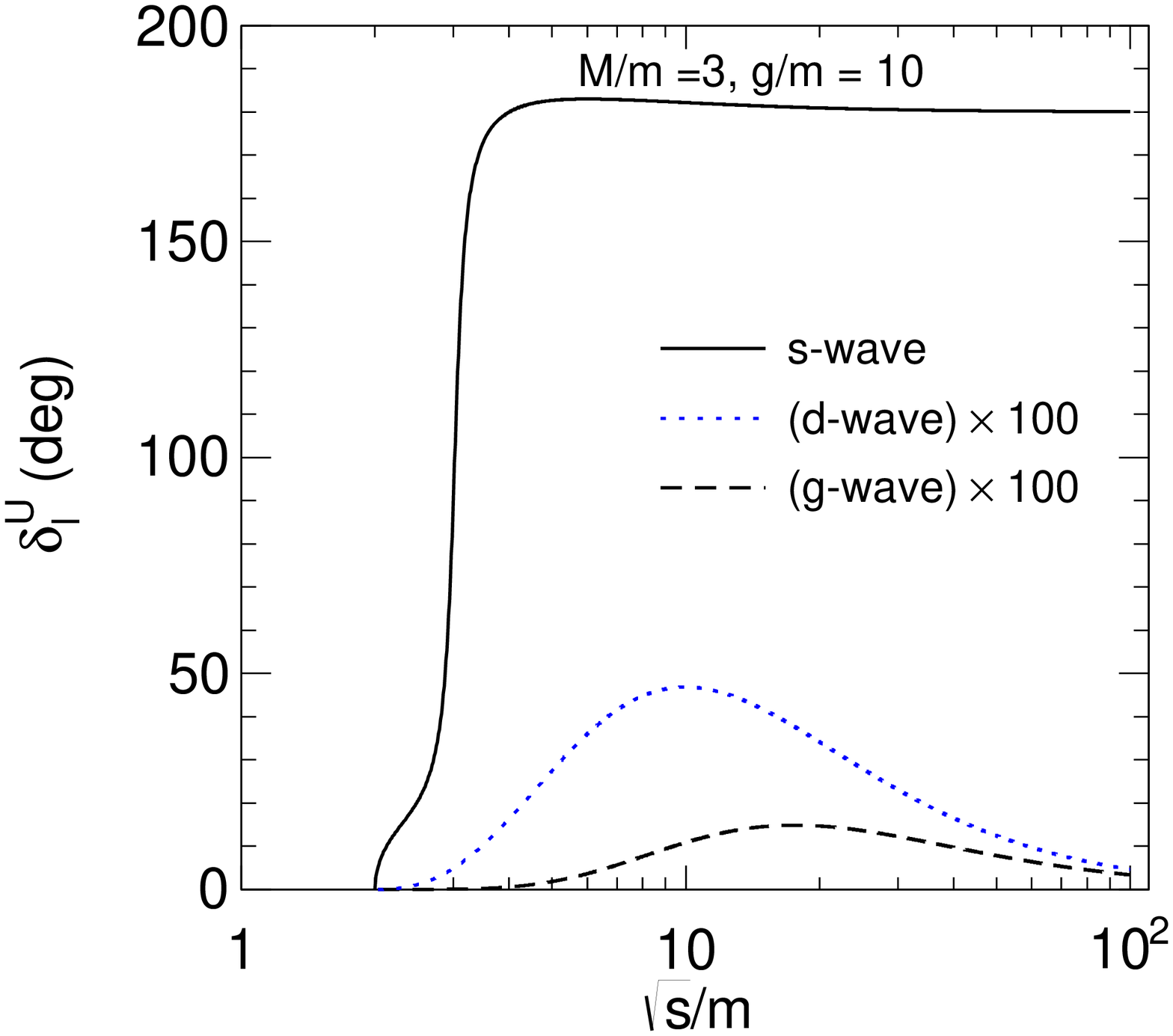}
\includegraphics[width=0.48\textwidth]{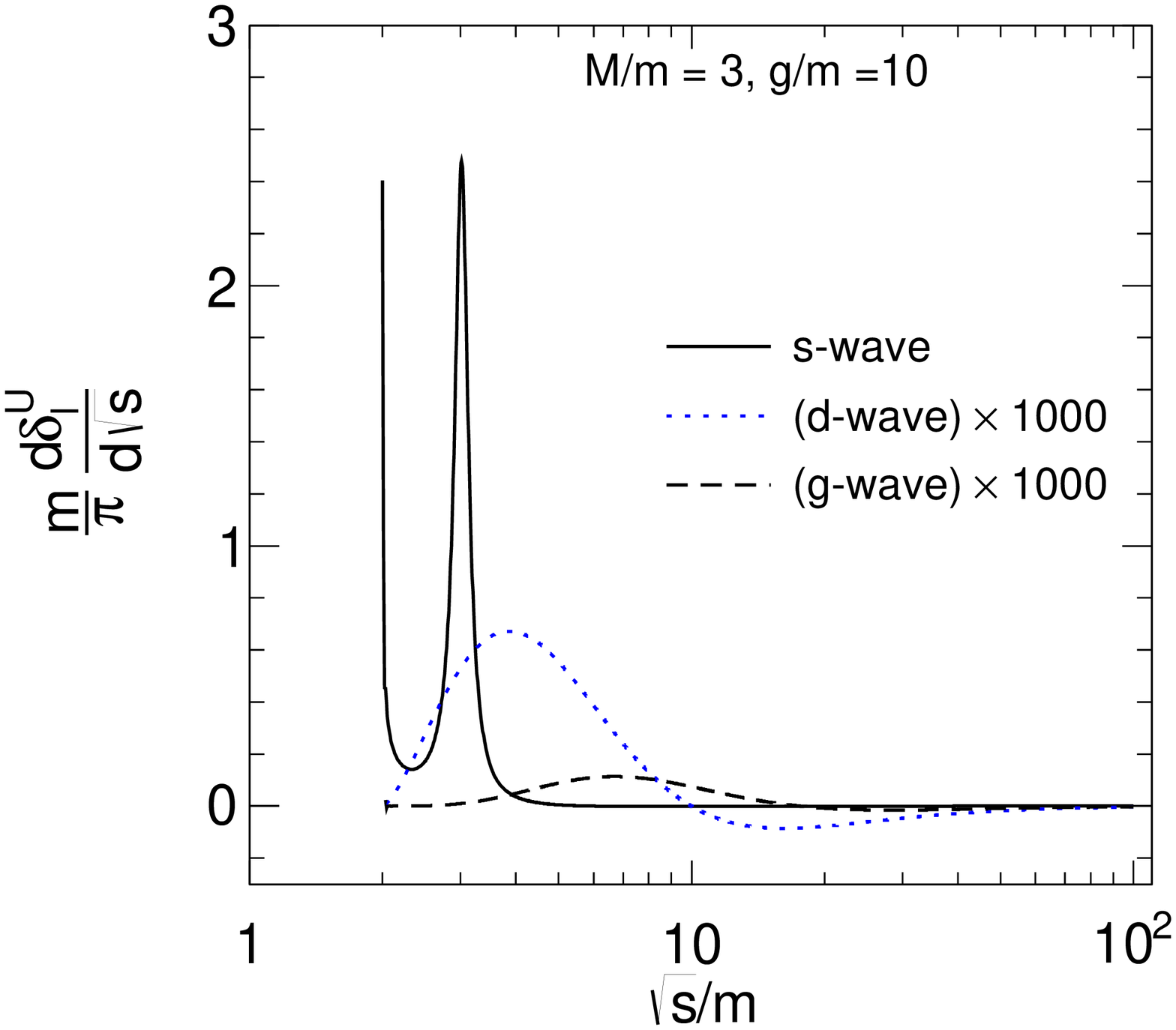}
\caption{(Left) Variation of phase shifts  with $\sqrt{s}/m$ for s-, d-, and g-waves at $g = 10 m<g_{S,c}$ when the resonance $S$ has mass $M = 3 m$. The d- and g-wave phase shifts are multiplied by a factor of 100 to make them visible. (Right) Energy dependence of s-,d-, g-wave phase shift derivatives. The d- and g-waves contributions are multiplied by a factor of 1000. }
\label{Fig:ps_der_ps_S_g_10}
\end{figure}

 The unitarized amplitudes are given by  
\begin{equation}
A_{k}^{U}(s)=\left[  A_{k}^{-1}(s)-\Sigma(s)\right]^{-1} \text{ ,}
\label{resum}
\end{equation}
just as in Eq. (\ref{eq:Au}).

Here, if the mass $M$ is sufficiently larger than the threshold $2m,$ the
$s$-channel can be regarded as dominant for values of $s$ of the order of $M^{2}$. In fact, the tree-level result (neglecting $t$ and $u$ channels)
reads $A_{l=0}(s)\simeq\frac{-g^{2}}{s-M^{2}}.$ Yet, the pole at $M^{2}$ is an
artifact of the tree-level result. The unitarized amplitude (keeping only the
$s$ channel) reads 
$A_{l=0}^{\text{U}}(s)\simeq\frac{-g^{2}}{s-M^{2}+g^{2}\Sigma(s)}$: the pole on the real axis moves to the complex plane
($s_{pole}\simeq M^{2}-g^{2}\operatorname{Im}\Sigma(M^{2})$) and the
unitarized amplitude is simply proportional to the one-loop resumed
propagator for the scalar resonance $S$ with the loop function $\Sigma(s)$
presented in\ Eq. (\ref{Sonesub}).

The critical value of  the coupling $g$ for obtaining a bound state in the s-channel is determined by
\begin{equation}
A_{0}^{-1}(4m^{2})-\Sigma(4m^{2})=0
\end{equation}
with
\begin{align}
A_{0}(s  & =4m^{2})=\frac{-g^{2}}{4m^{2}-M^{2}}+\frac{2g^{2}}{M^{2}} \text{ ;}\\
\Sigma(4m^{2})  & =\frac{\sqrt{M^{2}-4m^{2}}\ln\left[  -1+\frac{M\left(
M+\sqrt{M^{2}-4m^{2}}\right)  }{2m^{2}}\right]  }{32M\pi^{2}} \text{ .}
\end{align}
Note, the fact that a critical $g$ is needed to obtain a bound state is well known in the quantum mechanical counterpart of the Yukawa interaction, e.g. Refs. \cite{Luo:2004rj,Napsuciale:2020ehf}.

 The energy dependence of the unitarized phase shifts for s-, d- and g-waves and their derivatives are shown Fig. \ref{Fig:ps_der_ps_S_g_10} for the choice $M = 3 m$ and $g=10m<g_{M,c} \approx 22.83 m$.
 The s-wave phase increases rapidly near $\sqrt{s} = 3m=M$ and tends towards $\approx \pi$ (from above). Compared to the s-wave, magnitudes of the d- and g-wave phase shifts are significantly smaller and hence they are multiplied by a factor of 100 to show them in this plot.
 %At $\sqrt{s} = 2m$ both $d$ and g-wave phase shifts are zero. With increase of $\sqrt{s}/m$ both the phase shifts increase and then again decrease to zero at higher $\sqrt{s}/m$. 
 %The derivative of the corresponding phase shifts with respect to $\sqrt{s}/m$ are shown in the right panel of Fig. \ref{Fig:ps_der_ps_S_g_10}). 
 At the threshold, the derivative of the s-wave  phase shift is infinite, but the area under the curve is finite, hence there is no problem in evaluating thermodynamical quantities. As expected, a resonance peak is observed near $\sqrt{s} = 3m=M$. The reason for this peak is the presence of the particle $S$ of mass $3m$. 
 %For d- and g- waves, the derivatives of phase shifts are multiplied by a factor of 1000 to show them in the same plot since their magnitudes are smaller compared to that of s-wave.  

Note, in Fig. \ref{Fig:ps_der_ps_S_g_10} we display the (derivatives of the) phase shifts by using a logarithmic plot up to large values of $\sqrt{s}/m$ in order to verify that the phase shifts tend to a multiple of $\pi$. We recall, however, that inelastic channels are not taken into account. At nonzero $T$, the left parts of the plots are the relevant ones. 
  
  %Similar to s-wave peak structures are observed in d-wave as well. The peak is observed slightly below $\sqrt{s} = 4m$. Here, a peak is observed at higher $\sqrt{s}$ compared to s-wave. Derivative of the d-wave phase shift then decreases and even becomes negative $\sqrt{s} \gtrapprox 10 m$. Above $\sqrt{s} \approx 15m$, derivative of the phase shift again increases slowly and becomes zero at higher  $\sqrt{s}/m$. The derivative of g-wave phase shift is similar to d-wave but the maxima and minima are observed at relatively larger $\sqrt{s}/m$.
  
\begin{figure}[ptb]
\centering
\includegraphics[width=0.48\textwidth]{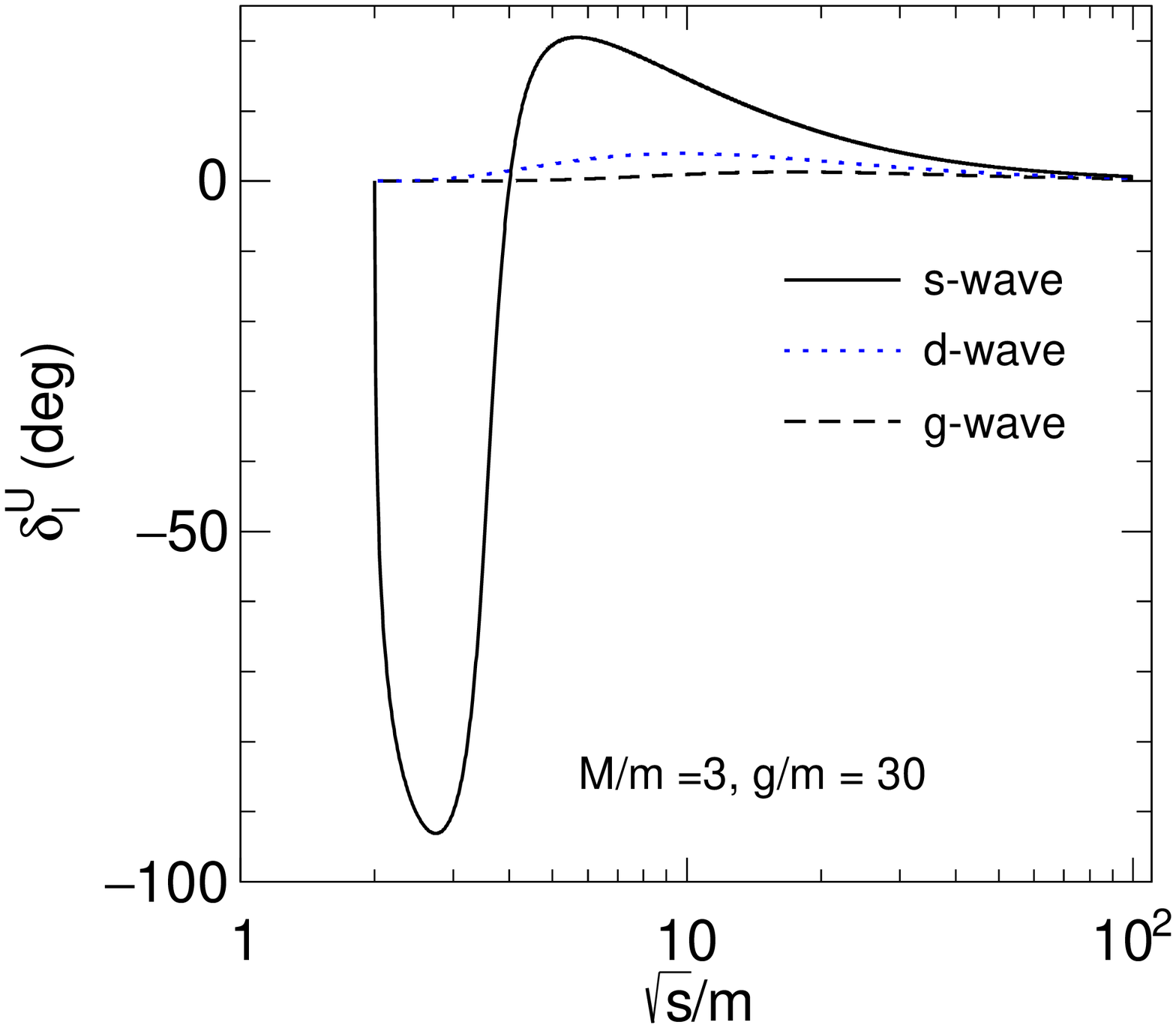}
\includegraphics[width=0.48\textwidth]{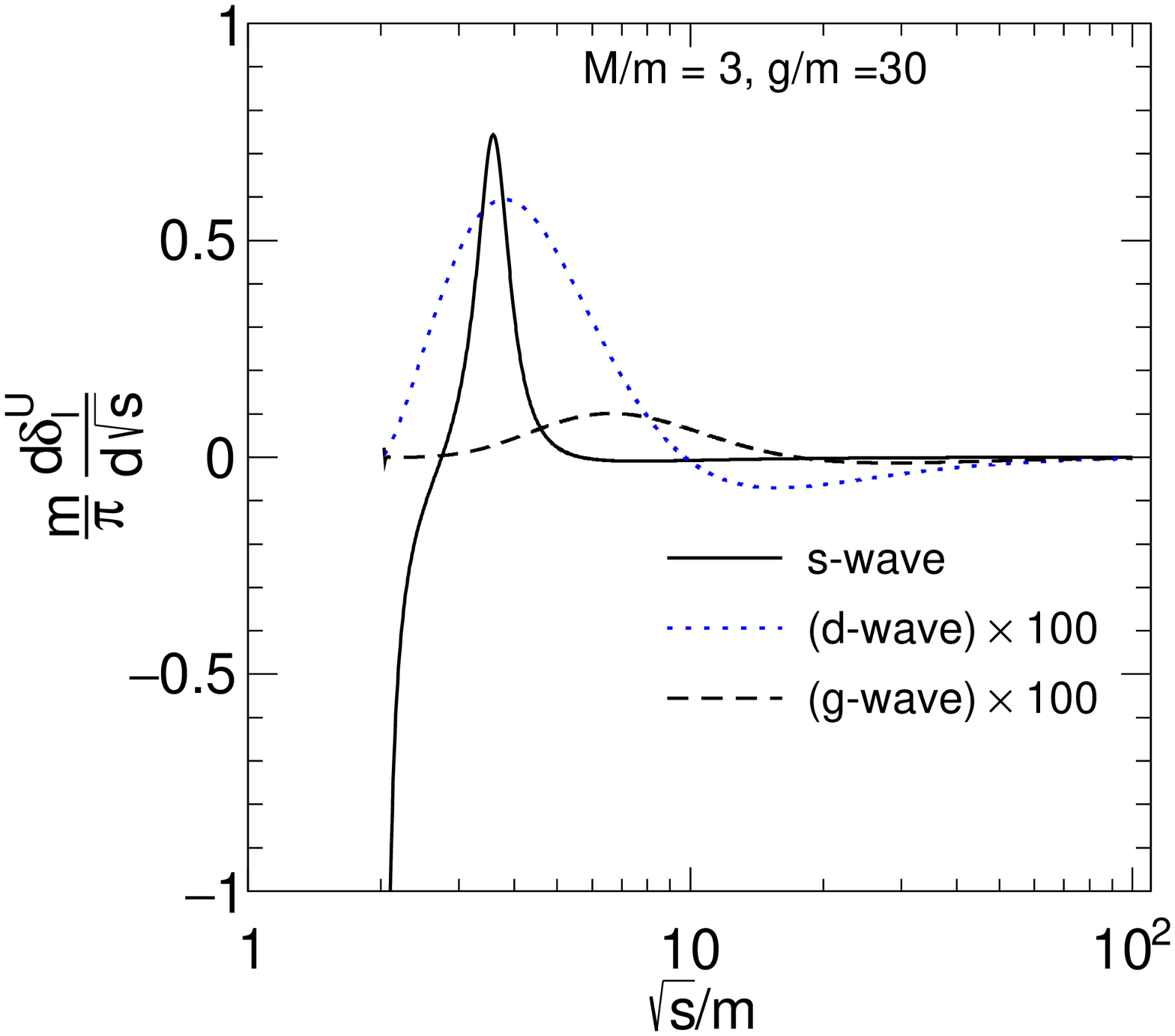}
\caption{Similar to Fig. \ref{Fig:ps_der_ps_S_g_10} but for $g = 30m>g_{S,c}$. The d- and g-wave phase shift derivatives are multiplied by a factor of 100.}
\label{Fig:ps_der_ps_S_g_30}
\end{figure}

In Fig. \ref{Fig:ps_der_ps_S_g_30} the case $g = 30 m>g_{M,c}$ is shown. A drastic change in the s-wave phase shift is observed, which is negative for $2m < \sqrt{s} \lessapprox 4m$, as a consequence of the presence of a bound state below the threshold. In this region, the s-wave phase shift decreases and reaches $-\pi/2$ around $\sqrt{s}/m \approx 3$, then starts increasing and becomes positive above $\sqrt{s}/m \approx 4$. It tends to zero at large $\sqrt{s}/m$. For the other two waves, the behavior is  similar to Fig. \ref{Fig:ps_der_ps_S_g_10}, but somewhat larger in magnitude. 
%The right panel of Fig. \ref{Fig:ps_der_ps_S_g_30} shows the derivatives of phase shifts. 
The derivative of the s-wave phase shift starts from $-\infty$ at the threshold, it then increases rapidly with the increase of $\sqrt{s}/m$ and becomes positive. Around $\sqrt{s}/m \approx 3.5$ it shows a peak and starts decreasing towards zero above that. The variation of derivatives of the d- and g-waves are similar to Fig. \ref{Fig:ps_der_ps_S_g_10}, but larger in magnitude.

Finally, we refer to Appendix C for the study of causality (Wigner condition and speed of sound) for this theory. Both of them confirm that it is not violated.

\subsection{Thermodynamic properties of the system in presence of $S$}

\begin{figure}[ptb]
\centering
\includegraphics[width=0.48\textwidth]{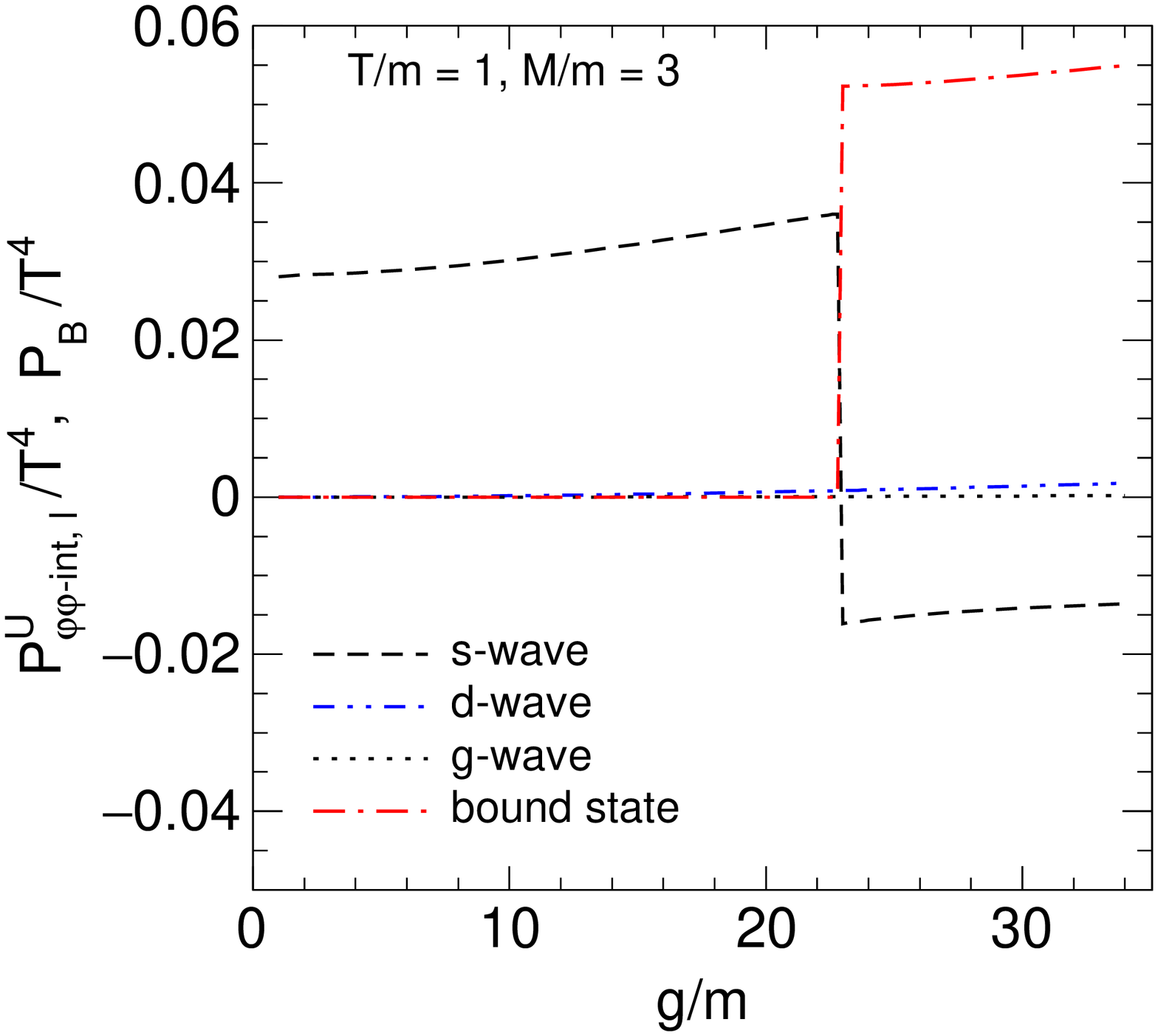}
\includegraphics[width=0.48\textwidth]{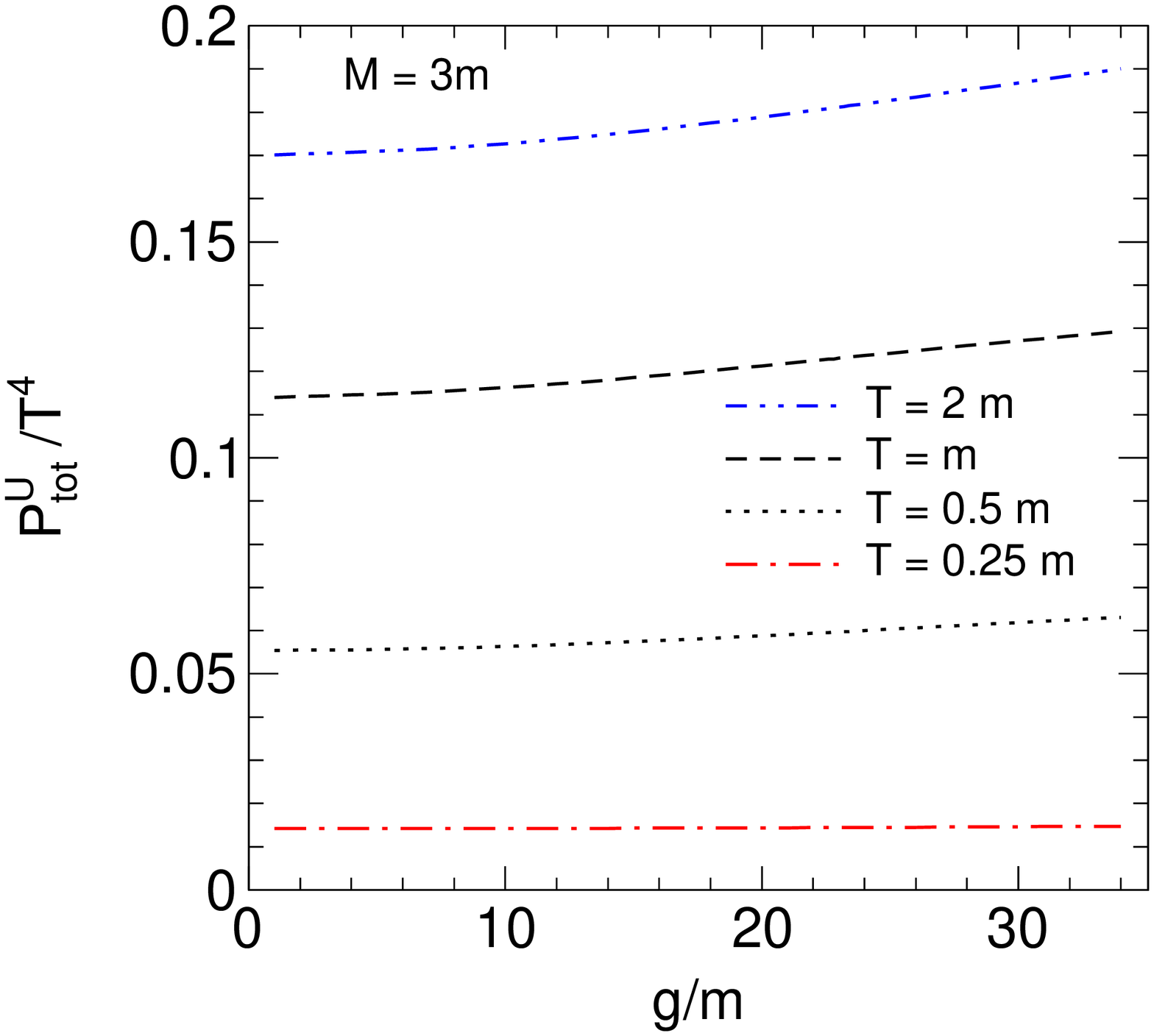}
\caption{Left panel shows the dependence of
%variation of
normalized pressure of s-, d- and g-waves with $g/m$ at $T = m $ in presence of an intermediate state $S$ of mass $M = 3 m$ The normalized pressure of the bound state is also shown. The right panel shows the normalized total pressure at four different temperatures.}
\label{Fig:P_int_reso_vs_g}
\end{figure}

%FGFG
In this subsection, we discuss the thermodynamical properties of the system in presence of an intermediate state $S$ with mass $M$. 
The pressure is evaluated via Eq. (\ref{eq:p_int}), which we report for convenience: 
\begin{equation}
P_{\varphi\varphi\text{-int}}^{U}+P_{B}=-T\int_{0}^{\infty}dx\frac{2l+1}{\pi
}\sum_{l=0}^{\infty}\frac{d\delta_{l}^{U}(s=x^{2})}{dx}\int_{k}\ln\left[
1-e^{-\beta\sqrt{k^{2}+x^{2}}}\right]  \text{ ,}\label{pbt}%
\end{equation}
which gives the overall interacting contribution. In particular, it should be stressed that:

(i) The pressure of the bound state $P_{B}$ is, as usual, nonzero only for
$g>g_{M,c}.$

(ii) The contribution of the state $S$ is contained in the term $P_{\varphi
\varphi\text{-int}}^{U}$; in the limit $g\rightarrow0,$ one has
\begin{equation}
\frac{1}{\pi}\frac{d\delta_{0}^{U}(s=x^{2})}{dx}=\delta(x-M)\text{ ,}%
\end{equation}
thus $P_{\varphi\varphi\text{-int}}^{U}$ reduces to the pressure of free $S$
particles with mass $M.$ 
In other words, the contribution of the resonance $S$ for thermodynamic
quantities is taken into account by the $\varphi\varphi\rightarrow$
$\varphi\varphi$ scattering process. In Ref. \cite{Lo:2019who} this point was discussed in detail in connection with the example of the $\rho$-meson, which is not added as an independent state to the thermodynamics but is reproduced by the $\pi\pi$ phase shift's derivative in the appropriate scattering channel.

(iii) Care is needed when $g=0.$ For this choice, the interaction contribution
is, clearly, exactly zero. The state $S$ is a free field that \textit{cannot}
be obtained from the interaction part of the $\varphi$ field.\ This issue is
discussed in detail in Ref. \cite{Lo:2019who}:  Eq. (\ref{pbt}) and is valid for nonzero (even if infinitesimal) $g$:
\begin{equation}
\lim_{g\rightarrow0}P_{\varphi\varphi\text{-int}}^{U}=P_{S,\text{free}}%
\neq\left(  P_{\varphi\varphi\text{-int}}^{U}\right)  _{g=0}=0\text{ .}%
\end{equation}

(iv) In connection to the discussion of Sec. III.A, the limit of small (but nonzero) $g,$ the set of
asymptotic states of the theory consists of the field $\varphi $ only. The
total pressure in this framework is $P_{\varphi \text{,free}}+P_{\varphi
\varphi \text{-int}}^{U}$, since the state $S$ is a resonance and not an
asymptotic state, no matter how small $g$ is.

Next, we turn to numerical examples for the specific choice $M = 3m$. 
The interacting part of the normalized pressure of s-, d- and g-waves as a function of $g/m$ is shown in the left panel of Fig. \ref{Fig:P_int_reso_vs_g}, in which the temperature is taken as $T = m$. When $g/m$ is small, the interacting part of the pressure of the s-wave is dominated by the free particle $S$ with mass $M = 3m$. As the coupling $g$ increases, the interacting part of the pressure of the s-wave contribution also increases. 
Up to the critical value of $g_{M,c}$ ($\approx 22.83 m$), the s-wave has a positive contribution to the pressure, while it is negative above it. 
The interacting part of the normalized pressure of d- and g-waves are always positive, but the magnitudes are much smaller compared to that of the s-wave. In this plot, we also show the normalized pressure for the bound state (which is clearly nonzero only for $g>g_{M,c}$). 
As for the $\varphi^3$ case, the pressure of the bound state exactly compensates for the abrupt jump in the s-wave pressure at the critical value $g_{M,c}$.
The normalized total pressure as a function of $g/m$ for four different $T$ is shown in the right panel of Fig. \ref{Fig:P_int_reso_vs_g}.
%in Fig. \ref{Fig:P_tot_reso_vs_g}. 
%Similar to Figs. \ref{Fig:ptotvsg} 
Similar to Fig. \ref{Fig:pvsg}, the total pressure is continuous in $g$, 
showing that also in this case the emergence of the bound state does not imply a discontinuity for the pressure.

%\begin{figure}[ptb]
%\centering
%\includegraphics[width=0.48\textwidth]{p_reso_tot_vs_g_diff_T.eps}
%\caption{Variation of the normalized total pressure (Eq. \ref{eq:ptot}) with $g/m$ in presence of an intermediate state of mass $M = 3 m$.}
%\label{Fig:P_tot_reso_vs_g}
%\end{figure}

\begin{figure}[ptb]
\centering
\includegraphics[width=0.48\textwidth]{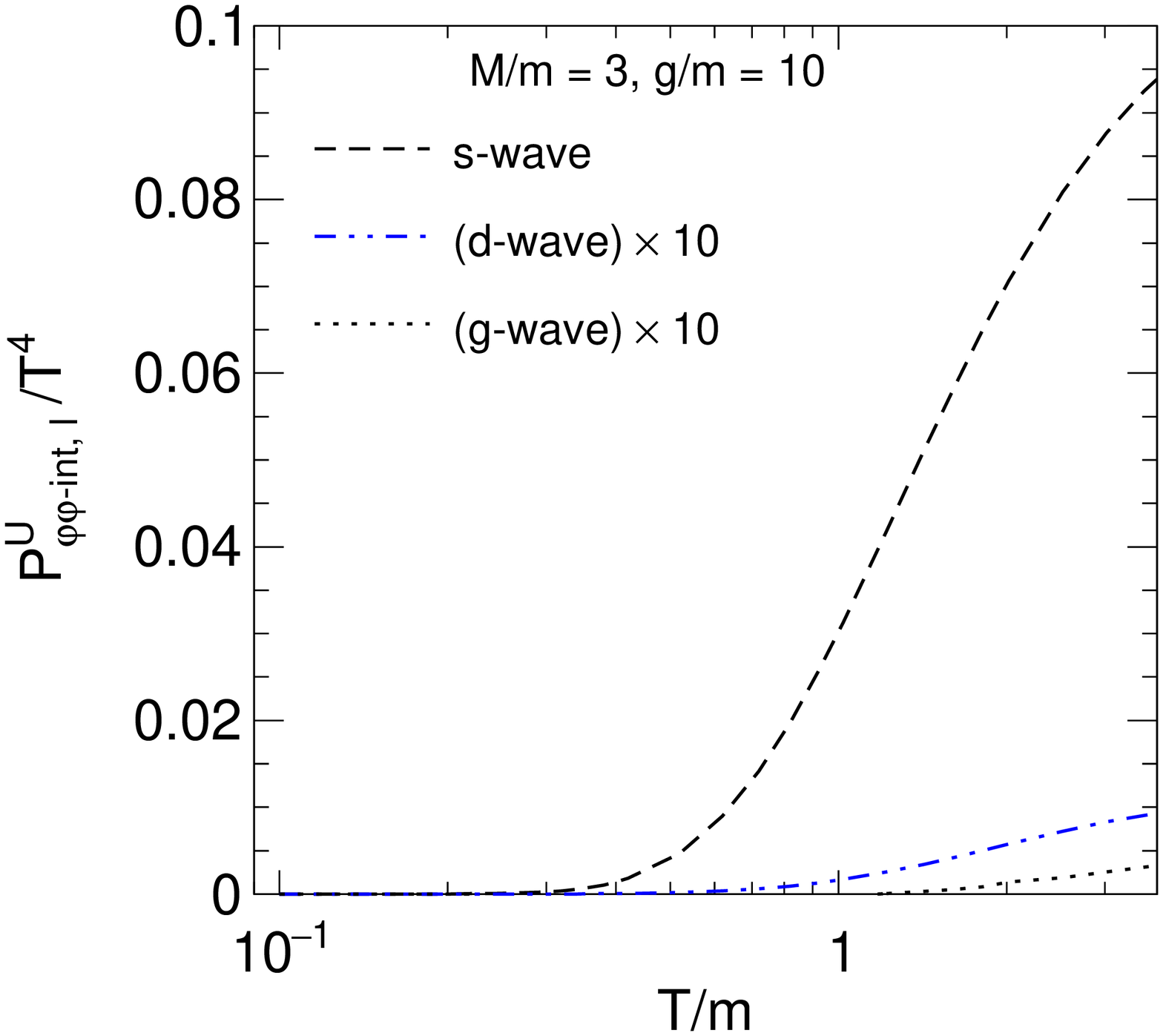}
\includegraphics[width=0.48\textwidth]{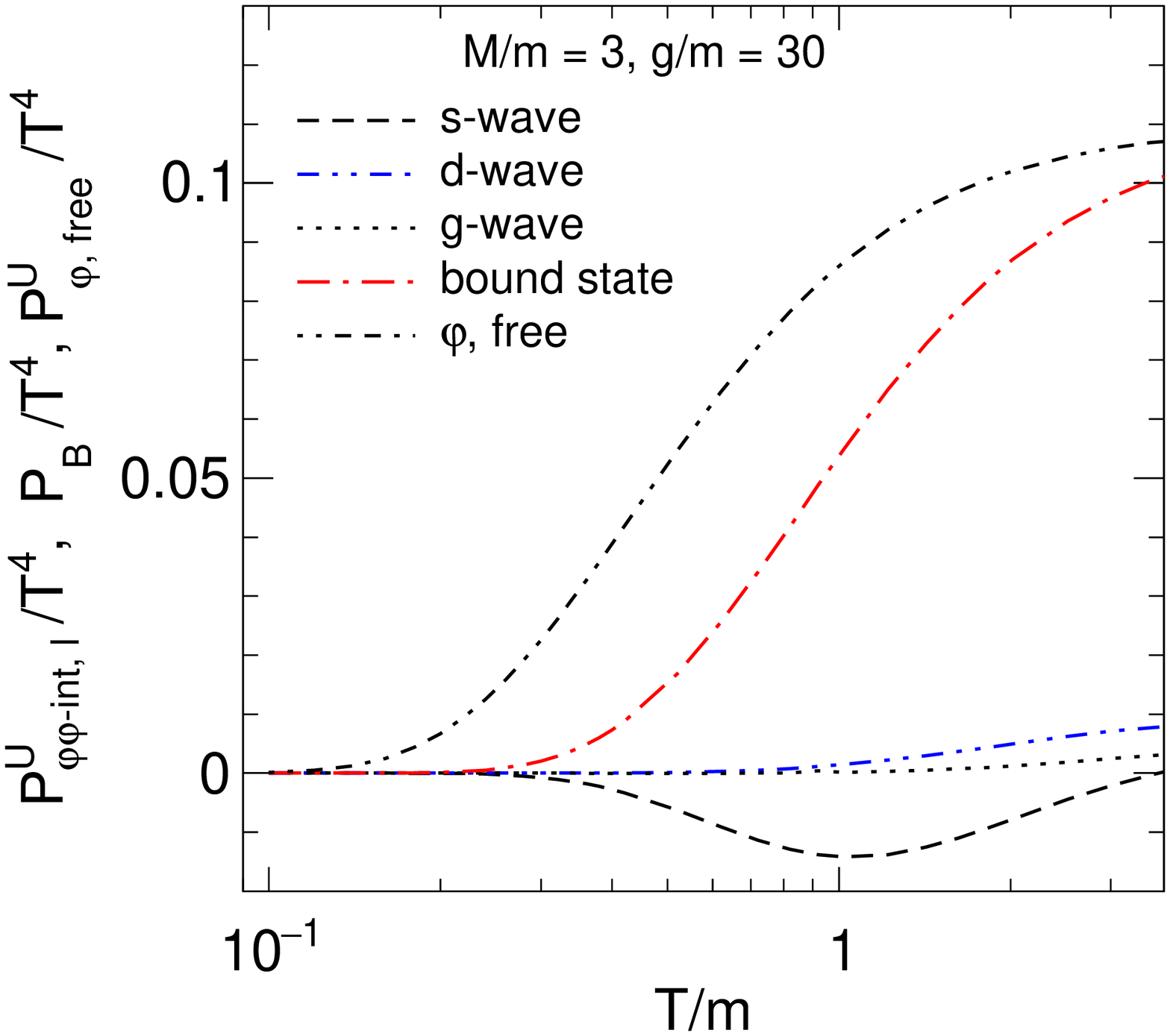}
\caption{The left panel shows the temperature dependence of the normalized pressure for s-, g-, and d-wave at $g = 10 m<g_{S,c}$ in presence of an intermediate state $S$ of mass $M = 3 m$. The results of d- and g-waves are multiplied by a factor of 10. A similar result for $g = 30 m>g_{S,c}$ is shown in the right panel of this figure. The pressure of the bound state and free particles$\varphi$ are also shown.}
\label{Fig:P_int_reso_vs_T_g_10}
\end{figure}
  
%The intermediate state $S$ is a resonance whose width is proportional to $g^2$. When $g$ is sufficiently small, the field $S$ can be treated as a free particle. However, when $g$ is large this approximation does not hold. In order to estimate the role of the interaction in the case without the bound state, we define the following quantity:

%In Fig. \ref{Fig:Ratio_P_int_reso_Ps} we show temperature dependence of $\zeta_{S}$ at $g = 10 m $ and $20 m$ (both below  $g_c$). For both cases the ratio is significantly larger than one, implying a large role of the interaction: one should not approximate such a gas bys simply adding $S$ as a free field. 
%As $T/m$ increases this ratio decreases and saturate to unity. This plot indicates that only when $T/m$ is large and $g$ is small, the interaction can be approximated by a free particle of mass $M$.

%\begin{figure}[ptb]
%\centering
%\includegraphics[width=0.48\textwidth]{p_int_s_d_g_bs_vs_T_uniterized_reso_g_30.eps}
%\caption{Similar to Fig. \ref{Fig:P_int_reso_vs_T_g_10} but for $g = 30 m>g_{S,c}$. The pressure of bound state and free particles$\varphi$ are also shown.}
%\label{Fig:P_int_reso_vs_T_g_30}
%\end{figure}

\begin{figure}[ptb]
\centering
\includegraphics[width=0.48\textwidth]{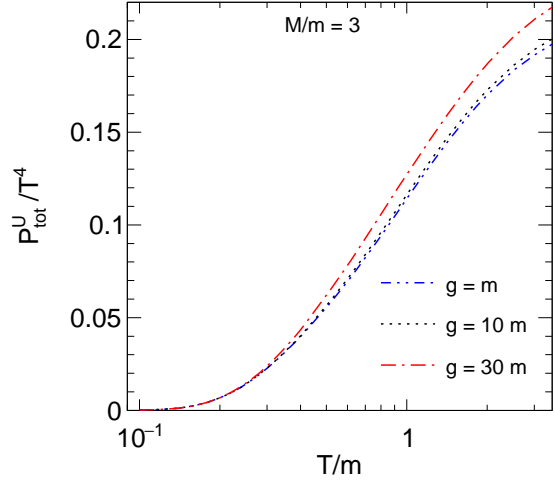}
\caption{Variation of normalized total pressure (Eq. \ref{eq:ptot}) with temperature in presence of an intermediate state of mass $M = 3 m$.}
\label{Fig:P_tot_reso_vs_T}
\end{figure}
 
Next, we calculate the temperature dependence of the pressure for s-, d- and g-waves in presence of an intermediate state $S$ of mass $M = 3 m$. In the left panel of Fig. \ref{Fig:P_int_reso_vs_T_g_10} we show the $P^{U}_{\varphi\varphi\text{-int}, l}/T^4$ for $g/m = 10$
(the d- and g-waves contributions are multiplied by 10 to make them visible).
For all three waves, the normalized pressure increases and saturates at large $T/m$. 
The right panel of Fig. \ref{Fig:P_int_reso_vs_T_g_10} shows the case $g = 30m> g_{M,c}$, for which a bound state forms. 
%Figure \ref{Fig:P_int_reso_vs_T_g_30} shows the case $g = 30m> g_{M,c}$, for which a bound state forms. 
%The  pressure of the bound state, as well as free particle $\varphi$, are also shown. Both of them  saturate at the value $\pi^2/90 \approx 0.109$ at high $T/m$. The normalized pressure for the s-wave is negative up to $T/m \approx 3.5$ and, beyond that, is slightly positive. The other two waves contribute positively. The magnitudes of three partial wave contributions are small compared to contributions of the free particles $\varphi$ and the bound state $B$.

The temperature dependence of the normalized total pressure as a function of $T/m$ is shown for three values of $g$ in Fig. \ref{Fig:P_tot_reso_vs_T}. When $g=m\ll g_{M,c}$, the attraction is very small and the resonance $S$ behaves almost like a free particle, thus basically the two free particles $\varphi$ and $S$ contribute to the pressure. At large $T$ the normalized pressure saturates to $2 \pi^2/90 \approx 0.219$. For the larger coupling $g = 10 m$, the state $S$ has a sizable width.  As a result, the normalized total pressure is larger than the one for $g = m$. For $g = 30 m$, there is, in addition, also a bound state.
Although the pressure of the s-wave is negative up to a certain $T/m$ (see the right panel of Fig. \ref{Fig:P_int_reso_vs_T_g_10}),
%Fig. \ref{Fig:P_int_reso_vs_T_g_30}), 
the overall effect of the interaction for $g = 30m$ is, as expected, positive. 
%The d- and g-waves also contribute positively, while the normalized total pressure for s-wave is slightly negative up to $T/m \approx 3.5$ and slightly positive beyond that. The overall effect of the interaction for $g/m = 30$ is, as expected, positive. 

Finally, in order to study the overall effect of the interaction, we define the following ratios:
\begin{equation}
 \eta_{S} = \frac{P^{U}_{\text{tot}}}{ P_{\varphi,\text{free}}+ P_{S,\text{free}}} \text{ ,}\label{eq:eta_S}
\end{equation}
and 
\begin{equation}
 \zeta_{S} = \frac{P^{U}_{\varphi\varphi-\text{int}} + P_{B}}{P_{S,\text{free}}+ P_{B}} \text{ .} \label{eq:zeta_S}
\end{equation}

\begin{figure}[ptb]
\centering
\includegraphics[width=0.48\textwidth]{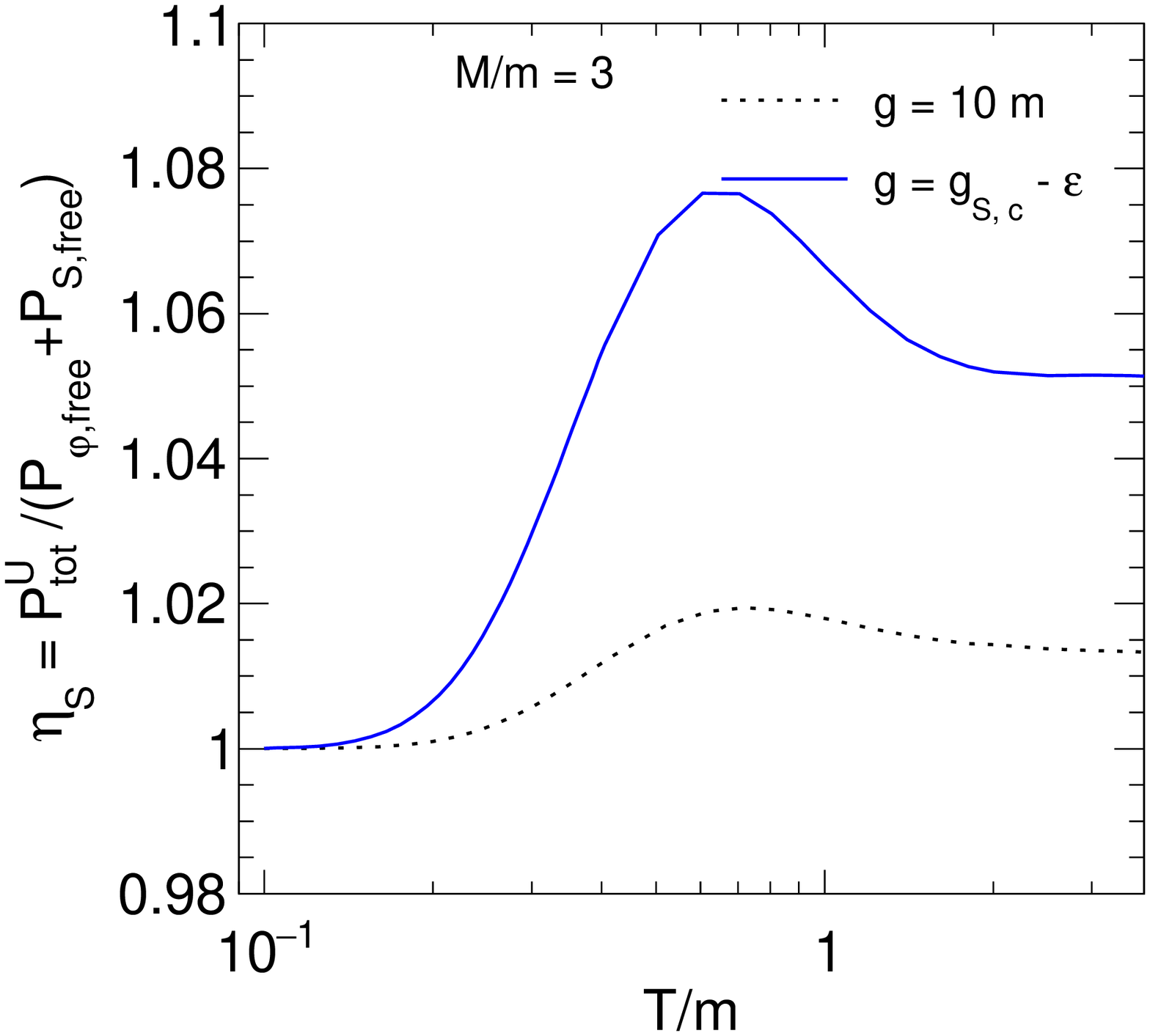}
\includegraphics[width=0.48\textwidth]{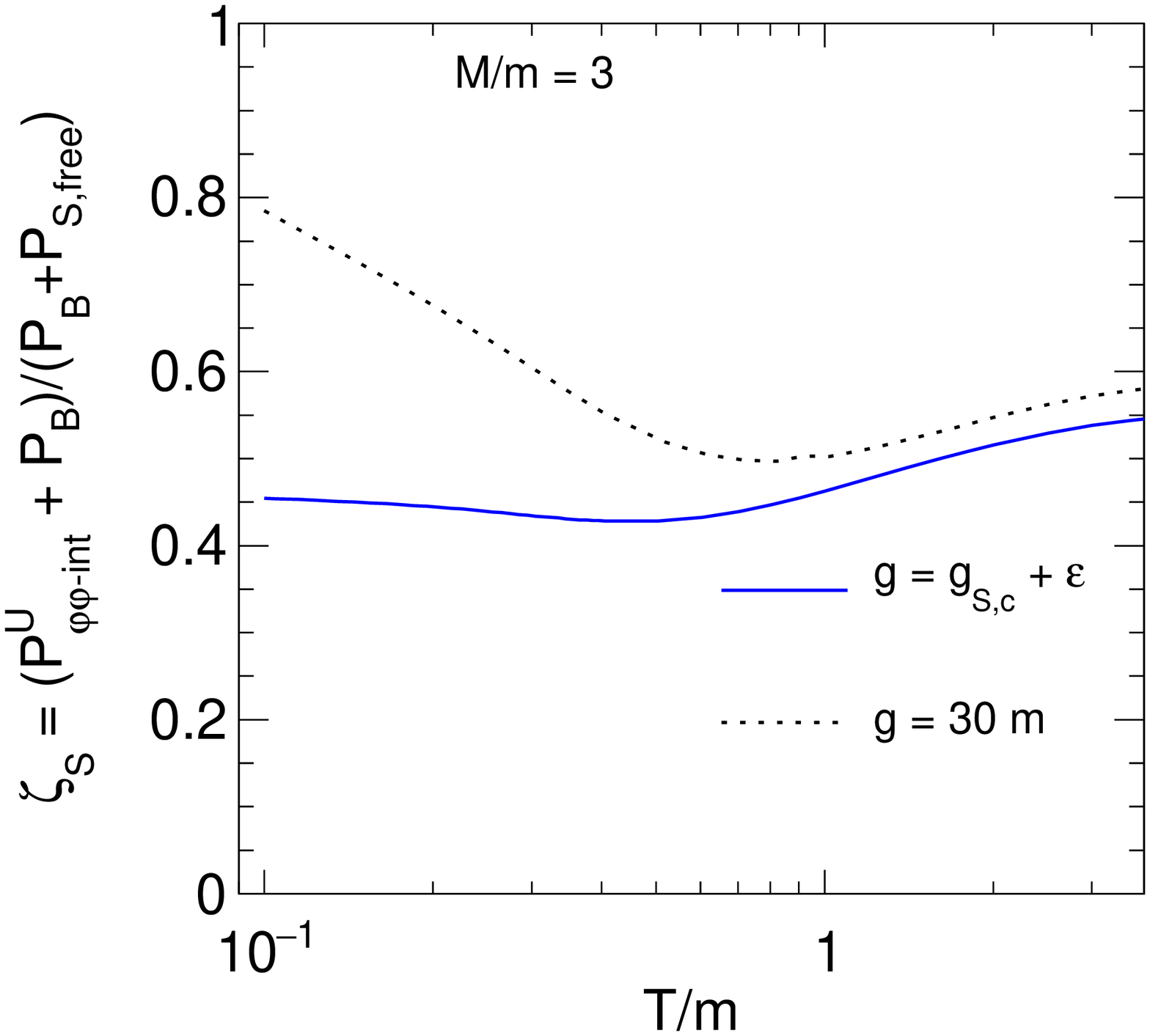}
\caption{Dependence of
%Variation of 
the ratio $\eta$ (Eq. \ref{eq:eta_S}) (Left) and $\zeta$ (Eq. \ref{eq:zeta_S}) (Right) with $T/m$ in presence of an intermediate state of mass $M = 3 m$ (no bound state present).}
\label{Fig:eta_reso_vs_T}
\end{figure}
   
%\begin{figure}[ptb]
%\centering
%\includegraphics[width=0.48\textwidth]{p_int_p_bs_vs_T_reso.eps}
%\caption{Variation of the ratio $\zeta$ (Eq. \ref{eq:zeta_S}) with temperature in presence of an intermediate state of mass $M = 3 m$ (a bound state is present).}
%\label{Fig:zeta_reso_vs_T}
%\end{figure}

In the left panel of Fig. \ref{Fig:eta_reso_vs_T} we show $\eta_{S}$ as a function of $T/m$ at two different $g$ below the critical value $g_{M,c}$. For both cases, $\eta_S$ is greater than one.  This implies that neglecting interaction would lead to an underestimation of the overall actual pressure.
In the right panel of Fig. \ref{Fig:eta_reso_vs_T} 
%Fig. \ref{Fig:zeta_reso_vs_T} 
we show $\zeta_{S}$ with $T$ for three different $g$, one just above the critical value $g_{M,c} \approx 22.83 m$ and for $g = 30 m$.
%and $g = 100 m$. 
For the first case, $\zeta_{S} \approx 1$ at low temperature but, as $T/m$ increases, it decreases and reaches a minimum value $\approx 0.43$ for $T/m \approx 0.5$. With a further increase of $T/m$, the ratio $\zeta_{S}$ slightly increases and saturates at a value $\approx 0.52$.
For $g = 30 m$, the behavior of $\zeta_{S}$ is qualitatively similar to the previous one, but at high $T/m$ it saturates to $\approx 0.55$. In both cases, neglecting the interaction would lead to a quite large overestimation of the actual results. 
%Finally, for the strong coupling $g = 100 m$, the ratio $\zeta_{S}$ is close to unity. Quite interestingly, the various effects compensate in such a way that the free gas with the resonance $S$ and the bound state $B$ reproduce quite well the interacting contribution to the pressure.

%It indicates that in presence of $S$ when the bound state is formed if we add the contributions of the bound state and $S$ just like free particles, then there will be overcounting of the actual effect which comes from the interaction. 
%Thus, if $T/m$ is sufficiently small, this approximation works quite well. 
%As $g$ increases, $\zeta_{B,S}$ increases at all temperature.
%$\zeta_{S}$ measures the overall effect of the interaction when a bound state is formed. 
%Namely, the naive inclusion of a particle $S$ and a bound state $B$ as free fields is not correct if $\zeta_{S}$ sizably deviates from unity. Whereas, $\eta_{S}$ indicates the effect of interaction when bound state is not formed. 

%\begin{longtable}{|p{.2\textwidth}|p{.3\textwidth}| p{.3\textwidth}|} 
% \hline

% \bf  Quantity  & \bf Expected value in absence of interaction & %\bf Estimated range \\ 
% \hline
% $\zeta$  & 1 & $\approx 0.09$ to $\approx 0.7$ \\ 
% \hline
 
% $\eta$  & 1 & \\
%\hline
% $\zeta_{\lambda}$  & 1 &  \\ 
% \hline
 
% $\eta_{\lambda}$  & 1 & $\approx 0.05$ to $\approx 1.2$\\
%\hline
% $\zeta_{B,S}$  & ? & $\approx 0.43$ to ? \\ 
% \hline
 
% $\eta_{S}$  & 1 & 1 to $\approx 1.08$ \\
% \hline

%\end{longtable} 
\section{Conclusions}\label{sec:Conclusions}
In this work, we have investigated the role of particle interaction in a thermal gas in the context of selected scalar QFTs that contain three-leg vertices and can lead to the formation of bound states, if the attraction is strong enough. 
To this end, we have calculated the scattering phase shifts of the s-, d-, and g-waves using the partial wave decomposition of two-body scattering and we implemented a non-perturbative unitarized one-loop resummed approach for which the theory is a unitary, finite and well defined for a large range of the three-leg coupling constant $g$.

\renewcommand*{\arraystretch}{1.6} 
\begin{table}[h] \centering
%EndExpansion
$%
\begin{tabular}
[c]{|c|c|c|c|}\hline
\textbf{QFT} & \textbf{Quantity} & \textbf{Total pressure }$P_{tot}^{U}$ &
\textbf{Figure}\\\hline
$\frac{g}{3!}\varphi^{3}$ & $\eta$ & $P_{tot}^{U}=\eta P_{\varphi,free}$ &  \ref{Fig:etavsT} (Left panel)
\\\hline
$\frac{g}{3!}\varphi^{3}+\frac{\lambda}{4!}\varphi^{4}$ & $\eta_{\lambda}$ &
$P_{tot}^{U}=\eta_{\lambda}P_{\varphi,free}$ & \ref{Fig:etalambdavsT} (Left Panel) \\\hline
$\frac{g}{2!}S\varphi^{2}$ & $\eta_{S}$ & $%
\begin{array}
[c]{c}%
P_{tot}^{U}=\eta_{S}\left(  P_{\varphi,free}+P_{S,free}\right)  \\
=P_{\varphi,free}+\zeta_{S}P_{S,free}%
\end{array}
$ & \ref{Fig:eta_reso_vs_T} (Left panel) \\\hline
\end{tabular}
\ \ \ $%
%TCIMACRO{\TeXButton{Caption}{\caption
%{Estimate of the interaction's role when no bound state forms ($g<g_c$).
%In each case, the expected value in absense of the interaction is one. \label
%{Table1}}}}%
%BeginExpansion
\caption
{Estimate of the interaction's role when no bound state forms ($g$ less than the respective critical values).
In each case, the expected value in absence of the interaction is one. \label
{Table1}}%
%EndExpansion%
%TCIMACRO{\TeXButton{E}{\end{table}}}%
%BeginExpansion
\end{table}%
%EndExpansion

\bigskip%

%TCIMACRO{\TeXButton{B}{\begin{table}[h] \centering}}%
%BeginExpansion
\renewcommand*{\arraystretch}{1.6} 
\begin{table}[h] \centering
%EndExpansion
$%
\begin{tabular}
[c]{|c|c|c|c|}\hline
\textbf{QFT} & \textbf{Quantity} & \textbf{Total pressure }$P_{tot}^{U}$ &
\textbf{Figure}\\\hline
$\frac{g}{3!}\varphi^{3}$ & $\zeta$ & $P_{tot}^{U}=P_{\varphi,free}+\zeta P_{B}$ & \ref{Fig:etavsT} (Right panel) \\\hline
$\frac{g}{3!}\varphi^{3}+\frac{\lambda}{4!}\varphi^{4}$ & $\zeta_{\lambda}$ &
$P_{tot}^{U}=P_{\varphi,free}+\zeta_{\lambda}P_{B}$ & \ref{Fig:etalambdavsT} (Right Panel)\\\hline
$\frac{g}{2!}S\varphi^{2}$ & $\zeta_{S}$ & $P_{tot}^{U}=P_{\varphi
,free}+\zeta_{S}\left(  P_{S,free}+P_{B}\right)  $ & \ref{Fig:eta_reso_vs_T} (Right panel) \\\hline
\end{tabular}
\ \ \ $%
%TCIMACRO{\TeXButton{Caption}{\caption
%{Estimate of the interaction's role when a bound state forms ($g>g_c$). In each case, the expected value in absense of the interaction is one. \label
%{Table2}}}}%
%BeginExpansion
\caption  
{Estimate of the  interaction's role when a bound state forms ($g$ greater than the respective critical values). In each case, the expected value in absence of the interaction is one. \label
{Table2}}%
%EndExpansion%
%TCIMACRO{\TeXButton{E}{\end{table}}}%
%BeginExpansion
\end{table}%

In all cases, we studied the role of the interaction and realized that, in general, it is non-negligible. 
Even in the case when no bound state is present, a sizable role of the interaction implies that the simple inclusion of a gas of free $\varphi$ particles maybe not be sufficient. In general, an attractive interaction (alias, a positive phase shift derivative close to threshold) implies an increase in pressure and vice-versa. This is understandable in gas at zero chemical potential determined only by the vacuum's density of states, but it should be stressed that this is quite different from the classical case, e.g. the van der Waals gas, in which (at a fixed number of particles) a repulsion induces an increase of the pressure, see also the discussion in Ref. \cite{Yen:1997rv}. 
   
Conversely, when a bound state forms, the simple inclusion of the bound state into the thermal gas is not enough for a correct description of the pressure of the system. The derivative of the phase shift switches sign and a partial cancellation between the corresponding negative contribution of the interaction with the positive one of the bound state occurs. 

%An interesting case is the QFT with interaction $S\varphi^2$, in which $S$ represents a resonance when it is heavy enough. 

We summarize our results in Table I and Table II. In table, I, the cases in which the interaction does not lead to the formation of a bound state are presented, while in Table II the cases for which a bound state appears are listed.  
It is then clear that the answer to our original questions about the role of the interaction as well as that of the bound state and resonances is not a simple one. The results depend on the coupling strength and eventually on other parameters and on the temperature range. Yet, as a general statement, our results show that the role of the interaction can be sizable and the consideration of a simple free gas is in most cases insufficient. 

In the end, we turn back to the original question formulated in the introduction  about the quite peculiar production of a bound state at a nonzero temperature in thermal models, according to which the multiplicity depends solely on the mass of the bound state but is not affect by the typically large dimension or the binding energy of the composite object. 
In our approach, the phase shift calculated in the vacuum is the quantity that is used to obtain the properties of the system at any temperature.  Since the  partition function
$Z=\sum_{n}e^{-E_{n}/T}$ is determined by the energy eigenvalues $E_n$ calculated in the vacuum, once these are known, the quantity $Z$ is fixed for each temperature. In other words, $Z$ is solely fixed by vacuum physics. 
In our work, the sum over the states is replaced by an integral over the derivatives of the spectral function, but the basic idea is the same since the pressure is still determined by vacuum quantities. One may then speculate that the production (alias the multiplicity) of a bound state is not related to the dimension of the composite objects but is solely controlled by the corresponding thermal integral as thermal models suggest, but one needs to investigate this issue more in detail in the future. 

As an additional important outlook, we mention the extension of the present work to particles with spins (both bosons and fermions), with particular attention to the study of nuclei as bound states of nucleons, the easiest of such systems being the deuteron, as well as to resonances that do not fit into the quarkonium picture, such as the famous exotic meson $X(3872)$.

\bigskip

\textbf{Acknowledgments: }
The authors thank W. Broniowski and S. Mrówczyński for useful discussions about the thermodynamics of bound states and A. Pilloni for scattering issues. 
F.G. acknowledges financial support from the Polish National Science Centre NCN
through OPUS project no. 2019/33/B/ST2/00613. 
S.S. acknowledges financial support from the Ulam Scholarship of the Polish
National Agency for Academic Exchange (NAWA) with agreement no: PPN/ULM/2019/1/00093/U/00001. 

\appendix
\section{Brief recall of the phase shift formula}
In this Appendix, for completeness, we recall a simple QM-based justification of the
phase shift formula \cite{Florkowski:2010zz} that shows how it is linked to the density of
states. The radial wave function with angular momentum $l$ of a particle
scattered by central potential $U(r)$ is
\begin{equation}
\psi_{l}(r)\propto\sin[kr-l\pi/2+\delta_{l}]\text{ ,}%
\end{equation}
where $k$ is the length of the three-momentum
and $\delta_{l}$ is the phase shift due to the interaction with the potential.

If we confine our system into a sphere of radius $R,$ the condition
$kR-l\pi/2+\delta_{l}=n\pi$ with $n=0,1,2,..$ holds because $\psi_{l}(r)$ must
vanish at the boundary. Conversely, the number of states $n_{0}$ when $k$
belongs to the range $(0,k_{0})$ is given by $n_{0}=\left(  k_{0}%
R-l\pi/2+\delta_{l}\right)  /\pi.$ Then, the density of states that one can
place between $k$ and $k+dk$ is given by%
\begin{equation}
\frac{dn_{l}}{dk}=\frac{R}{\pi}+\frac{1}{\pi}\frac{d\delta_{l}}{dk}\text{ ,}%
\end{equation}
where the first term describes the density of states $\frac{dn_{l}^{free}}%
{dk}$ in absence of interactions, while the second term $\frac{1}{\pi}%
\frac{d\delta_{l}}{dk}$ describes the effect of the interacting potential.
When translating these results from QM to QFT,
we replace the momentum $k$ by the invariant mass $x=\sqrt{s}.$ For
instance, in the case of the $\varphi^{3}$-QFT and for $l=0$ and no bound state,
one has:
\begin{equation}
\frac{dn_{0}}{dx}=\delta(x-m)+\frac{1}{\pi}\frac{d\delta_{0}}{dx}\text{ ,}%
\end{equation}
where the phase shift is nonzero for $x\geq2m.$ If a bound state in the s-wave
channel forms, one has:
\begin{equation}
\frac{dn_{0}}{dx}=\delta(x-m)+\delta(x-M_{B})+\frac{1}{\pi}\frac{d\delta_{0}%
}{dx}\text{ .}%
\end{equation}
Indeed, the formal extension outlined in Sec. II.C amounts to
\begin{equation}
\frac{dn_{0}}{dx}=\frac{1}{\pi}\frac{d\delta_{0}^{\text{(extended)}}}%
{dx}\text{ .}%
\end{equation}

\section{Loop function for small but nonzero $\varepsilon$}

The loop function $\Sigma $ is, in general, a complex function which (on its
first Riemann sheet) is regular everywhere apart from a cut that starts
from a certain threshold $s_{th}$ to $\infty .$ Upon considering (at first)
the case without subtractions, its general form as a function of the complex
variable $z$ is given by: 
\begin{equation}
\Sigma (z)=-\frac{1}{\pi }\int_{s_{th}}^{\infty }ds^{\prime }\frac{%
f(s^{\prime })}{z-s^{\prime }}\text{ ,}  \label{loopgeneral}
\end{equation}%
which is well-defined for each $z\in 
%TCIMACRO{\U{2102} }%
%BeginExpansion
\mathbb{C}
%EndExpansion
$ except for the cut $(s_{th},\infty ).$ Above, $f(s)$ is a well-defined
function (that can be continued to the whole complex plane: $f(s)\rightarrow
f(z);$ this is necessary to go to the second Riemann sheet, yet we will not
deepen this matter here). In our specific case (see also below) we have: 
\begin{equation}
f(s)=\frac{\sqrt{s-s_{th}}}{32\pi \sqrt{s}}\text{ , }s_{th}=4m^{2}\text{ .}
\end{equation}%
When convergence is not guaranteed (as for the $f(s)$ above), subtractions
are necessary. For instance, with two subtractions (as in the main text) one
has%
\begin{equation}
\Sigma _{\text{2S}}(z)=-\frac{1}{\pi }(z-s_{1})(z-s_{2})\int_{s_{th}}^{%
\infty }ds^{\prime }\frac{f(s^{\prime })}{\left( z-s^{\prime }\right) \left(
s^{\prime }-s_{1}\right) (s^{\prime }-s_{2})}\text{ ,}
\end{equation}%
for $s_{1,2}<s_{th}.$ It is, however, important to stress that subtractions
do not affect the behavior of the imaginary part that we shall describe
below. We shall then omit them in the following.

As it is common (being an outcome of the Feynman prescription, as we
show explicitly below in the case of scalar QFT) the cut is
moved to the negative axis by a small amount, leading to: 
\begin{equation}
\Sigma (z)=-\frac{1}{\pi }\int_{s_{th}}^{\infty }ds^{\prime }\frac{%
f(s^{\prime })}{z-s^{\prime }+i\varepsilon } ]text{ .}
\end{equation}%
Namely, the cut is now located at $(s_{th}-i\varepsilon ,\infty
-i\varepsilon ),$ where $\varepsilon $ is an infinitesimal (but nonzero)
number. 

Since the cut is below the real axis, we can now consider $z=s\in $ $%
%TCIMACRO{\U{211d} }%
%BeginExpansion
\mathbb{R}
%EndExpansion
$ (including $(s_{th},\infty )$). The imaginary part for $z=s\in $ $%
%TCIMACRO{\U{211d} }%
%BeginExpansion
\mathbb{R}
%EndExpansion
$ takes the form:%
\begin{equation}
\operatorname{Im}\Sigma (z=s)=\int_{s_{th}}^{\infty }ds^{\prime }f(s^{\prime })\frac{%
1}{\pi }\frac{\varepsilon }{(s-s^{\prime })^{2}+\varepsilon ^{2}}.
\end{equation}%
It is useful to distinguish three regions for this object.

(i) For $s-s_{th}\ll -\varepsilon ,$ the previous expression reads:%
\begin{equation}
\operatorname{Im}\Sigma (z=s)=\varepsilon \int_{s_{th}}^{\infty }ds^{\prime
}f(s^{\prime })\frac{1}{\pi }\frac{1}{(s-s^{\prime })^{2}}=\varepsilon
h(s)\sim \varepsilon 
\end{equation}%
where $h(s)$ is a certain finite function for the considered range of
interest; thus $\operatorname{Im}\Sigma (z=s)$ is an infinitesimal (but nonzero) quantity 
for a non-vanishing $\varepsilon $.

(ii) For $-\varepsilon \lesssim s-s_{th}\lesssim \varepsilon $ (for $s$
close to threshold), one has: 
\begin{equation}
\int_{s_{th}}^{\infty }ds^{\prime }f(s^{\prime })\frac{1}{\pi }\frac{%
\varepsilon }{(s-s^{\prime })^{2}+\varepsilon ^{2}}\sim \sqrt{\varepsilon }
\text{ ,}
\end{equation}%
as can be seen by a lengthy but explicit calculation. A simple verification
can be obtained by setting $s=s_{th}$ and by approximating $f(s)$ as $%
f(s)\simeq c\sqrt{s-s_{th}}$ (valid close to threshold, where the dominant
part of the integral comes from). In this case, the integral reads $c\sqrt{%
\varepsilon /2}$ (with $c$ being a positive constant).

(iii) For $s-s_{th}\gg \varepsilon ,$ the quantity $\frac{1}{\pi }\frac{%
\varepsilon }{(s-s^{\prime })^{2}+\varepsilon ^{2}}$ can be safely replaced
by $\delta (s-s^{\prime })$ when $\varepsilon $ is sufficiently small, leading
to 
\begin{equation}
\operatorname{Im}\Sigma (z=s)=f(s) \text{ .}
\end{equation}%
By going to the the next order in $\varepsilon$, one has 
\begin{equation}
\operatorname{Im}\Sigma (z=s)=f(s)+\varepsilon r(s),
\end{equation}%
where $r(s)$ is a certain finite function given by $r(s)=\lim_{\varepsilon \rightarrow 0^{+}}\int_{s_{th}}^{\infty }ds^{\prime }f(s^{\prime })\frac{1}{\pi }\frac{%
(s-s^{\prime })^{2}-\varepsilon ^{2}}{[(s-s^{\prime })^{2}+\varepsilon^{2}]^{2}}$.
[Note, the function $r(s)$ can be calculated by taking the Fourier transform of the
Lorentzian $\delta $-function, which then leads to 
$r(s)=\int_{-\infty}^{+\infty}\frac{dq}{\sqrt{2\pi }}(-\left\vert q\right\vert )F(q,s)$, where 
$F(q,s)=\int_{s_{th}}^{\infty }\frac{ds^{\prime }}{\sqrt{2\pi }}f(s^{\prime})e^{-iq\cdot (s^{\prime }-s)}$.]

Finally, in the transition regions between (i) and (ii) the small quantity
rises from $\varepsilon $ to $\sqrt{\varepsilon },$ while between (ii) and
(iii) from $\sqrt{\varepsilon }$ to a finite number.

\begin{figure}[ptb]
\centering
\includegraphics[width=0.48\textwidth]{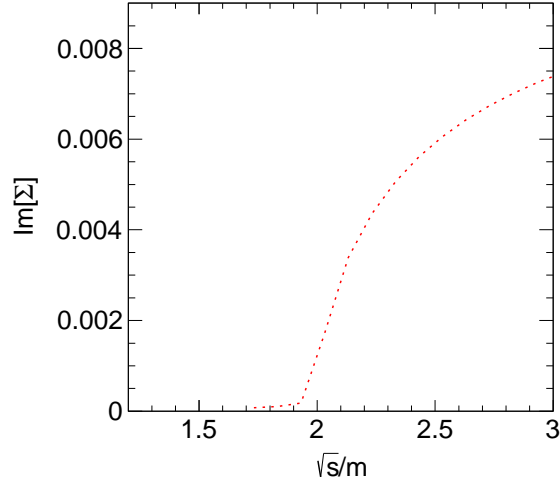}\caption{Energy dependence of the function Im$\Sigma(z=s)$ for a nonzero $\varepsilon$ (Eq. \ref{eq:ImSigma}).}
\label{Fig:ImSigma}
\end{figure}

Thus, we may summarize the outcome as%
\begin{equation}
\operatorname{Im}\Sigma (z=s)\left\{ 
\begin{array}{c}
\epsilon \propto \varepsilon \text{ for }s-s_{th}\ll -\varepsilon  \\ 
\epsilon \propto \sqrt{\varepsilon }\text{ for }-\varepsilon \lesssim
s-s_{th}\lesssim \varepsilon  \\ 
f(s)+\epsilon \text{ with }\epsilon \propto \varepsilon \text{ for }%
s-s_{th}\gg \varepsilon 
\end{array}%
\right. \text{ ,}
\label{eq:ImSigma}
\end{equation}%
whose schematic and illustrative behavior is reported in Fig.\ref{Fig:ImSigma}.
In particular, one may appreciate that the function is continuous and very
small (but nonzero) below threshold \footnote{%
When subtractions are considered, $\operatorname{Im}\Sigma (z=s)$ is indeed exactly
zero at the subtraction points. Moreover, $\operatorname{Im}\Sigma (z=s)\propto
\varepsilon $ for $s<m^{2},$ $-\varepsilon $ for $m^{2}<s<3m^{2}$, and $%
\varepsilon $ for $3m^{2}<s\ll 4m^{2}-\varepsilon $. This feature shows also why the
bound state is expected above $3m^{2}$. Nevertheless, the qualitative
discussion remains unchanged.}. By expressing the part below threshold by
an infinitesimal quantity $\epsilon $ (to be distinguished from the
original $\varepsilon $) we obtain Eq. (\ref{imsigma}) presented in the main
text. Of course, in this respect $\epsilon $ is in general a rather
complicated function of $\varepsilon $ and $s$, but the important point here
is that it is a very small number that approaches $0$ when $\varepsilon $
goes to zero.

As a final step, we show how the loop function of Eq. (\ref{loopgeneral})
emerges from the standard Feynman rules. We start with the very well known
expression of the loop of two scalar particles with mass $m$ as a function of
the overall momentum $p^{2}$ ($p$ being the overall momentum of the
two-particle system):%
\begin{equation}
-i\Sigma (p^{2})=\frac{1}{2}\int \frac{d^{2}q}{(2\pi )^{4}}\frac{i}{%
(p/2+q)^{2}-m^{2}+i\frac{\varepsilon }{2}}\frac{i}{(p/2-q)^{2}-m^{2}+i\frac{%
\varepsilon }{2}}
\end{equation}%
where the infinitesimal quantity $\varepsilon$ is chosen according to the Feynman prescription
(roughly speaking, positive energy solutions propagate forward in time and
negative energy solutions backward). In other words, we add a small
imaginary part to the particle $\varphi ,$ thus making itself unstable. The
factor $1/2$ in front of the integral is due to identical particles. The
choice of $\varepsilon /2$ is for future convenience. Taking for simplicity $p=(\sqrt{s},\mathbf{0})$
(rest frame for the colliding particles), we have: 
\begin{equation}
\Sigma (s)=-i\frac{1}{2}\int \frac{d^{4}q}{(2\pi )^{4}}\frac{1}{(\sqrt{s}%
/2+q^{0})^{2}-\mathbf{q}^{2}-m^{2}+i\frac{\varepsilon }{2}}\frac{1}{(\sqrt{s}%
/2-q^{0})^{2}-\mathbf{q}^{2}-m^{2}+i\frac{\varepsilon }{2}}
\end{equation}%
The integral over $q^{0}$ can be performed by a standard residuum calculus
(here, we keep track of the $\varepsilon $, since this is important for our
purposes):%
\begin{equation}
\Sigma (s)=-\frac{1}{2}\int \frac{d^{3}q}{(2\pi )^{2}}\frac{1}{E-i\frac{%
\varepsilon }{4E}}\frac{1}{s-4E^{2}+i\varepsilon }\text{ .}
\end{equation}%
with $E=\sqrt{\mathbf{q}^{2}+m^{2}}$ (note, formally $\varepsilon $ has
dimension $[$Energy$^{2}]$). Since $E>0$ (we do not consider the case of
zero masses), the previous expression simplifies as 
\begin{equation}
\Sigma (s)=-\frac{1}{2}\int \frac{d^{3}q}{(2\pi )^{2}}\frac{1}{E}\frac{1}{%
s-4E^{2}+i\varepsilon }\text{ .}
\end{equation}%
As a last step, upon introducing $s^{\prime }=4E^{2}$ as a variable, we get 
\begin{equation}
\Sigma (s)=-\frac{1}{2\cdot 16\pi ^{2}}\int_{s_{th}=4m^{2}}^{\infty
}ds^{\prime }\frac{\sqrt{s^{\prime }-4m^{2}}}{\sqrt{s^{\prime }}}\frac{1}{%
s-s^{\prime }+i\varepsilon }\text{ }
\end{equation}%
which coincides with the loop of Eq. (\ref{loopgeneral}). 

\bigskip

\section{$N/D$ unitarization scheme}
In this Appendix, we present an alternative unitarization scheme for the case of the $\varphi^3$-QFT. To this end, we choose the well-known
$N/D$ scheme \cite{Frazer:1969euo,Hayashi:1967bjx,Gulmez:2016scm,Oller:2020guq,Cahn:1983vi,Mai:2022eur,Giacosa:2021brl}. At the lowest order, the unitarized amplitude reads \cite{Cahn:1983vi,Gulmez:2016scm}
\begin{equation}
A_{l}^{N/D}(s)=\frac{N_{l}(s)}{D_{l}(s)},
\end{equation}
where $N_{l}(s)=A_{l}(s)$ (tree-level result, see Sec. II), and the denominator
takes the form%
\begin{equation}
D_{l}(s)=1-\dfrac{(s-m^{2})}{\pi}\int_{4m^{2}}^{\infty}\frac{\rho(s_{1}%
)N_{l}(s_{1})}{(s_{1}-s-i\varepsilon)(s_{1}-m^{2})}ds_{1}\text{ .}\label{den}%
\end{equation}
The right-hand-cut is
contained in $D_{l}(s)$ ($\operatorname{Im}D_{l}=-\rho(s)N_{l}(s)\theta(s-4m^{2})$, while the left-hand cut is contained in $N_{l}(s).$    
Here, a single subtraction (at the particle pole $s=m^2$) is implemented: this is enough to guarantee
convergence and, within this method, there is no problem with emerging
ghosts.

Quite interestingly, the once-subtracted on-shell
unitarization scheme is obtained if we approximate $D_{l}(s)$ as
\begin{equation}
D_{l}(s) \approx 1-N_{l}(s)\dfrac{(s-m^{2})}{\pi}\int_{4m^{2}}^{\infty}\frac{\rho(s_{1}%
)}{(s_{1}-s-i\varepsilon)(s_{1}-m^{2})}ds_{1}\text{ .}
\end{equation}
Roughly
speaking, the quantity $N_{l}(s)$ is taken outside the integral. Yet, as
discussed in the main text, this scheme leads to issues related to the
emergence of a ghost pole.

As a consequence of Eq. (\ref{den}), the bound state in the $N/D$ approach, in any given wave, is realized by
$D_{l}(s)=0$ for $s<4m^{2},$ but only the case $l=0$ is relevant for our
purposes. In the present unitarization, the bound state mass (if existent)
belongs to the interval $(m,2m)$, thus the range is different from the one of
the on-shell approximation $(\sqrt{3}m,2m)$. This aspect shows explicitly what
was discussed in\ Secs. II.B and II.C: different unitarization schemes typically agree
when the bound state mass is not too far from the threshold, but may
be quite different when the coupling constant becomes too large.

Let us now discuss in more detail the s-wave. The numerator reads explicitly%

\begin{equation}
N_{0}(s)=A_{0}(s)=\frac{-g^{2}}{s-m^{2}}+2g^{2}\frac{\ln\left[  1-\frac
{s-4m2}{m^{2}}\right]  }{s-4m^{2}}\text{ ,}%
\end{equation}
thus the left-hand cut with branch point at $s=3m^{2}$ as well as a
single-particle pole is encoded in the numerator. The denominator $D_{0}(s)$ can be then
evaluated numerically from Eq. (\ref{den}).

\begin{figure}
    \centering
    \includegraphics[width=0.48\textwidth]{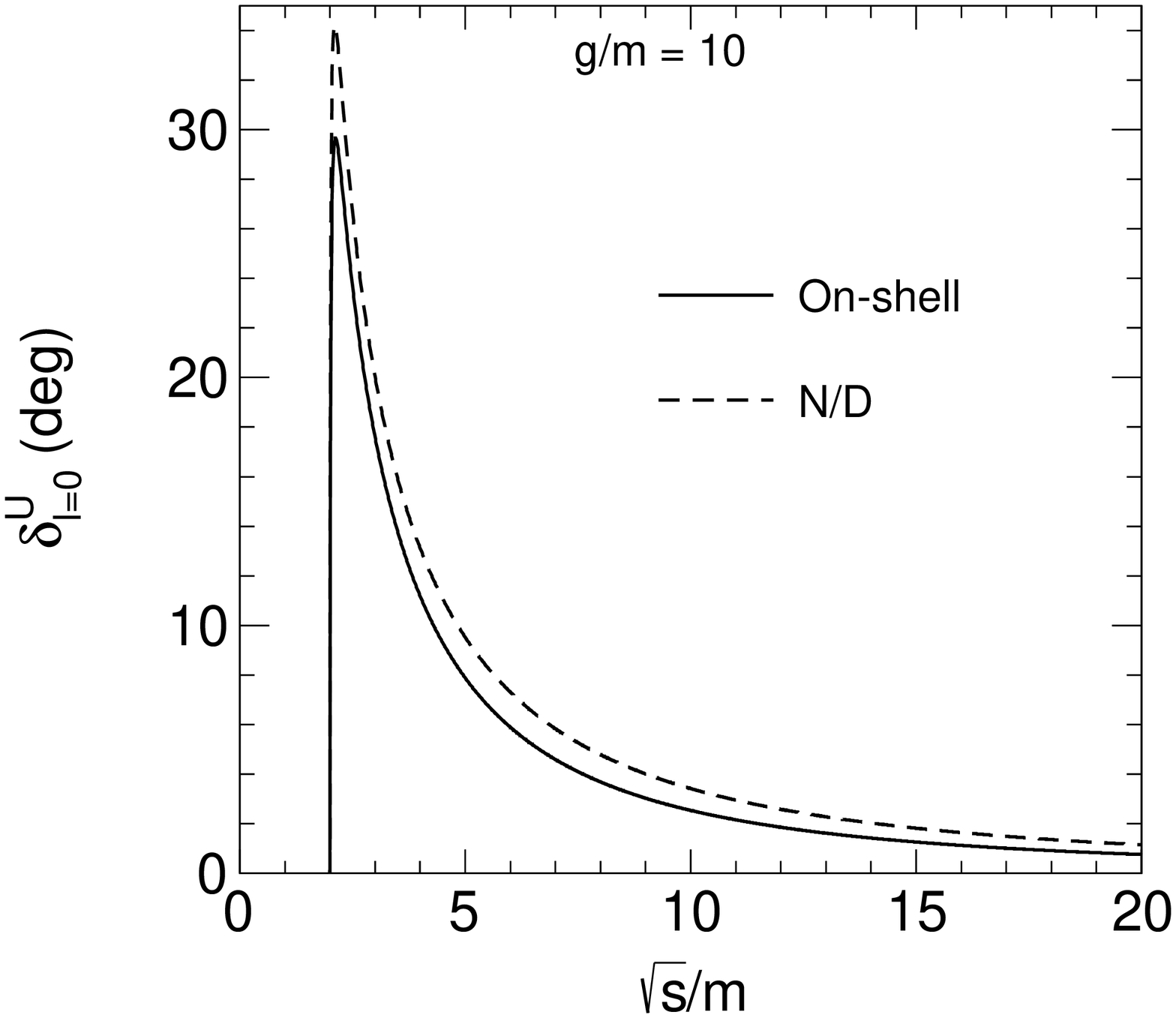}
     \includegraphics[width=0.48\textwidth]{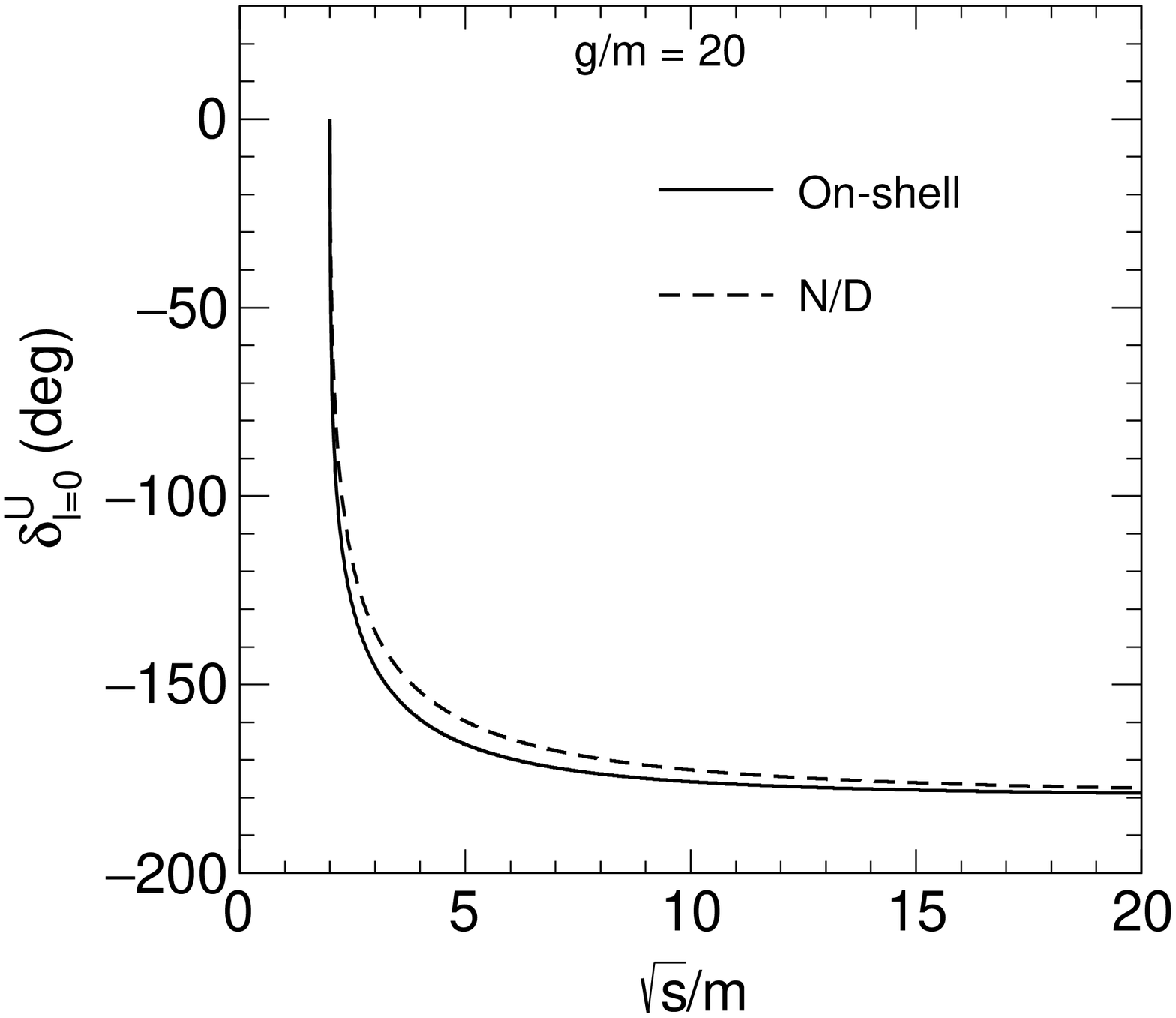}
    \caption{Comparisons of the energy dependence of s-wave phase shifts in two different unitarization approaches. The left panel is for $g/m =10$ and the right panel is for $g/m =20$.  }
    \label{fig:ps_unitarization_com}
\end{figure}
 
in Fig. \ref{fig:ps_unitarization_com} we present the s-wave phase shift for $g/m=10$ (no bound state in both $N/D$ and on-shell schemes) and $g/m=20$ (bound state present in both approaches),
where it is compared to the result of the on-shell scheme. As it is visible, both results are very similar.

\begin{figure}
    \centering
    \includegraphics[width=0.48\textwidth]{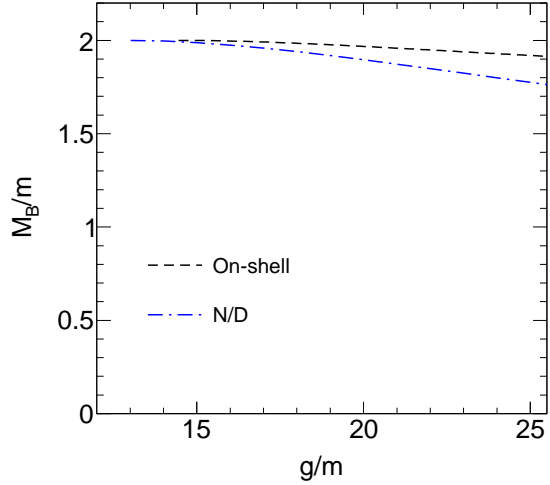}
    \caption{Comparisons of bound state masses in two different unitarization approaches. }
    \label{fig:mB_unitarization_com}
\end{figure}

Next, the critical value of $g_{c}$ for obtaining a bound state reads
$g_{c}^{N/D}/m=13.0177,$ that compares well to $g_{c}/m=14.4551$
discussed in Sec. II.B. The behavior of the mass of the bound state as a function
of $g$ is depicted in Fig. \ref{fig:mB_unitarization_com} for both approaches, displaying a comparable behavior.

\begin{figure}
    \centering
    \includegraphics[width=0.48\textwidth]{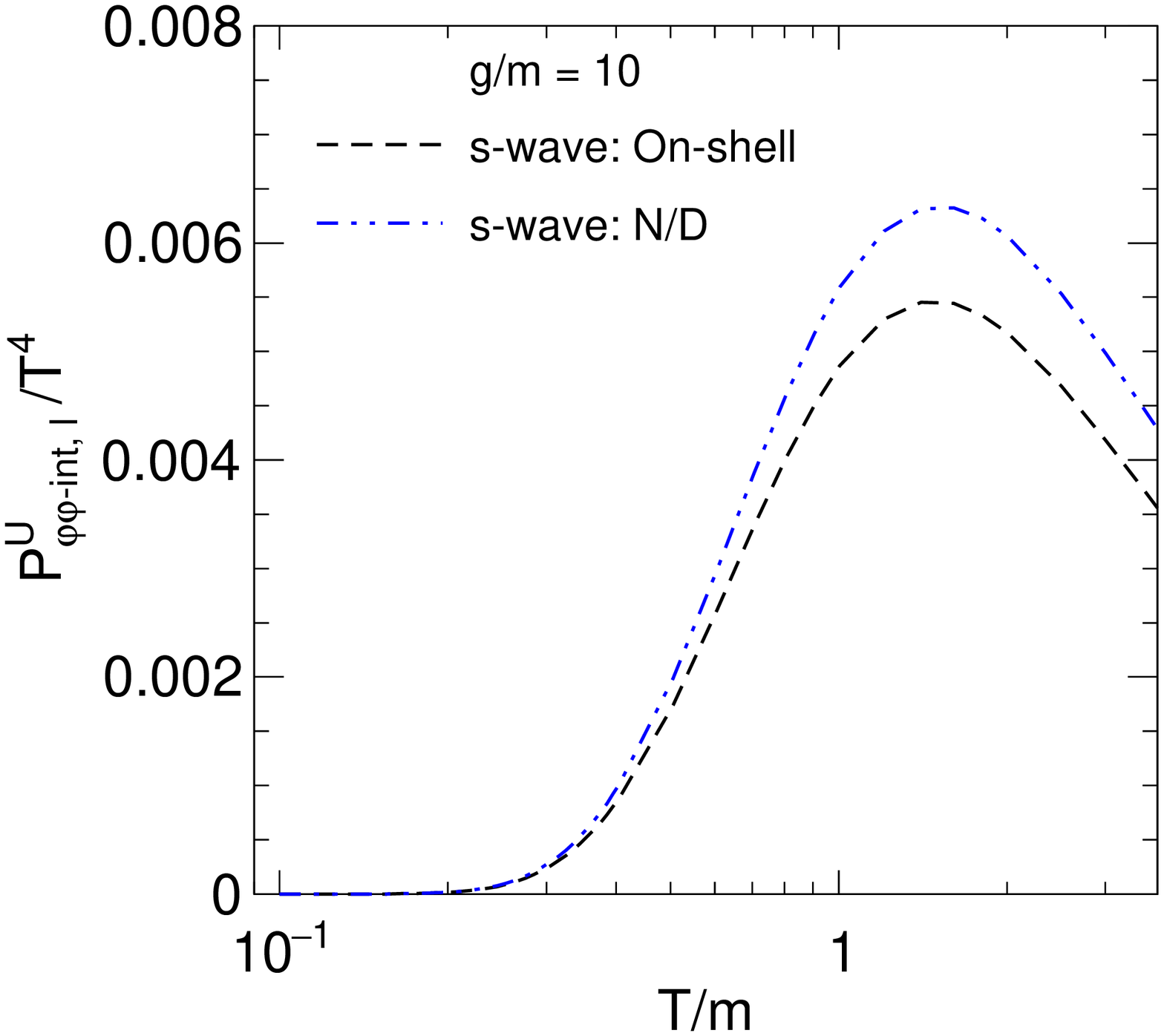}
     \includegraphics[width=0.48\textwidth]{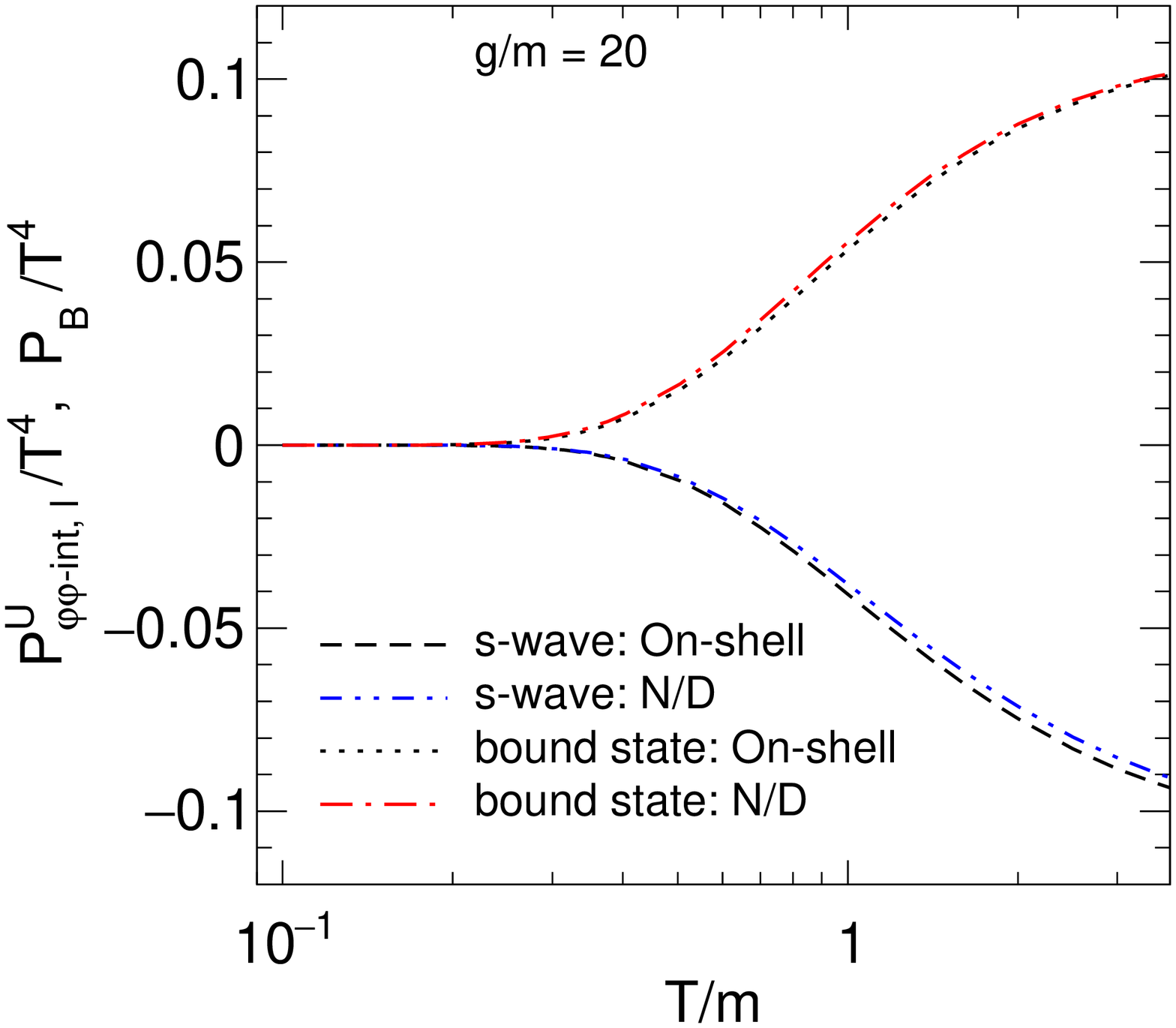}
  The l  \caption{Left panel compares the temperature dependence of the normalized pressure for s-wave in two different unitarization approaches for $g = 10m$. The right panel shows a similar plot for $g =20m$ where the bound state contribution is also included. }
    \label{fig:P_T_unitarization_com}
\end{figure}   
 
Finally, in Fig. \ref{fig:P_T_unitarization_com} we show the s-wave contribution of the pressure for both
unitarization schemes. Also, in this case, the results are numerically close to each other for the considered range of temperatures.

In conclusion, the results of the $N/D$ approach confirm qualitatively the
ones shown in the main part of the manuscript. Of course, one could go beyond
the lowest order in $N/D$ and/or also attempt other unitarizations. This
task is left for the future.

\section{Causality of the $\varphi^3$ and $S\varphi^2$ QFTs}
\label{sec:causality}

\begin{figure}[ptb]
\centering
\includegraphics[width=0.48\textwidth]{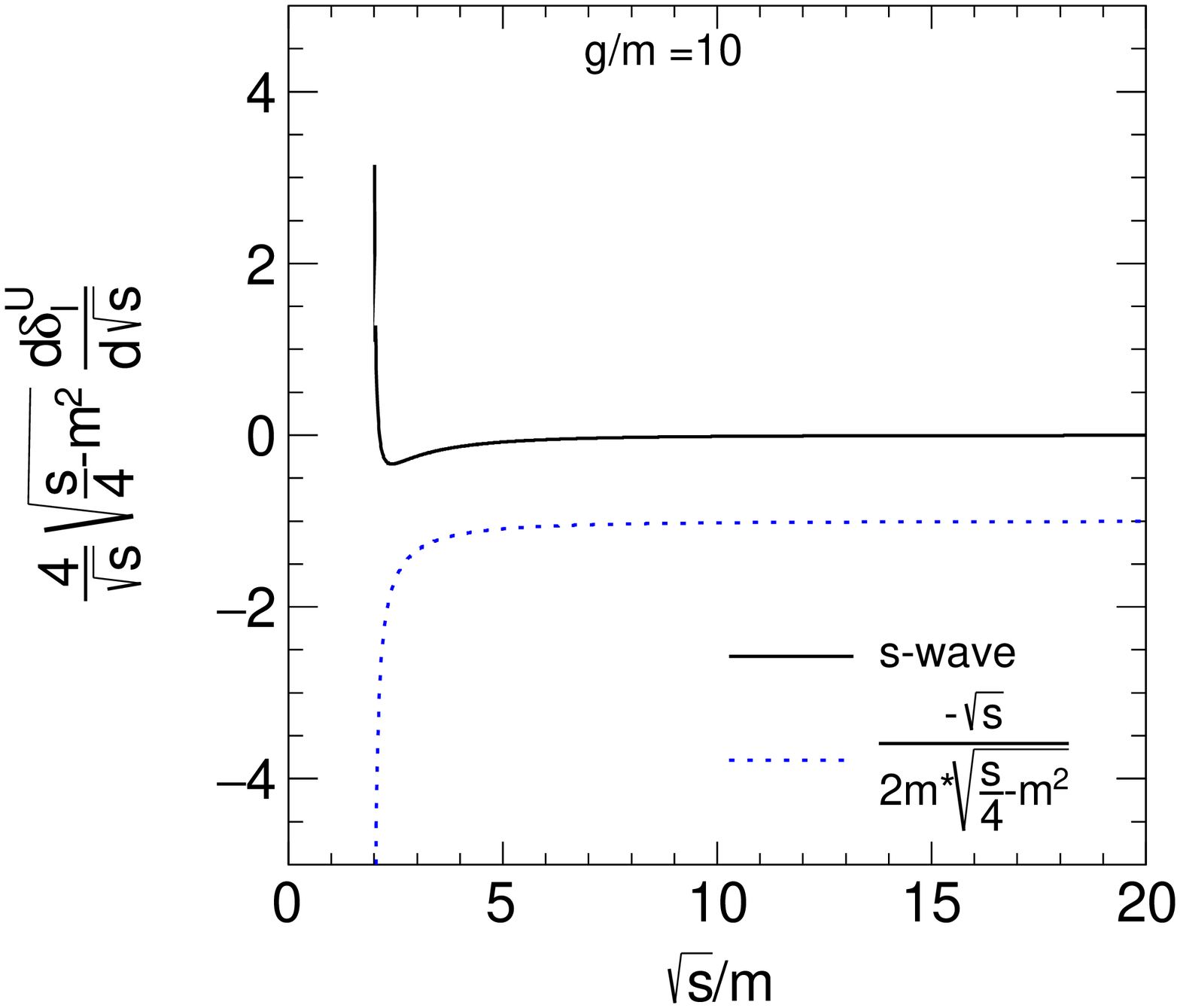}
 \includegraphics[width=0.48\textwidth]{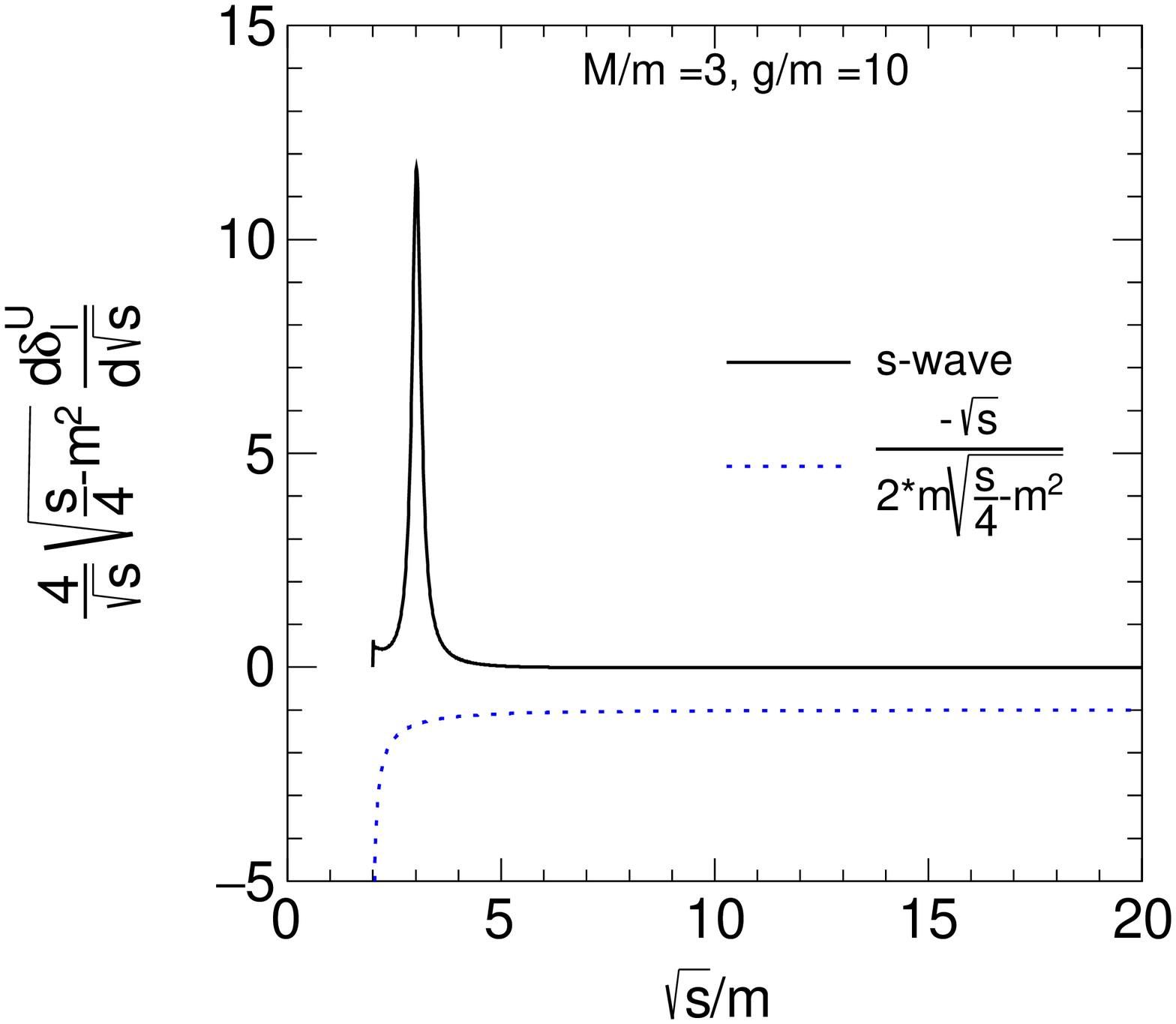}
\caption{Wigner's causality condition for $\varphi^3$ (left) type interaction and $S\varphi^2$ (right) type interaction. }
\label{Fig:causality} 
\end{figure} 
  
\begin{figure}[ptb]
\centering
\includegraphics[width=0.48\textwidth]{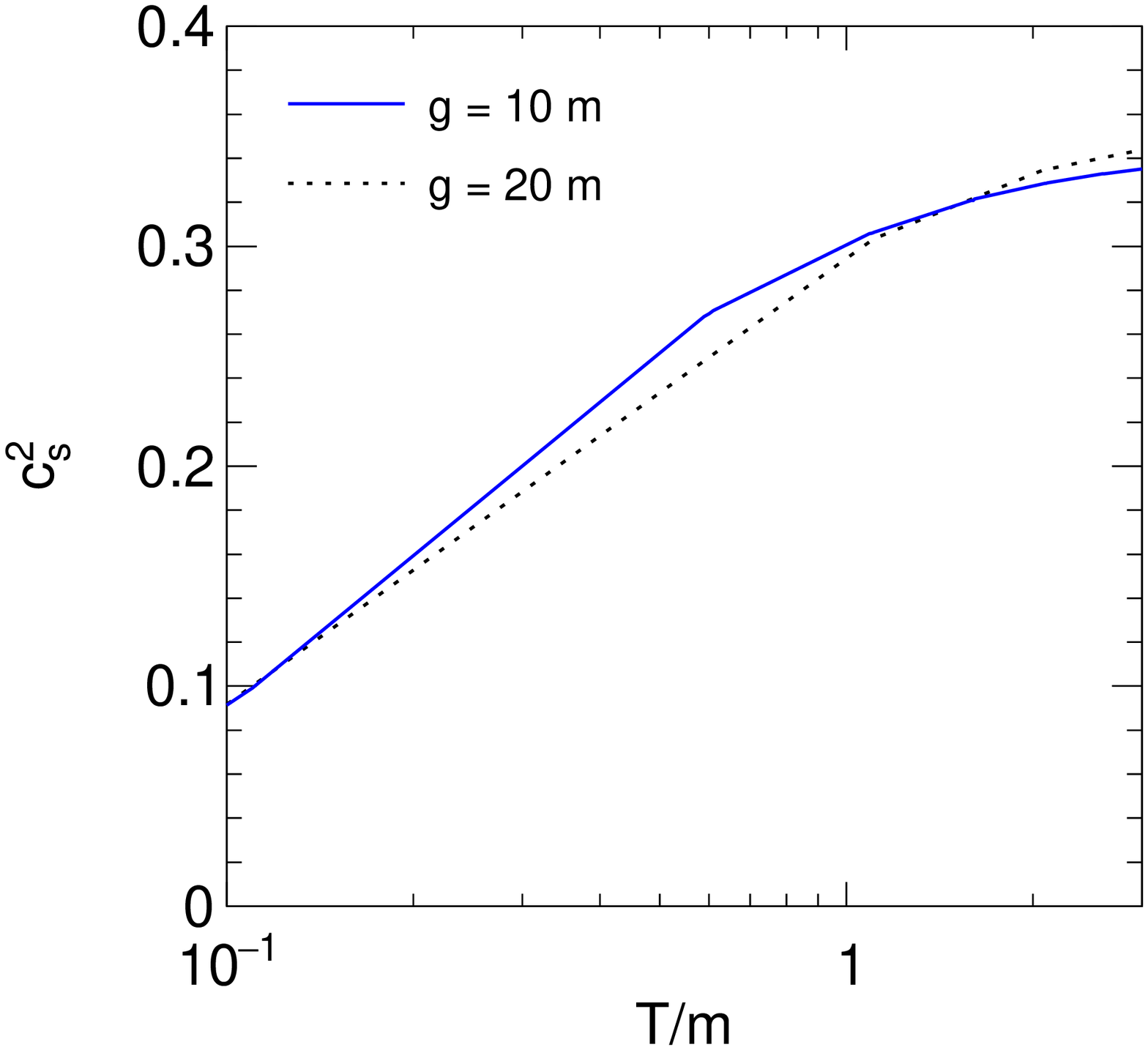}
 \includegraphics[width=0.48\textwidth]{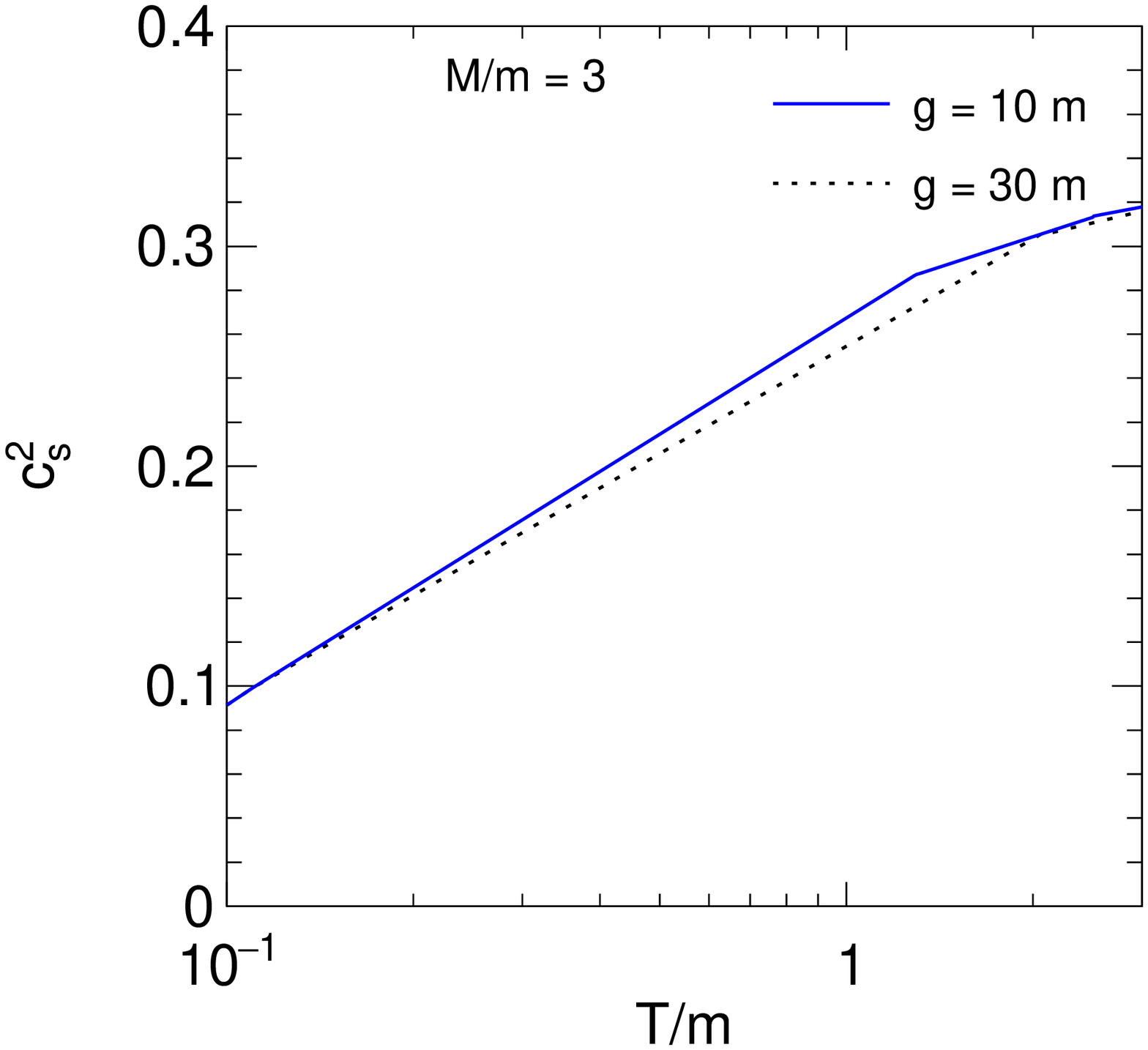}
\caption{Temperature dependence of $c_s^2$ for $\varphi^3$ (left) type interaction and $S\varphi^2$ (right) type interaction. }
\label{Fig:cs2}
\end{figure}
 
In this Appendix, we verify that causality is fulfilled for the $\varphi^3$ and $S\varphi^2$ QFTs by evaluating the Wigner condition  for the phase shifts in the vacuum as well and as the speed of sound $c_s$ in the medium.

In Fig. \ref{Fig:causality} we show the Wigner's causality condition of Eq. \ref{wc} (see Ref. \cite{Boglione:2002vv}) for $\varphi^3$ and the $S\varphi^2$ interactions at $g/m =10$, which show that this condition is always fulfilled. Test with different values of $g$ confirms this result. 
%Only the s-wave contribution is shown in this figure. It can be seen that the solid black line is greater than blue dotted line at all $\sqrt{s}/m$, which indicated the fulfillment of causality condition. We have also checked that the same condition is also satisfied for d- and g-waves. This condition is also satisfied for $g > g_c$. In the right panel we show similar plot for $S\varphi^2$ type interaction.

Next, for both theories, we show in Fig. \ref{Fig:cs2} the temperature dependence of the square of the speed of sound $c_s^2$ (Eq. \ref{speedsound}), which is less than one in the whole considered range of the $T/m$. 

\section{Adding a four-leg interaction}
\label{sec:phi4}
In this Appendix, we show how the results change when adding an interaction term proportional to $\varphi^4$:

\begin{equation}
\mathcal{L}=\frac{1}{2}\left(  \partial_{\mu}\varphi\right)  ^{2}-\frac{1}%
{2}m^{2}\varphi^{2}-\frac{g}{3!}\varphi^{3}-\frac{\lambda}{4!}\varphi^{4},
\end{equation}
whose tree-level scattering amplitude reads%
\begin{align}
A(s,t,u)  &  =-\lambda-\frac{g^{2}}{s-m^{2}+i\epsilon}-\frac{g^{2}}%
{t-m^{2}+i\epsilon}-\frac{g^{2}}{u-m^{2}+i\epsilon}\text{ .}%
\end{align}
Only the s-wave amplitude is modified by including the $\varphi^4$ term, hence we concentrate on the lowest wave.  
The new expression for the tree-level s-wave amplitude reads:
\begin{equation}
A_{0}(s)=\frac{1}{2}\int_{-1}^{+1}d\xi A(s,\theta)=-\lambda-\frac{g^{2}%
}{s-m^{2}}+2g^{2}\frac{\ln\left[  1+\frac{s-4m^{2}}{m^{2}}\right]  }%
{s-4m^{2}}\text{ ,}\label{eq:A0_lambda}
\end{equation}
out of which the tree-level scattering length takes the form:
\begin{equation}
a_{0}^{\text{SL}}=\frac{1}{2}\frac{A_{0}(s=4m^{2})}{8\pi\sqrt{4m^{2}}}%
=\frac{-\lambda}{32\pi m}+\frac{1}{32\pi m}\frac{5g^{2}}{3m^{2}}\text{ .}%
\end{equation}
The loop function allows calculating the unitarized amplitudes in the
$s$-channel as:
\begin{equation}
A_{0}^{U}(s)=\left[  A_{0}^{-1}(s)-\Sigma(s)\right]  ^{-1}\text{ ,}%
\end{equation}
where $\Sigma(s)$ is the loop function reported in Eq. (\ref{loop}), see also Ref. \cite{Giacosa:2021brl}. 
The unitarized scattering length reads:
\begin{align}
a_{0}^{U,\text{SL}}  &  =a_{0}^{U}(s=4m^{2})=\frac{1}{2}\frac{1}{16\pi m}%
\frac{1}{A_{0}^{-1}(4m^{2})-\Sigma(4m^{2})} =\frac{1}{32\pi m}\frac{1}{\left[  -\lambda+\frac{5g^{2}}{3m^{2}}\right]
^{-1}- \frac{1}{64\sqrt{3}\pi}\text{.}}\text{.}%
\end{align}
%\begin{align}
%a_{0}^{U,\text{SL}}  &  =a_{0}^{U}(s=4m^{2})=\frac{1}{2}\frac{1}{16\pi m}%
%\frac{1}{A_{0}^{-1}(4m^{2})-\Sigma(4m^{2})}\\
%&  =\frac{1}{32\pi m}\frac{1}{\left[  -\lambda+\frac{5g^{2}}{3m^{2}}\right]
%^{-1}-0.0028715\text{.}}\text{.}%
%\end{align}
For a certain fixed value of $\lambda,$ the new critical value of $g$ is given
by:
\begin{equation}
-\lambda+\frac{5g^{2}}{3m^{2}}=64\sqrt{3}\pi \text{ ,  } g_{\lambda,c}^{2}=g_{c}^{2}+\frac{3m^{2}}{5}\lambda \text{ ,}
\label{glambdac}
\end{equation}

\begin{figure}[ptb]
\centering
\includegraphics[width=0.48\textwidth]{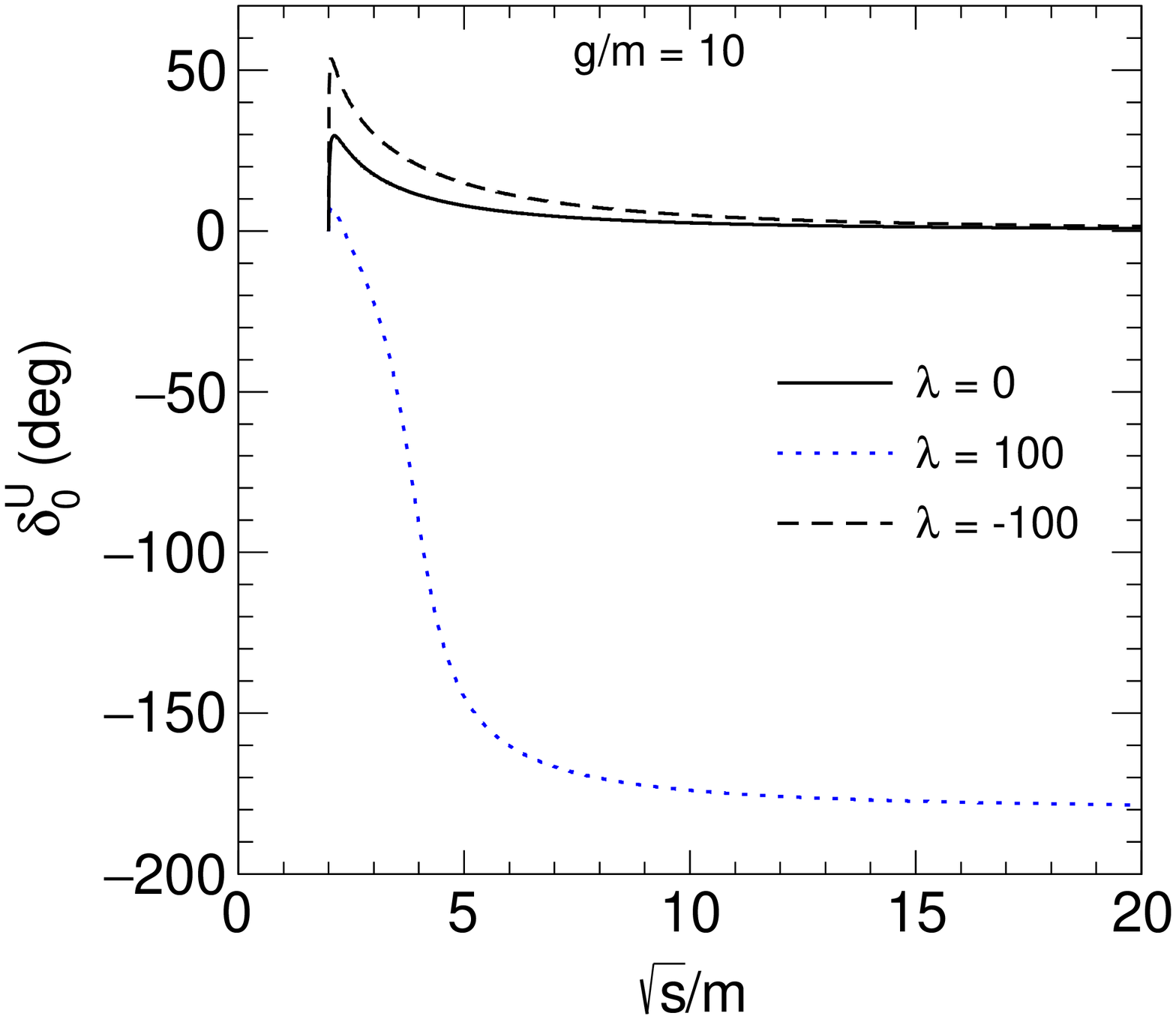}
\includegraphics[width=0.48\textwidth]{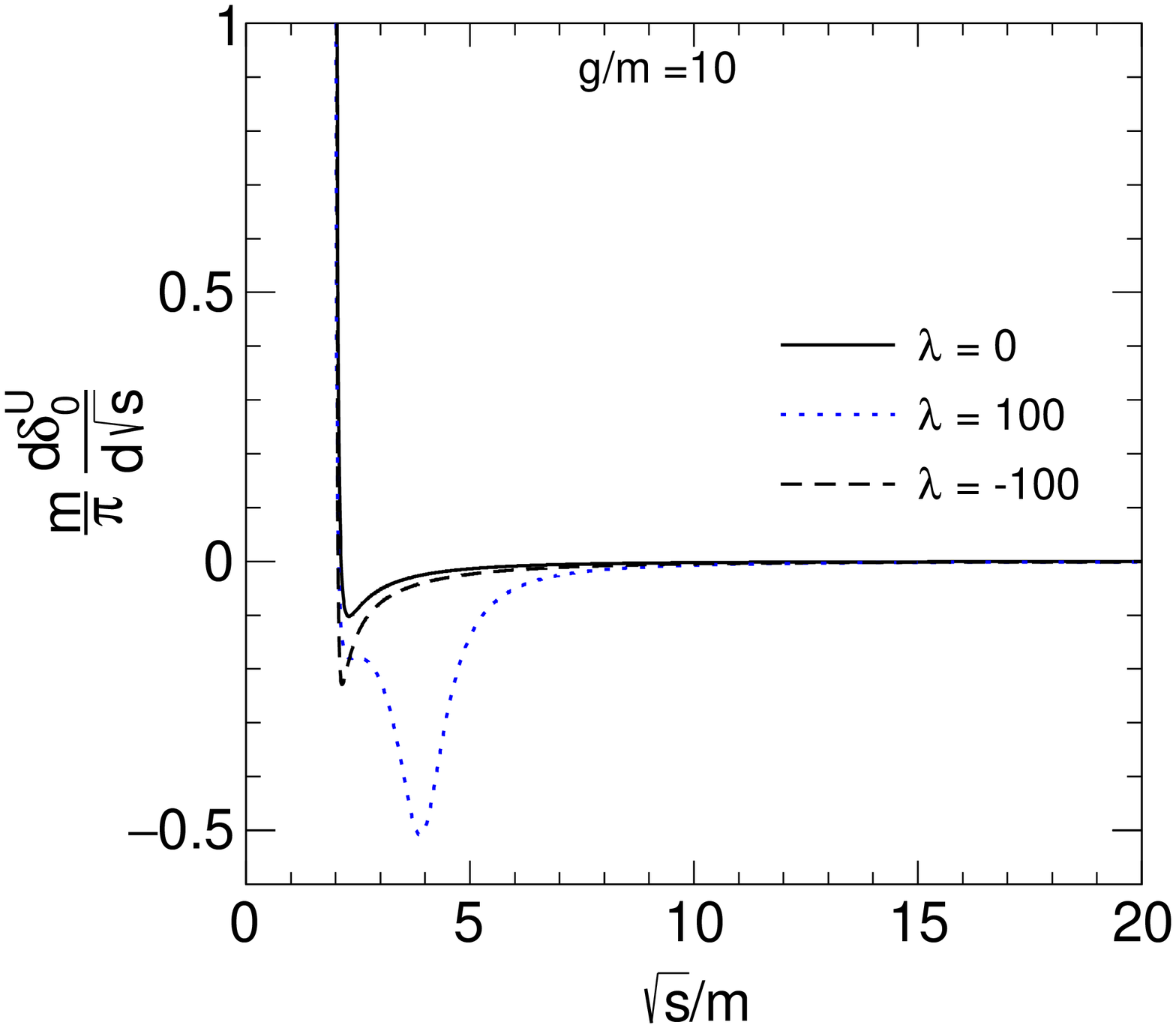}
\caption{(Left) The plot of the s-wave phase shift (using Eqs. \ref{eq:A0_lambda} and \ref{eq:ps_unitarized}) as a function of $\sqrt{s}/m$ at $g = 10m$ when the additional $\varphi^4$-term is present. (Right) The corresponding derivatives with respect to $\sqrt{s}/m$. }
\label{Fig:ps_g_10_lambda}
\end{figure}
  
\begin{figure}[ptb]
\centering
\includegraphics[width=0.32\textwidth]{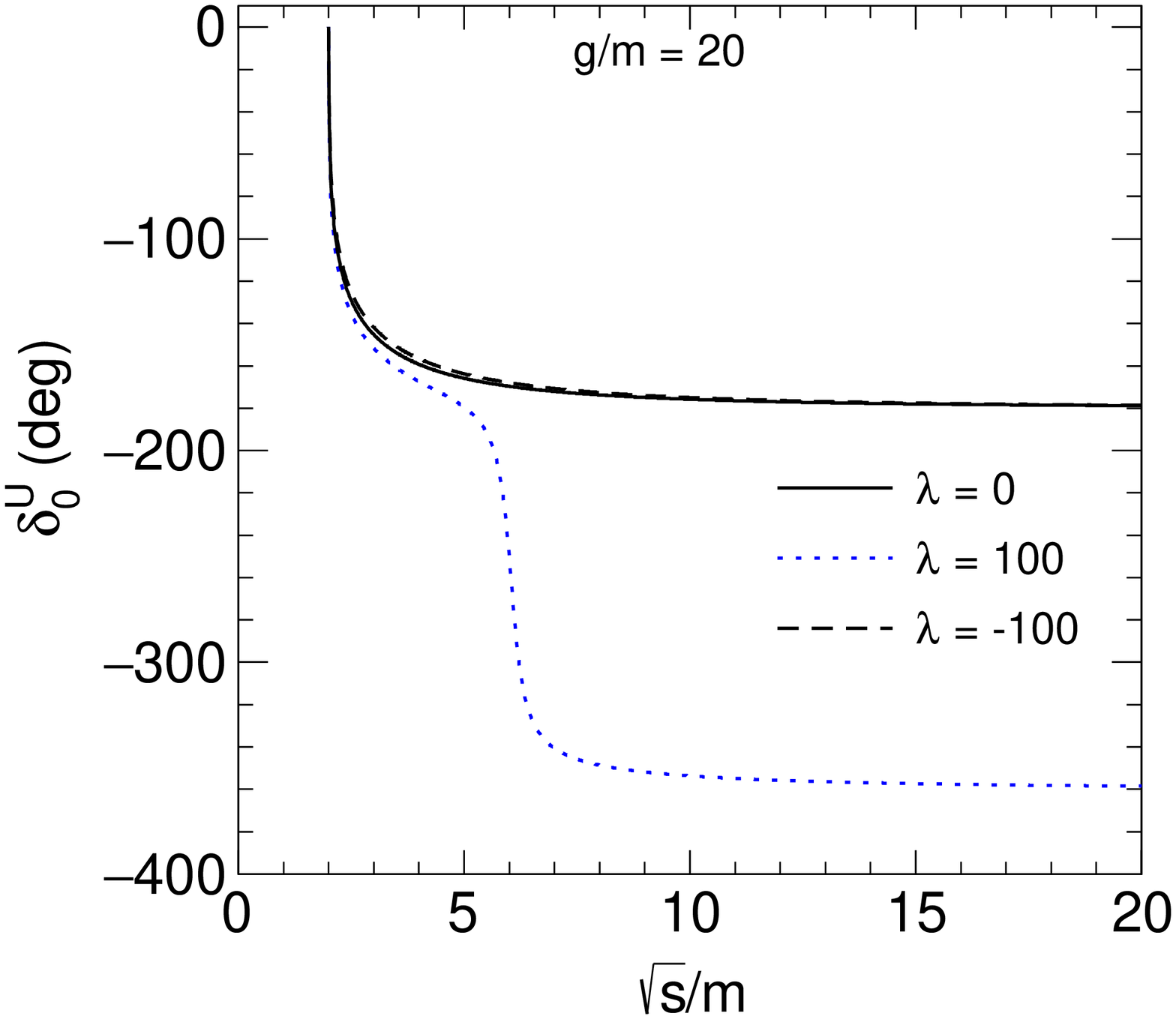}
 \includegraphics[width=0.32\textwidth]{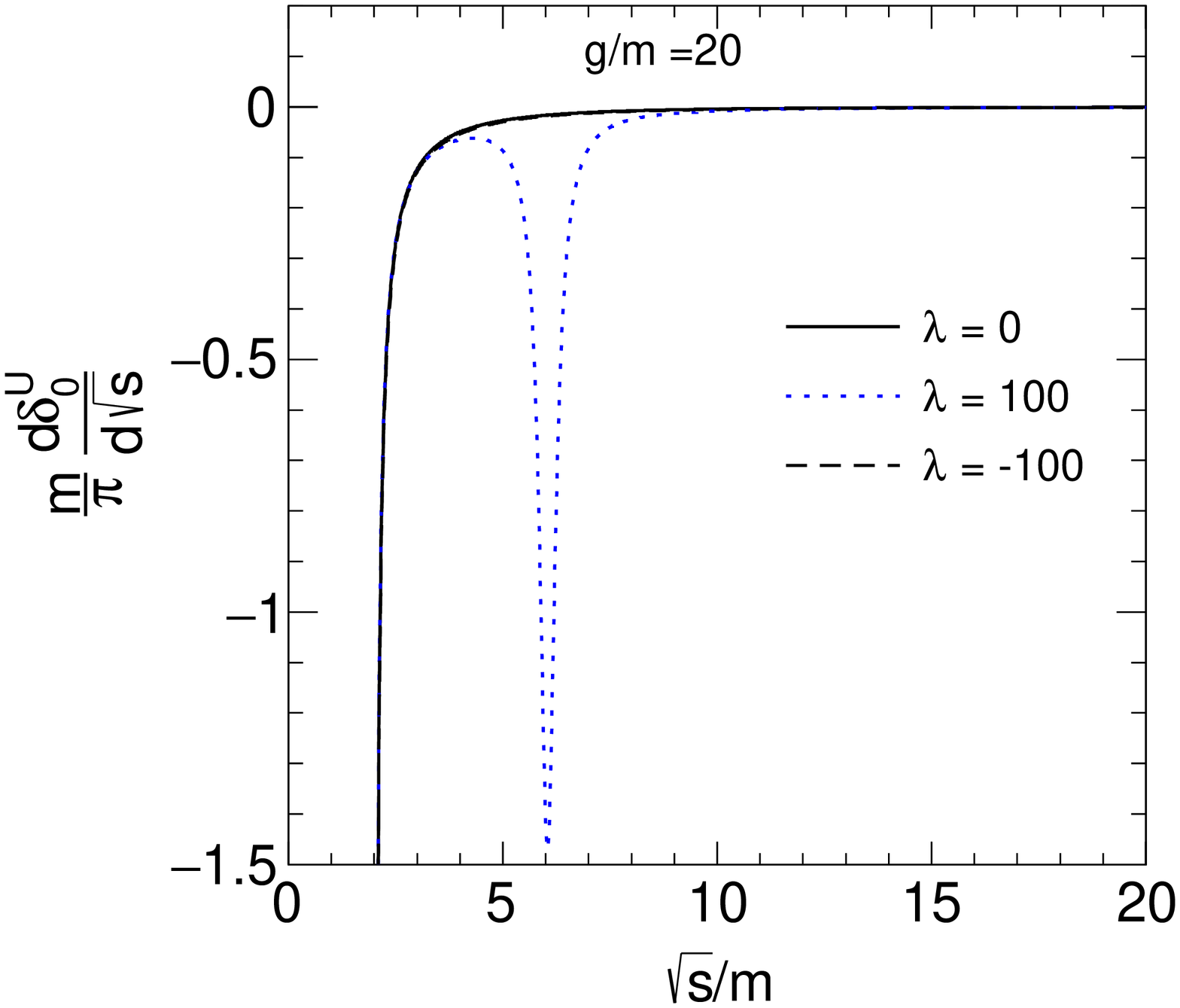}
 \includegraphics[width=0.32\textwidth]{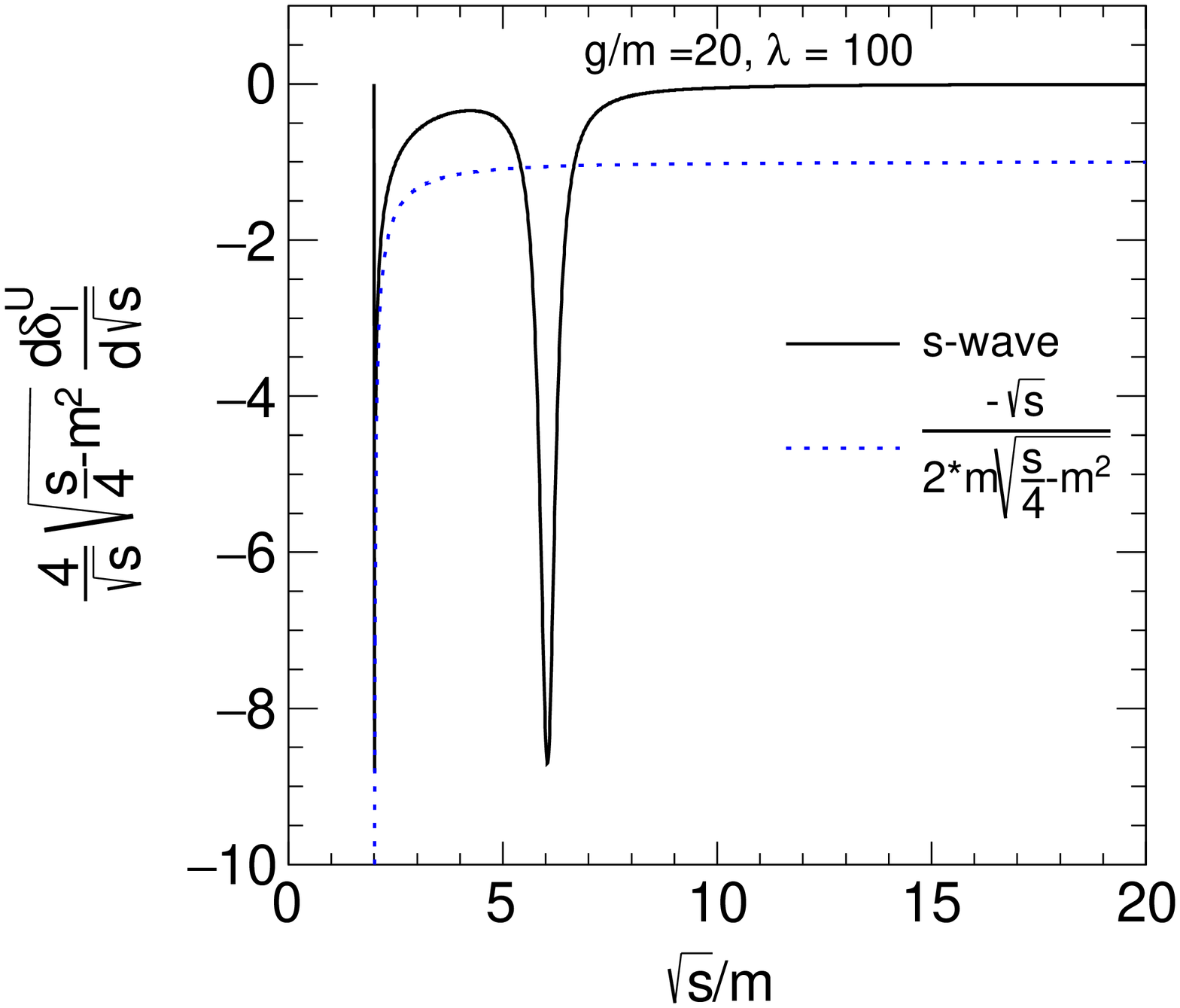}
\caption{Left and middle panels: same as Fig. \ref{Fig:ps_g_10_lambda} but for $g = 20m$. Right panel: Wigner condition.}
\label{Fig:ps_g_20_lambda}
\end{figure}

%\begin{figure}[ptb]
%\centering
%\includegraphics[width=0.48\textwidth]{causality_check_phi3_g_20_lambda_100.eps}
%\includegraphics[width=0.48\textwidth]{phi4_g_20_lambda_100_cs2.eps}
%\caption{Verification of causality for $g = 20m$ and $\lambda = 100$.}
%\label{Fig:causality_g_20_lambda_100}
%\end{figure}

\begin{figure}[ptb]
\centering
\includegraphics[width=0.32\textwidth]{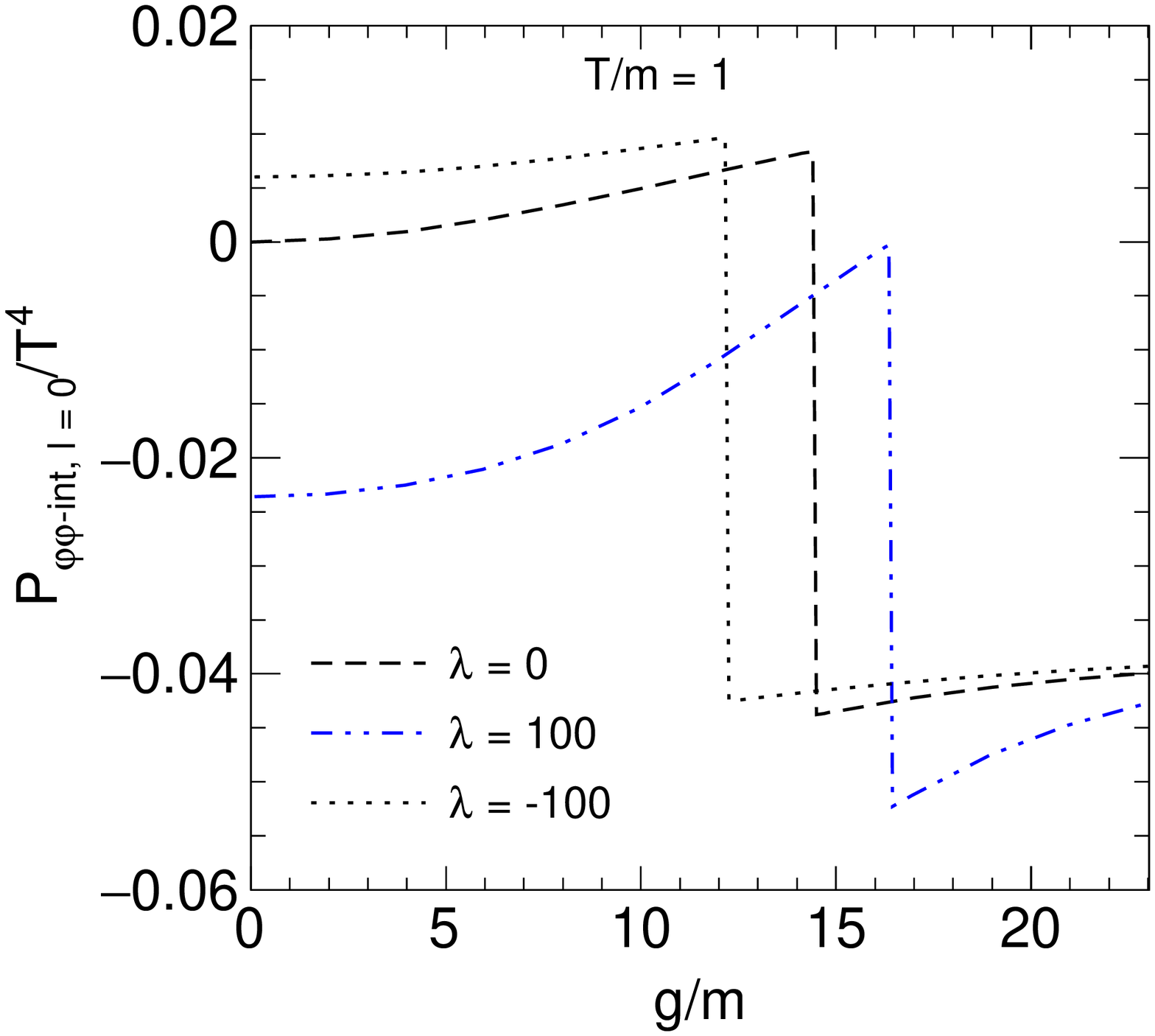}
 \includegraphics[width=0.32\textwidth]{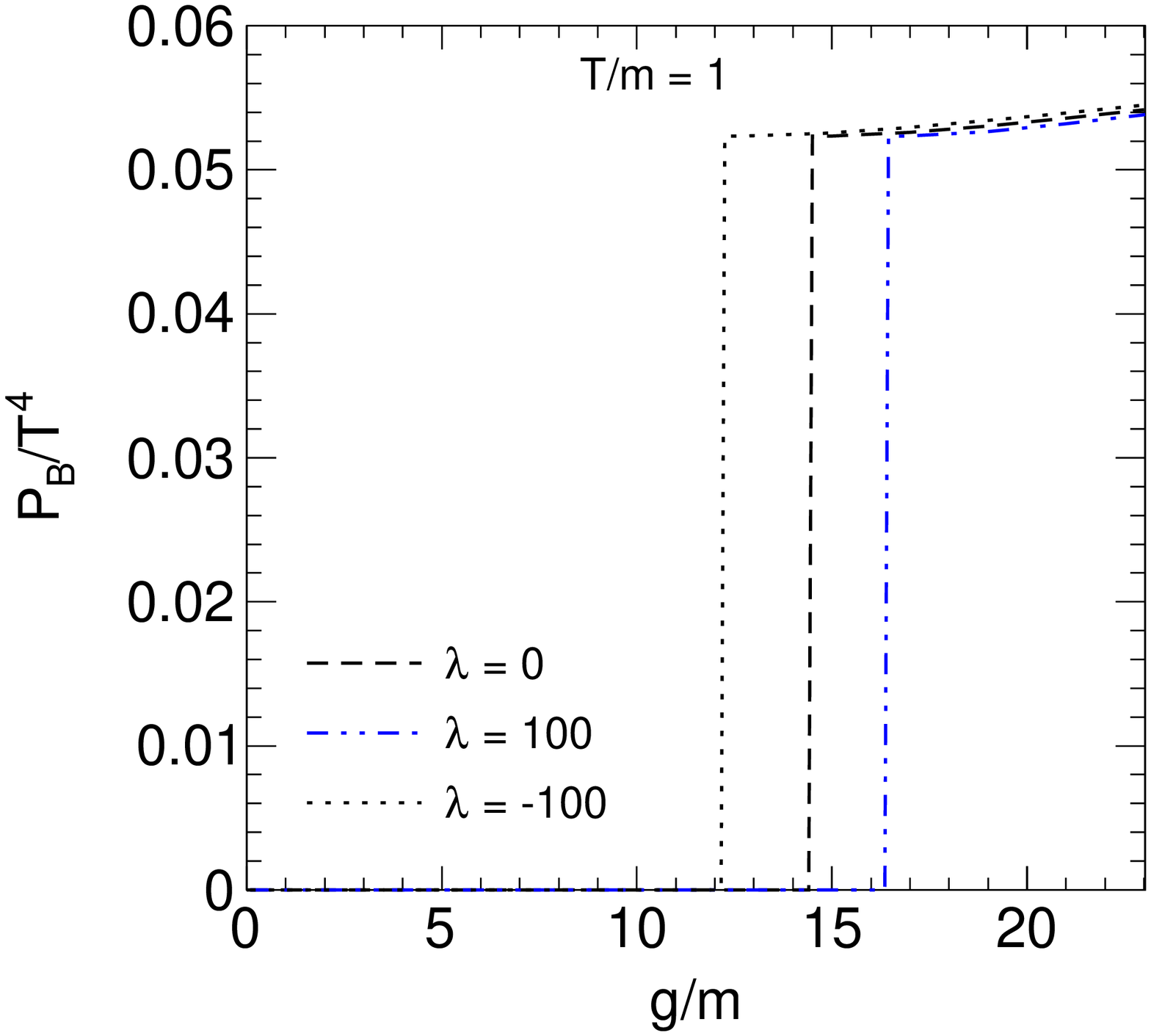}
 \includegraphics[width=0.32\textwidth]{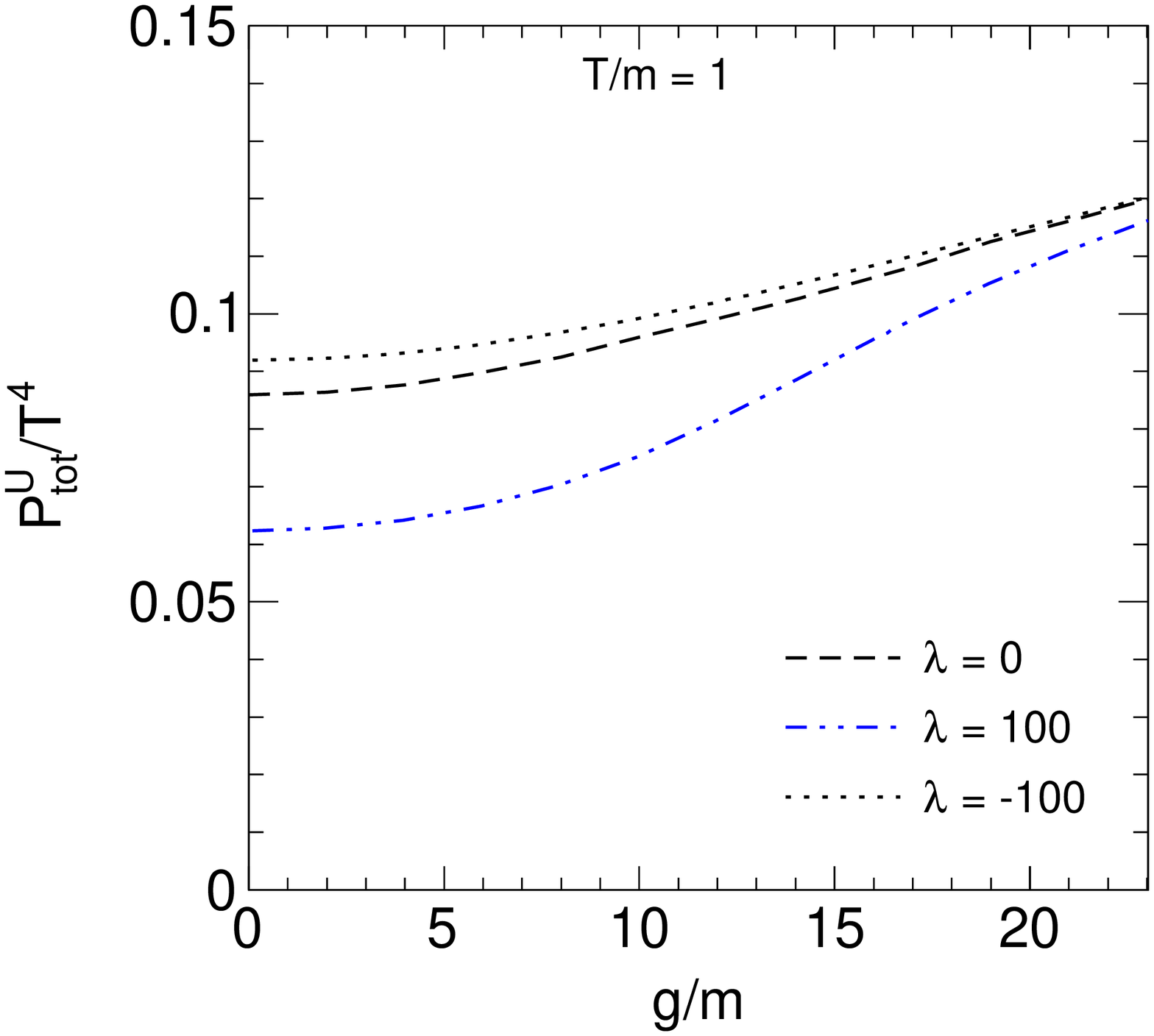}
\caption{Normalized s-wave contribution to the pressure (left) (Eq. \ref{eq:p_int}) and bound state  (middle) (Eq. \ref{eq:pb}) as a function of $g/m$ when the additional $\varphi^4$-term is present. Right: total pressure as a function of $g$.}
\label{Fig:pintvsg_lambda}
\end{figure}

In Fig. \ref{Fig:ps_g_10_lambda} we show the s-wave phase shift as a function of $\sqrt{s}/m$ for $g/m=10$ and for three different values of $\lambda$ ($\lambda = 0$ as a reference, $\lambda = 100$ (repulsive) for which $g_{\lambda,c} /m\approx 16.4$, and $\lambda = - 100$ (attractive) for which  $g_{\lambda,c}/m\approx 12.2$, thus no bound state forms in any of these cases). 
Interestingly, the case $\lambda = -100$ is similar to  $\lambda = 0$. Yet, the situation is completely different for $\lambda = 100$, where the phase shift decreases and saturates to $-\pi$ at large energies. 
%Note, this result is not in disagreement with Levinson's theorem. Even if no physical pole is realized, there is a pole for negative $s$ values which causes the phase shift to tend to $-\pi$, see also Ref. \cite{Samanta:2020pez} where this feature is discussed for the $\varphi^4$-theory. [indeed, this result is expected since for large energies the $\varphi^4$-term dominates].  Finally, the phase shift derivatives are shown in the right panel of Fig. \ref{Fig:ps_g_10_lambda}. 
%For $\lambda = 0$ and -100, derivatives are large positive just above the threshold. Above the threshold, these derivatives become slightly negative and then go to zero at large $\sqrt{s}/m$. While for $\lambda = 100$, the magnitude of the negative part is a little bit more than the other two.

The figure \ref{Fig:ps_g_20_lambda}is obtained for $g = 20m >g_{\lambda,c}$, for which a bound state forms. 
%Correspondingly, the phase shift tends to $-\pi$ for high energies. 
Also here, the case $\lambda = -100$ is similar to $\lambda = 0$. 
%In both the cases, s-wave phase shift starts from zero and saturates to $-\pi$  at large $\sqrt{s}/m$. 
Quite different behavior is observed in the case of $\lambda = 100$ where at large $\sqrt{s}/m$ the phase shift saturates to $-2\pi$.
%in agreement with the presence of two poles below threshold (one of which is the physical  bound state). Note that the mass of the bound state is different for different $\lambda$ values. For $\lambda = 0, 100$ and -100, bound state masses are respectively $1.968m$, $1.98 m$ and $1.955m$.
The phase shift derivatives are plotted in the right panel. Interestingly, a deep near $\sqrt{s}/m = 6$ is observed for $\lambda = 100$. 

 The case $\lambda>0$ needs some care. There is a certain value of $s=s_{0}\simeq6m^{2}$ for which
the tree-level and the unitarized amplitudes vanish: $A(s_{0})=A^{U}(s_{0})=0$.
At this point, the attractive $\varphi^{3}$-term cancels the repulsive
$\varphi^{4}$-term. For this energy, $\delta^{U}(x=\sqrt{s_{0}})=n\pi$,
where the choice $n=-1$ is dictated by the requirement of continuity.
Indeed, the dip in Fig. \ref{Fig:ps_g_20_lambda} corresponds to the $s_{0}$ value. For $s>s_{0}$ the $\varphi^{4}$-term
becomes dominant and the phase shift tends slowly to $-2\pi$. Yet, the Wigner
condition of Eq. (\ref{wc}) is violated around $s_{0}$, as the right plot of Fig. \ref{Fig:ps_g_20_lambda} shows. This violation takes place at energy for which inelastic channels and higher waves are also expected to be relevant and is a further indication that  a limited range of $s$ should be considered. Indeed, the proper treatment of inelasticities and/or the use of other unitarization approaches could solve this issue.    

%These features imply that the results for the present model should be regarded with care. Namely, while the results seem appropriate to determine the emergence or not of the bound state in the vacuum, they are not such for the further behavior of the phase shift and consequently for the implications at nonzero $T.$ Since this model (for certain values of $\lambda>0$) may have the typical Mexican-hat form, its study is relevant for symmetry restauration, but one has to go beyond the unitarization used in this work. This is definitely an interesting task for the future.

Next, we turn to the pressure. In the left panel of Fig. \ref{Fig:pintvsg_lambda} we show the normalized s-wave interacting pressure as a function of $g/m$ for three values of $\lambda$ and for $T/m =1$.  
%This case is already shown in Fig. \ref{Fig:pvsg}. Positive value of $\lambda$ implies an repulsive interaction. Therefore, for $\lambda = 100$ normalized pressure for s-wave is negative when $g/m$ is zero. As $g/m$ increases pressure increases because of the attractive $\varphi^3$ interaction. 
Note, the discontinuity in the s-wave pressure corresponds to $g_{\lambda,c}$: the larger $\lambda$, the larger $g_{\lambda,c}$ needed to form a bound state, see Eq. (\ref{glambdac}).
%We recall that only the s-wave is considered; If we add components of the other two partial waves, just below the critical $g$ the total interaction is really attractive as required for the formation of a bound state. The discontinuity in s-wave pressure  will occur at an even higher value if we further increase $\lambda$. For $\lambda = -100$ effect is just opposite. The bound state is also formed here.
The normalized pressure of the bound state is shown in the middle panel of Fig. \ref{Fig:pintvsg_lambda}, which -just as before- starts abruptly at $g_{\lambda,c}$. 
%Note that only s-wave is affected by the $\varphi^4$ interaction. Therefore d and g-waves results are not shown in this figure. The total normalized pressure as a function of $g/m$ for different $\lambda$ is shown in Fig. \ref{Fig:ptotvsg_lambda}. Total pressure becomes positive for all the three $\lambda$ values although s-wave contribution is negative for $\lambda = 100$. In this case, the free part of the pressure is larger in magnitude compared to the interacting part.
The total pressure is, as expected, continuous in $g$, which is shown in the right panel.

%\begin{figure}[ptb]
%\includegraphics[width=0.48\textwidth]{p_tot_vs_g_T_1_lambda_uniterized.eps}
%\caption{Variation of the normalized total pressure (Eq. \ref{eq:ptot}) with $g/m$ when both the additional $\varphi^4$-terms are present.}%
%\label{Fig:ptotvsg_lambda}
%\end{figure}

%\begin{figure}[ptb]
%\centering
%\includegraphics[width=0.48\textwidth]{p_int_s_wave_vs_T_g_10_uniterized.eps}
%\includegraphics[width=0.48\textwidth]{p_int_s_wave_vs_T_g_20_uniterized.eps}
%\caption{Temperature dependence of the s-wave normalized pressure (Eq. \ref{eq:p_int}) at %different $\lambda$. The left panel is for $g = 10 m<g_c$ (no bound state) and the right panel %for $g = 20 m>g_c$ (bound state builds).}
%\label{Fig:pintvsT_lambda}
%\end{figure}

%In Fig. \ref{Fig:pintvsT_lambda} we show the temperature dependence of the s-wave normalized pressure for different $\lambda$ (left panel for $g = 10 m$ (no bound state), right panel for $g = 20 m$ (bound state present). For $g =10 m$, the s-wave contribution to the pressure is positive for $\lambda = 0$ and $\lambda=-100$, but is negative for $\lambda = 100$. On the other hand, for $g = 20m$ the s-wave pressure is negative for all the three $\lambda$'s. 

\begin{figure}[ptb]
\centering
\includegraphics[width=0.32\textwidth]{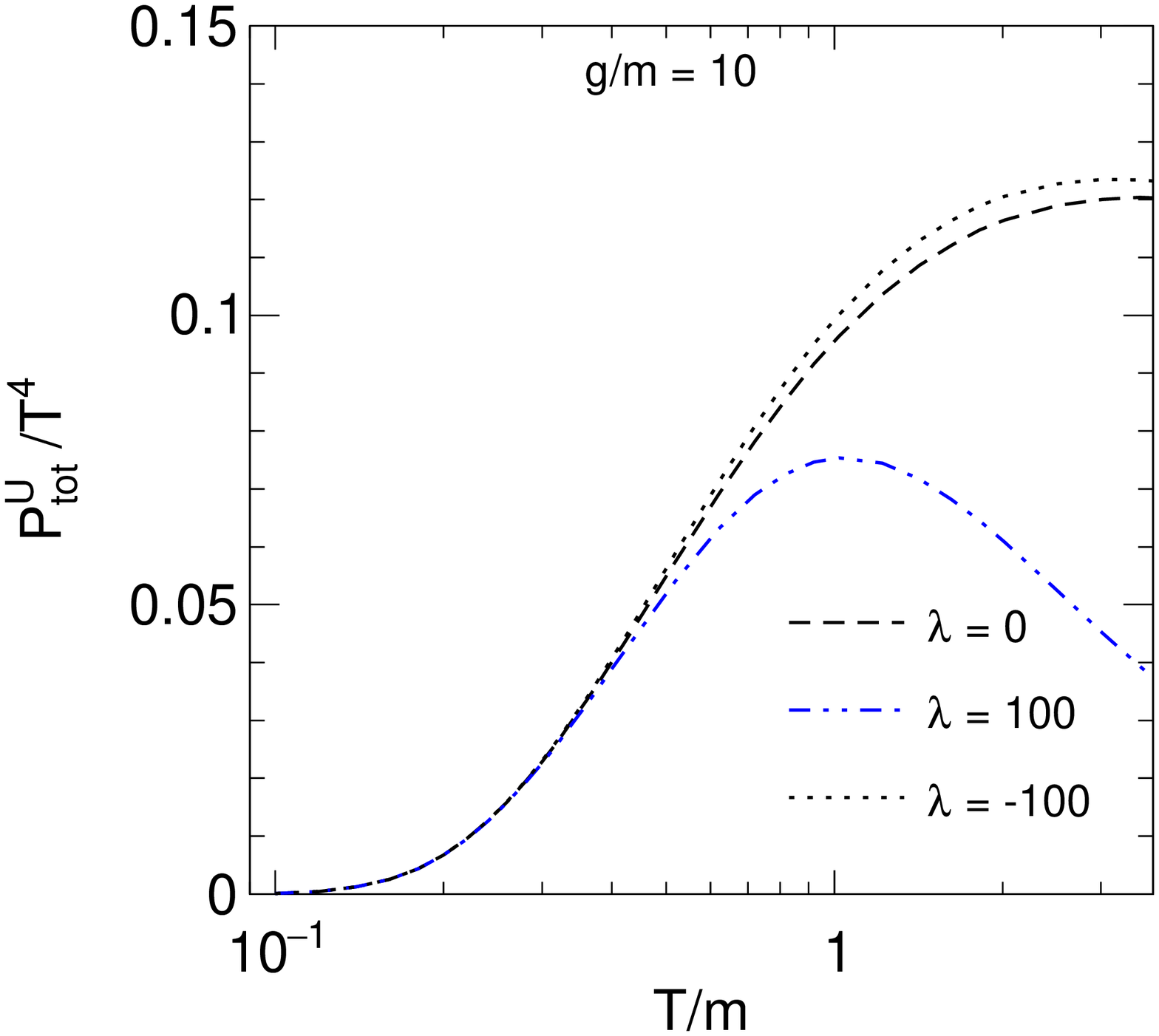}
\includegraphics[width=0.32\textwidth]{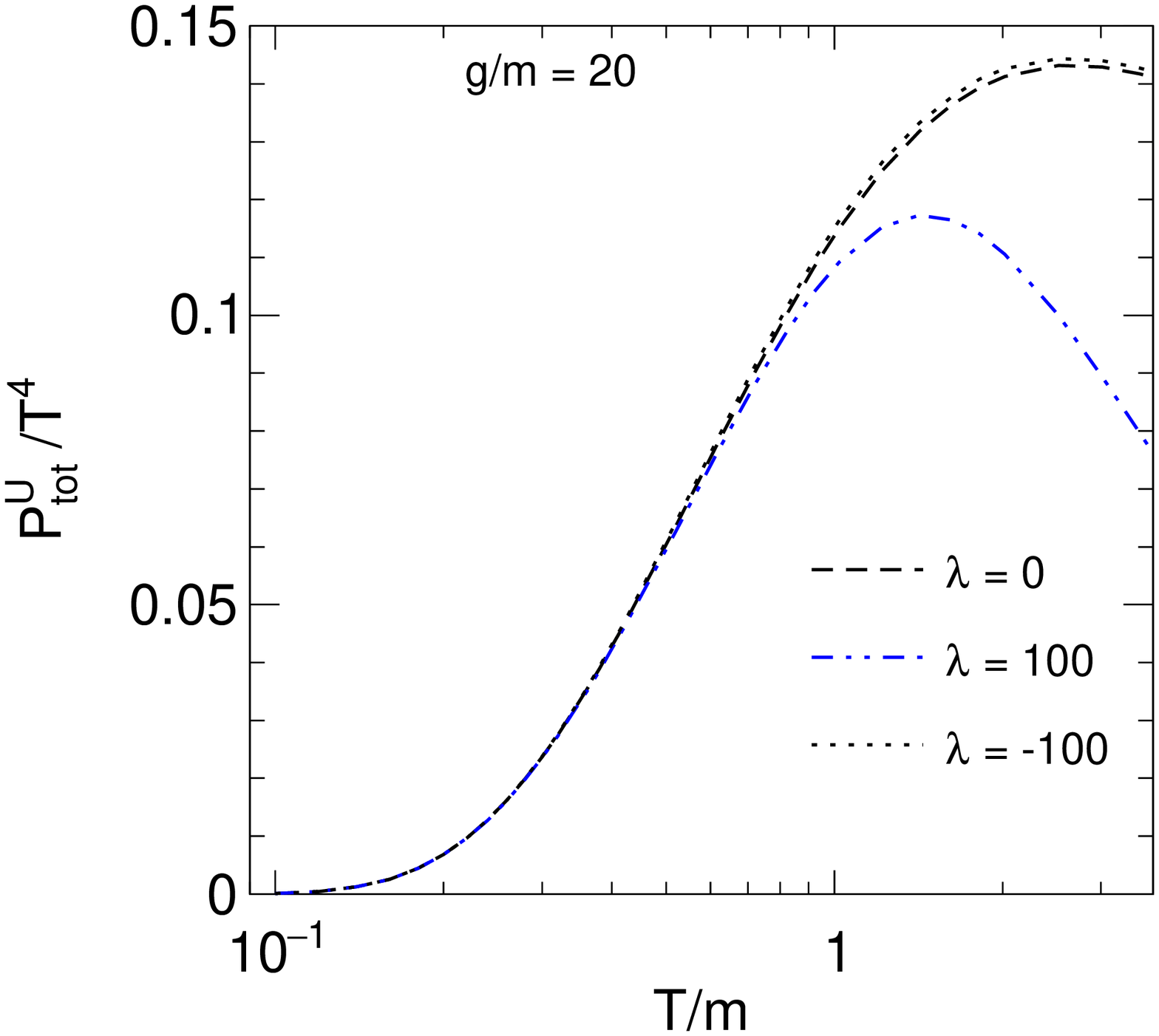}
\includegraphics[width=0.32\textwidth]{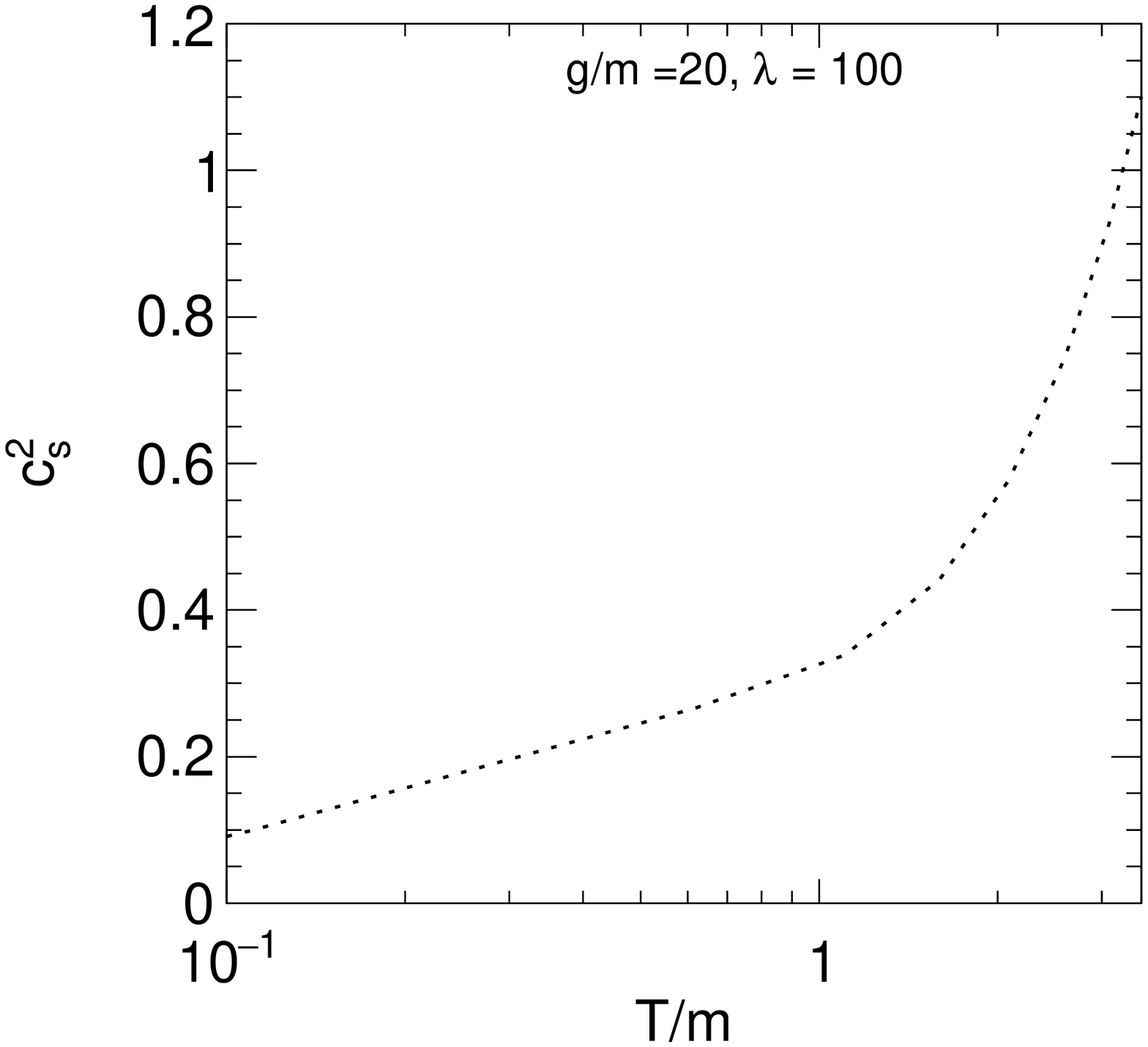}
\caption{ Left: Normalized total pressure (Eq. \ref{eq:ptot}) as function of $T/m$ at different $\lambda$ and for $g = 10 m$. Middle: the same  for $g = 20 m$. Right: Speed of sound as function of $T$.}
\label{Fig:ptotvsT_lambda}
\end{figure}

The temperature dependence of total normalized pressure for $g = 10m$ and $20 m$ is shown in Fig. \ref{Fig:ptotvsT_lambda} (left and middle plots). %Total pressure for $\lambda = -100$ is slightly larger than that of $\lambda = 0$ for both the cases because of attractive $\varphi^4$ interaction. 
Note, for $\lambda = 100$ total pressure is significantly reduced w.r.t. the case $\lambda = 0$.

In the right panel, we present the speed of sound for $\lambda = 100$. It overshoots one at about $T/m \sim 3.3$, which then represents a clear upper limit for the present model. As mentioned above, other effects are expected to be relevant for these temperatures.

%\begin{figure}[ptb]
%\centering
%\includegraphics[width=0.48\textwidth]{phi4_g_20_lambda_100_cs2.eps}
%\caption{Temperature dependence of $c_s^2$ for $g = 20m$ and $\lambda = 100$.}
%\label{Fig:cs2_g_20_lambda_100}
%\end{figure} 

\begin{figure}[ptb]
\centering
\includegraphics[width=0.48\textwidth]{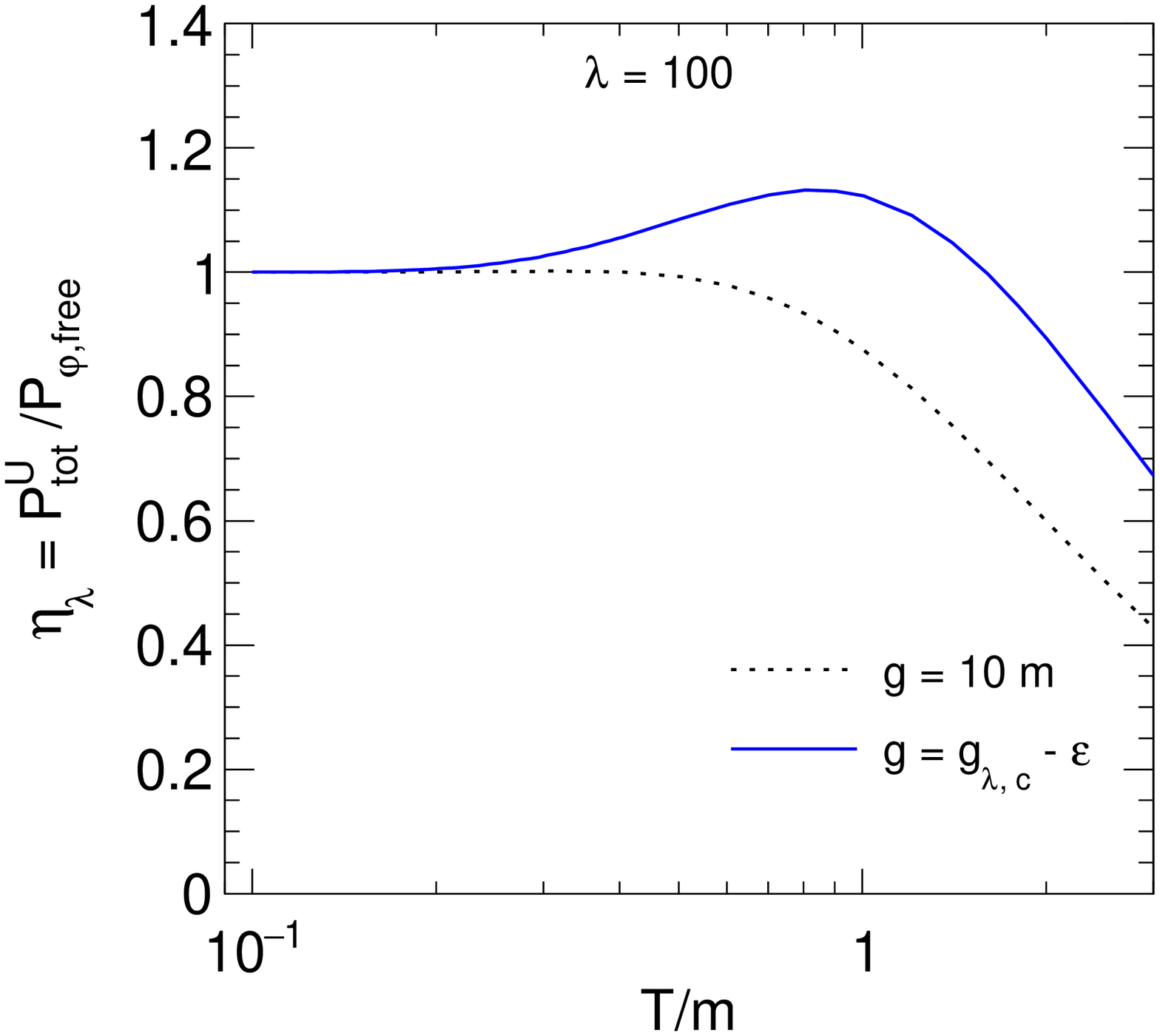}
\includegraphics[width=0.48\textwidth]{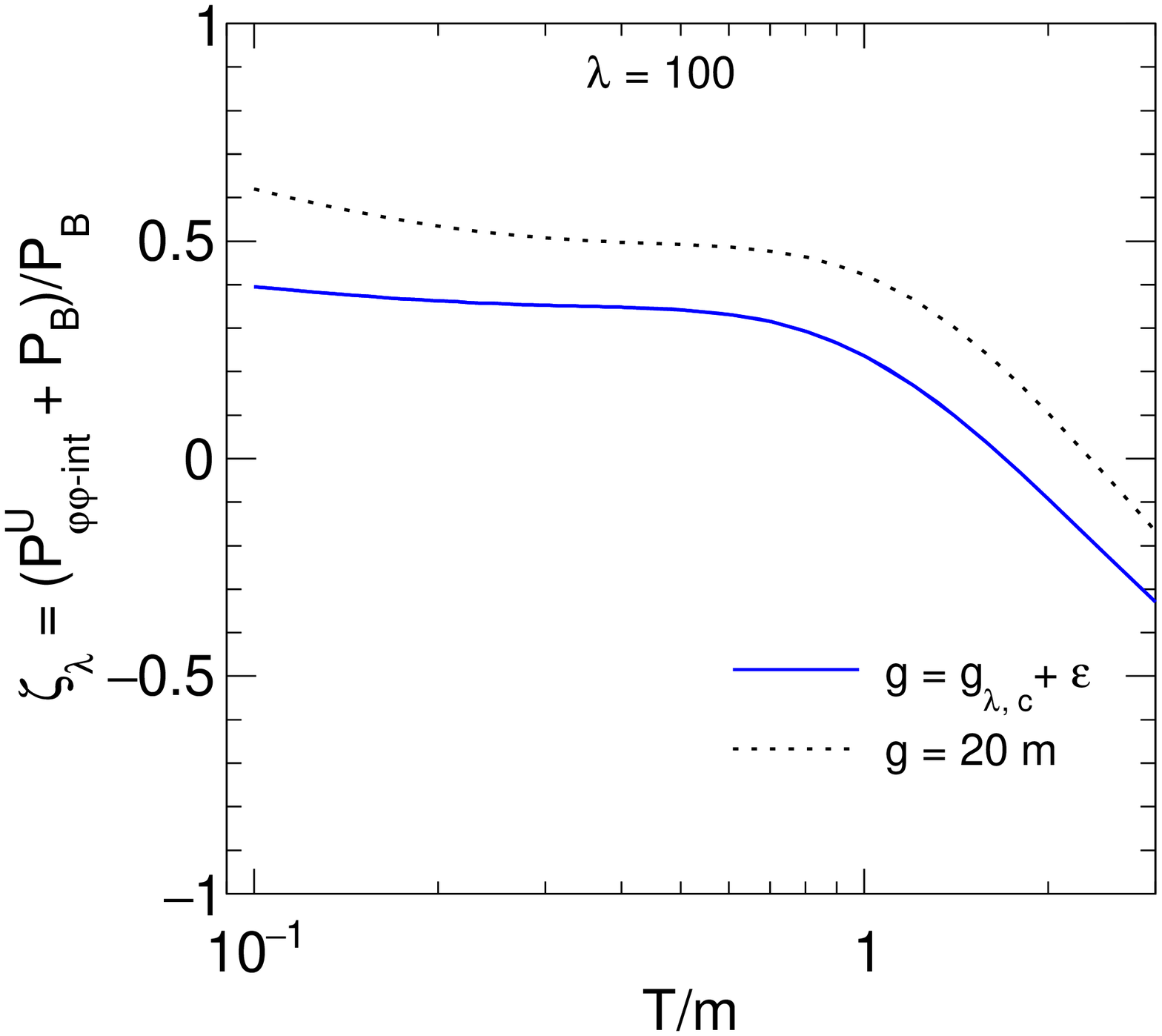}
\caption{Temperature dependence of the ratios $\eta_{\lambda}$ and $\zeta_{\lambda}$ for $\lambda = 100$ (Eq. \ref{eq:eta_lambda}) for different values of $g$.}
\label{Fig:etalambdavsT}
\end{figure}
 
%\begin{figure}[ptb]
%\centering
%\includegraphics[width=0.48\textwidth]{p_int_p_b_vs_T_lambda_100.eps}
%\caption{Temperature dependence of the ratio $\zeta_{\lambda}$ for $\lambda = 100$ (Eq. %\ref{eq:zeta_lambda}).}
%\label{Fig:zetalambdavsT}
%\end{figure}

In the end, we define quantities similar to Eqs. \ref{eq:eta} and \ref{eq:zeta} as:
\begin{equation}
 \eta_{\lambda} = \frac{P^{U}_{\text{tot}} }{P_{\varphi\text{,free}}} \text{ ,}
 \zeta_{\lambda} = \frac{P^{U}_{\varphi\varphi-\text{int}} + P_{B} }{P_{B}} \text{ ,}
\label{eq:eta_lambda} 
\end{equation} 
which are used to quantify the effect of the interaction.  Figure \ref{Fig:etalambdavsT} shows the variation of $\eta_{\lambda}$ and $\zeta_{\lambda}$ with $T/m$ at $\lambda = 100$ and for different $g$ values. The quantity $\eta_{\lambda}$ is close to one at low $T$, then it increases and reaches a maximum: the further decrease takes place in a region of $T$ where caution is needed. The quantity $\zeta_{\lambda}$ implies a partial cancellation of the single bound state contribution, in agreement with the other QFTs studied in this work.
%So for both the cases, effect of the interaction is negligible at low $T/m$. At high $T/m$ region, interaction is repulsive (because of positive $\lambda$; see Fig. \ref{Fig:pintvsg_lambda}). Note that the high $T/m$ part will be completely different if we take a negative $\lambda$.

%{\bf In Fig. \ref{Fig:zetalambdavsT} we show temperature dependence $\zeta_{\lambda}$ for $\lambda = 100$ at $g = g_{\lambda, c}+ \epsilon$ ($g_{\lambda, c} \approx 16.4 m$), and $g = 20 m$.
%and $g = 100 m$.
%Again, a partial cancellation of the single bound state contribution is visible for  $g = g_{\lambda, c}+ \epsilon$ and $g = 20 m$. 
%For the strong coupling limit one observes also a ratio larger than one for intermediate temperatures.

%At lower $T/m$, $\zeta_{\lambda}$ positive and it becomes negative at larger $T/m$. Further, the magnitude of $\zeta_{\lambda}$ increases as $g$ is increased. This is because of the increasing trend of the magnitude of $P^{U}_{\varphi\varphi-\text{int}}$ with $g/m$ (see Fig. \ref{Fig:pintvsg_lambda}).
%The $\zeta_{\lambda}$ could be even more than one at a larger $g$. 

\bigskip

\bigskip

\bibliographystyle{jhep}

\bibliography{RefFile}

\providecommand{\href}[2]{#2}\begingroup\raggedright\begin{thebibliography}{10}

\bibitem{Cocconi:1960zz}
V.~T. Cocconi, T.~Fazzini, G.~Fidecaro, M.~Legros, N.~H. Lipman, and A.~W.
  Merrison, {\it {Mass Analysis of the Secondary Particles Produced by the
  25-Gev Proton Beam of the Cern Proton Synchrotron}},  {\em Phys. Rev. Lett.}
  {\bf 5} (1960) 19--21.

\bibitem{Abelev:2010rv}
{\bf STAR} Collaboration, B.~I. Abelev et~al., {\it {Observation of an
  Antimatter Hypernucleus}},  {\em Science} {\bf 328} (2010) 58--62,
  [\href{http://arxiv.org/abs/1003.2030}{{\tt arXiv:1003.2030}}].

\bibitem{Agakishiev:2011ib}
{\bf STAR} Collaboration, H.~Agakishiev et~al., {\it {Observation of the
  antimatter helium-4 nucleus}},  {\em Nature} {\bf 473} (2011) 353,
  [\href{http://arxiv.org/abs/1103.3312}{{\tt arXiv:1103.3312}}]. [Erratum:
  Nature475,412(2011)].

\bibitem{Adam:2015vda}
{\bf ALICE} Collaboration, J.~Adam et~al., {\it {Production of light nuclei and
  anti-nuclei in pp and Pb-Pb collisions at energies available at the CERN
  Large Hadron Collider}},  {\em Phys. Rev.} {\bf C93} (2016), no.~2 024917,
  [\href{http://arxiv.org/abs/1506.08951}{{\tt arXiv:1506.08951}}].

\bibitem{Adam:2019phl}
{\bf STAR} Collaboration, J.~Adam et~al., {\it {Measurement of the mass
  difference and the binding energy of the hypertriton and antihypertriton}},
  {\em Nature Phys.} {\bf 16} (2020), no.~4 409--412,
  [\href{http://arxiv.org/abs/1904.10520}{{\tt arXiv:1904.10520}}].

\bibitem{Acharya:2019xmu}
{\bf ALICE} Collaboration, S.~Acharya et~al., {\it {Production of (anti-)$^3$He
  and (anti-)$^3$H in p-Pb collisions at $\sqrt{s_{\rm{NN}}}$ = 5.02 TeV}},
  {\em Phys. Rev.} {\bf C101} (2020), no.~4 044906,
  [\href{http://arxiv.org/abs/1910.14401}{{\tt arXiv:1910.14401}}].

\bibitem{Acharya:2020sfy}
{\bf ALICE} Collaboration, S.~Acharya et~al., {\it {(Anti-)Deuteron production
  in pp collisions at $\sqrt{s}=13$ TeV}},
  \href{http://arxiv.org/abs/2003.03184}{{\tt arXiv:2003.03184}}.

\bibitem{Esposito:2014rxa}
A.~Esposito, A.~L. Guerrieri, F.~Piccinini, A.~Pilloni, and A.~D. Polosa, {\it
  {Four-Quark Hadrons: an Updated Review}},  {\em Int. J. Mod. Phys. A} {\bf
  30} (2015) 1530002, [\href{http://arxiv.org/abs/1411.5997}{{\tt
  arXiv:1411.5997}}].

\bibitem{Aaij:2019vzc}
{\bf LHCb} Collaboration, R.~Aaij et~al., {\it {Observation of a narrow
  pentaquark state, $P_c(4312)^+$, and of two-peak structure of the
  $P_c(4450)^+$}},  {\em Phys. Rev. Lett.} {\bf 122} (2019), no.~22 222001,
  [\href{http://arxiv.org/abs/1904.03947}{{\tt arXiv:1904.03947}}].

\bibitem{Siemens:1979dz}
P.~J. Siemens and J.~I. Kapusta, {\it {Evidence for a soft nuclear matter
  equation of state}},  {\em Phys. Rev. Lett.} {\bf 43} (1979) 1486--1489.

\bibitem{Andronic:2010qu}
A.~Andronic, P.~Braun-Munzinger, J.~Stachel, and H.~Stocker, {\it {Production
  of light nuclei, hypernuclei and their antiparticles in relativistic nuclear
  collisions}},  {\em Phys. Lett.} {\bf B697} (2011) 203--207,
  [\href{http://arxiv.org/abs/1010.2995}{{\tt arXiv:1010.2995}}].

\bibitem{Andronic:2012dm}
A.~Andronic, P.~Braun-Munzinger, K.~Redlich, and J.~Stachel, {\it {The
  statistical model in Pb-Pb collisions at the LHC}},  {\em Nucl. Phys.} {\bf
  A904-905} (2013) 535c--538c, [\href{http://arxiv.org/abs/1210.7724}{{\tt
  arXiv:1210.7724}}].

\bibitem{Cleymans:2011pe}
J.~Cleymans, S.~Kabana, I.~Kraus, H.~Oeschler, K.~Redlich, and N.~Sharma, {\it
  {Antimatter production in proton-proton and heavy-ion collisions at
  ultrarelativistic energies}},  {\em Phys. Rev.} {\bf C84} (2011) 054916,
  [\href{http://arxiv.org/abs/1105.3719}{{\tt arXiv:1105.3719}}].

\bibitem{Ortega:2017hpw}
P.~G. Ortega, D.~R. Entem, F.~Fernandez, and E.~Ruiz~Arriola, {\it {Counting
  states and the Hadron Resonance Gas: Does X(3872) count?}},  {\em Phys.
  Lett.} {\bf B781} (2018) 678--683,
  [\href{http://arxiv.org/abs/1707.01915}{{\tt arXiv:1707.01915}}].

\bibitem{Ortega:2019fme}
P.~G. Ortega and E.~Ruiz~Arriola, {\it {Is X(3872) a bound state?}},  {\em
  Chin. Phys. C} {\bf 43} (2019), no.~12 124107,
  [\href{http://arxiv.org/abs/1907.01441}{{\tt arXiv:1907.01441}}].

\bibitem{Butler:1963pp}
S.~T. Butler and C.~A. Pearson, {\it {Deuterons from High-Energy Proton
  Bombardment of Matter}},  {\em Phys. Rev.} {\bf 129} (1963) 836--842.

\bibitem{Schwarzschild:1963zz}
A.~Schwarzschild and C.~Zupancic, {\it {Production of Tritons, Deuterons,
  Nucleons, and Mesons by 30-GeV Protons on A-1, Be, and Fe Targets}},  {\em
  Phys. Rev.} {\bf 129} (1963) 854--862.

\bibitem{Gutbrod:1988gt}
H.~H. Gutbrod, A.~Sandoval, P.~J. Johansen, A.~M. Poskanzer, J.~Gosset, W.~G.
  Meyer, G.~D. Westfall, and R.~Stock, {\it {Final State Interactions in the
  Production of Hydrogen and Helium Isotopes by Relativistic Heavy Ions on
  Uranium}},  {\em Phys. Rev. Lett.} {\bf 37} (1976) 667--670.

\bibitem{Sato:1981ez}
H.~Sato and K.~Yazaki, {\it {On the coalescence model for high-energy nuclear
  reactions}},  {\em Phys. Lett.} {\bf 98B} (1981) 153--157.

\bibitem{Mrowczynski:1992gc}
S.~Mrowczynski, {\it {On the neutron proton correlations and deuteron
  production}},  {\em Phys. Lett.} {\bf B277} (1992) 43--48.

\bibitem{Csernai:1986qf}
L.~P. Csernai and J.~I. Kapusta, {\it {Entropy and Cluster Production in
  Nuclear Collisions}},  {\em Phys. Rept.} {\bf 131} (1986) 223--318.

\bibitem{Mrowczynski:2016xqm}
S.~Mrowczynski, {\it {Production of light nuclei in the thermal and coalescence
  models}},  {\em Acta Phys. Polon.} {\bf B48} (2017) 707,
  [\href{http://arxiv.org/abs/1607.02267}{{\tt arXiv:1607.02267}}].

\bibitem{Bazak:2018hgl}
S.~Bazak and S.~Mrowczynski, {\it {$^4{\rm He}$ vs. $^4{\rm Li}$ and production
  of light nuclei in relativistic heavy-ion collisions}},  {\em Mod. Phys.
  Lett.} {\bf A33} (2018), no.~25 1850142,
  [\href{http://arxiv.org/abs/1802.08212}{{\tt arXiv:1802.08212}}].

\bibitem{Dong:2018cye}
Z.-J. Dong, G.~Chen, Q.-Y. Wang, Z.-L. She, Y.-L. Yan, F.-X. Liu, D.-M. Zhou,
  and B.-H. Sa, {\it {Energy dependence of light (anti)nuclei and
  (anti)hypertriton production in the Au-Au collision from $\sqrt{s_{NN}} =
  11.5$ to 5020 GeV}},  {\em Eur. Phys. J.} {\bf A54} (2018), no.~9 144,
  [\href{http://arxiv.org/abs/1803.01547}{{\tt arXiv:1803.01547}}].

\bibitem{Sun:2016rev}
K.-J. Sun and L.-W. Chen, {\it {Production of $\Lambda\Lambda$ and
  $\overline{\Lambda \text{n}}$ in central Pb+Pb collisions at $\sqrt{s_{NN}}$
  = 2.76 TeV within a covariant coalescence model}},  {\em Phys. Rev.} {\bf
  C94} (2016), no.~6 064908, [\href{http://arxiv.org/abs/1607.04037}{{\tt
  arXiv:1607.04037}}].

\bibitem{Sun:2018jhg}
K.-J. Sun, L.-W. Chen, C.~M. Ko, J.~Pu, and Z.~Xu, {\it {Light nuclei
  production as a probe of the QCD phase diagram}},  {\em Phys. Lett.} {\bf
  B781} (2018) 499--504, [\href{http://arxiv.org/abs/1801.09382}{{\tt
  arXiv:1801.09382}}].

\bibitem{Polleri:1997bp}
A.~Polleri, J.~P. Bondorf, and I.~N. Mishustin, {\it {Effects of collective
  expansion on light cluster spectra in relativistic heavy ion collisions}},
  {\em Phys. Lett.} {\bf B419} (1998) 19--24,
  [\href{http://arxiv.org/abs/nucl-th/9711011}{{\tt nucl-th/9711011}}].

\bibitem{Mrowczynski:2019yrr}
S.~Mr\'owczy\'nski and P.~S\l{}o\'n, {\it {Hadron\textendash{}Deuteron
  Correlations and Production of Light Nuclei in Relativistic Heavy-ion
  Collisions}},  {\em Acta Phys. Polon. B} {\bf 51} (2020), no.~8 1739--1755,
  [\href{http://arxiv.org/abs/1904.08320}{{\tt arXiv:1904.08320}}].

\bibitem{Bazak:2020wjn}
S.~Bazak and S.~Mrowczynski, {\it {Production of $^4\mathrm{Li}$ and
  $p\!-\!^3\mathrm{He}$ correlation function in relativistic heavy-ion
  collisions}},  {\em Eur. Phys. J. A} {\bf 56} (2020), no.~7 193,
  [\href{http://arxiv.org/abs/2001.11351}{{\tt arXiv:2001.11351}}].

\bibitem{Danielewicz:1991dh}
P.~Danielewicz and G.~F. Bertsch, {\it {Production of deuterons and pions in a
  transport model of energetic heavy ion reactions}},  {\em Nucl. Phys.} {\bf
  A533} (1991) 712--748.

\bibitem{Oliinychenko:2018ugs}
D.~Oliinychenko, L.-G. Pang, H.~Elfner, and V.~Koch, {\it {Microscopic study of
  deuteron production in PbPb collisions at $\sqrt{s} = 2.76 TeV$ via
  hydrodynamics and a hadronic afterburner}},  {\em Phys. Rev.} {\bf C99}
  (2019), no.~4 044907, [\href{http://arxiv.org/abs/1809.03071}{{\tt
  arXiv:1809.03071}}].

\bibitem{Samanta:2020pez}
S.~Samanta and F.~Giacosa, {\it {QFT treatment of a bound state in a thermal
  gas}},  {\em Phys. Rev. D} {\bf 102} (2020) 116023,
  [\href{http://arxiv.org/abs/2009.13547}{{\tt arXiv:2009.13547}}].

\bibitem{Dashen:1969ep}
R.~Dashen, S.-K. Ma, and H.~J. Bernstein, {\it {S Matrix formulation of
  statistical mechanics}},  {\em Phys. Rev.} {\bf 187} (1969) 345--370.

\bibitem{Venugopalan:1992hy}
R.~Venugopalan and M.~Prakash, {\it {Thermal properties of interacting
  hadrons}},  {\em Nucl. Phys. A} {\bf 546} (1992) 718--760.

\bibitem{Broniowski:2015oha}
W.~Broniowski, F.~Giacosa, and V.~Begun, {\it {Cancellation of the $\sigma$
  meson in thermal models}},  {\em Phys. Rev. C} {\bf 92} (2015), no.~3 034905,
  [\href{http://arxiv.org/abs/1506.01260}{{\tt arXiv:1506.01260}}].

\bibitem{Lo:2017ldt}
P.~M. Lo, B.~Friman, M.~Marczenko, K.~Redlich, and C.~Sasaki, {\it {Repulsive
  interactions and their effects on the thermodynamics of a hadron gas}},  {\em
  Phys. Rev. C} {\bf 96} (2017), no.~1 015207,
  [\href{http://arxiv.org/abs/1703.00306}{{\tt arXiv:1703.00306}}].

\bibitem{Lo:2017sde}
P.~M. Lo, {\it {S-matrix formulation of thermodynamics with N-body
  scatterings}},  {\em Eur. Phys. J. C} {\bf 77} (2017), no.~8 533,
  [\href{http://arxiv.org/abs/1707.04490}{{\tt arXiv:1707.04490}}].

\bibitem{Lo:2017lym}
P.~M. Lo, B.~Friman, K.~Redlich, and C.~Sasaki, {\it {S-matrix analysis of the
  baryon electric charge correlation}},  {\em Phys. Lett. B} {\bf 778} (2018)
  454--458, [\href{http://arxiv.org/abs/1710.02711}{{\tt arXiv:1710.02711}}].

\bibitem{Dash:2018can}
A.~Dash, S.~Samanta, and B.~Mohanty, {\it {Interacting hadron resonance gas
  model in the K -matrix formalism}},  {\em Phys. Rev. C} {\bf 97} (2018),
  no.~5 055208, [\href{http://arxiv.org/abs/1802.04998}{{\tt
  arXiv:1802.04998}}].

\bibitem{Dash:2018mep}
A.~Dash, S.~Samanta, and B.~Mohanty, {\it {Thermodynamics of a gas of hadrons
  with attractive and repulsive interactions within an S -matrix formalism}},
  {\em Phys. Rev. C} {\bf 99} (2019), no.~4 044919,
  [\href{http://arxiv.org/abs/1806.02117}{{\tt arXiv:1806.02117}}].

\bibitem{Lo:2019who}
P.~M. Lo and F.~Giacosa, {\it {Thermal contribution of unstable states}},  {\em
  Eur. Phys. J. C} {\bf 79} (2019), no.~4 336,
  [\href{http://arxiv.org/abs/1902.03203}{{\tt arXiv:1902.03203}}].

\bibitem{Lo:2020phg}
P.~M. Lo, {\it {Density of states of a coupled-channel system}},  {\em Phys.
  Rev. D} {\bf 102} (2020), no.~3 034038,
  [\href{http://arxiv.org/abs/2007.03392}{{\tt arXiv:2007.03392}}].

\bibitem{Giacosa:2007bn}
F.~Giacosa and G.~Pagliara, {\it {On the spectral functions of scalar mesons}},
   {\em Phys. Rev. C} {\bf 76} (2007) 065204,
  [\href{http://arxiv.org/abs/0707.3594}{{\tt arXiv:0707.3594}}].

\bibitem{Weinhold:1996ts}
W.~Weinhold, B.~Friman, and W.~Noerenberg, {\it {Thermodynamics of an
  interacting pi N system}},  {\em Acta Phys. Polon. B} {\bf 27} (1996)
  3249--3253.

\bibitem{Weinhold:1997ig}
W.~Weinhold, B.~Friman, and W.~Norenberg, {\it {Thermodynamics of Delta
  resonances}},  {\em Phys. Lett. B} {\bf 433} (1998) 236--242,
  [\href{http://arxiv.org/abs/nucl-th/9710014}{{\tt nucl-th/9710014}}].

\bibitem{Florkowski:2010zz}
W.~Florkowski, {\em {Phenomenology of Ultra-Relativistic Heavy-Ion
  Collisions}}.
\newblock World Scientific, Singapore, 2010.

\bibitem{Lo:2015cca}
P.~M. Lo, M.~Marczenko, K.~Redlich, and C.~Sasaki, {\it {Matching the Hagedorn
  mass spectrum with Lattice QCD results}},  {\em Phys. Rev. C} {\bf 92}
  (2015), no.~5 055206, [\href{http://arxiv.org/abs/1507.06398}{{\tt
  arXiv:1507.06398}}].

\bibitem{Peskin:1995ev}
M.~E. Peskin and D.~V. Schroeder, {\em {An Introduction to quantum field
  theory}}.
\newblock Addison-Wesley, Reading, USA, 1995.

\bibitem{Messiah}
A.~Messiah, {\em {Quantum Mechanics}}.
\newblock Dover Publication, New York, USA, 1999.

\bibitem{Note1}
Note, in order to avoid confusion, we denote with the Latin alphabet s-, d-, g-
  ...as the partial waves with increasing $l$. On the other hand, we employ
  calligraphic $s$, $t$, and $u$ when referring to the Mandelstam variables.
  Finally, the calligraphic $g$ refers to the three-leg coupling constant
  entering into the various Lagrangian(s).

\bibitem{Frazer:1969euo}
W.~R. Frazer, {\it {Dynamical Models Based on Unitarity and Analyticity}},  in
  {\em {Summer School in Elementary Particle Physics}: {Theories of strong
  interactions at high energies}} (H.~J. Yesian, ed.), 1969.

\bibitem{Taylor}
J.~R.~E. Taylor, {\em {Scattering Theory: The Quantum Theory of Nonrelativistic
  Collisions}}.
\newblock John Wiley, USA, 1972.

\bibitem{Dobado:1992ha}
A.~Dobado and J.~R. Pelaez, {\it {A Global fit of pi pi and pi K elastic
  scattering in ChPT with dispersion relations}},  {\em Phys. Rev. D} {\bf 47}
  (1993) 4883--4888, [\href{http://arxiv.org/abs/hep-ph/9301276}{{\tt
  hep-ph/9301276}}].

\bibitem{Oller:1997ng}
J.~A. Oller, E.~Oset, and J.~R. Pelaez, {\it {Nonperturbative approach to
  effective chiral Lagrangians and meson interactions}},  {\em Phys. Rev.
  Lett.} {\bf 80} (1998) 3452--3455,
  [\href{http://arxiv.org/abs/hep-ph/9803242}{{\tt hep-ph/9803242}}].

\bibitem{Giacosa:2021brl}
F.~Giacosa, A.~Pilloni, and E.~Trotti, {\it {Glueball\textendash{}glueball
  scattering and the glueballonium}},  {\em Eur. Phys. J. C} {\bf 82} (2022),
  no.~5 487, [\href{http://arxiv.org/abs/2110.05582}{{\tt arXiv:2110.05582}}].

\bibitem{Donoghue:2019fcb}
J.~F. Donoghue and G.~Menezes, {\it {Unitarity, stability and loops of unstable
  ghosts}},  {\em Phys. Rev. D} {\bf 100} (2019), no.~10 105006,
  [\href{http://arxiv.org/abs/1908.02416}{{\tt arXiv:1908.02416}}].

\bibitem{Nieves:1998hp}
J.~Nieves and E.~Ruiz~Arriola, {\it {Bethe-Salpeter approach for meson meson
  scattering in chiral perturbation theory}},  {\em Phys. Lett. B} {\bf 455}
  (1999) 30--38, [\href{http://arxiv.org/abs/nucl-th/9807035}{{\tt
  nucl-th/9807035}}].

\bibitem{Black:2000qq}
D.~Black, A.~H. Fariborz, S.~Moussa, S.~Nasri, and J.~Schechter, {\it
  {Unitarized pseudoscalar meson scattering amplitudes in three flavor linear
  sigma models}},  {\em Phys. Rev. D} {\bf 64} (2001) 014031,
  [\href{http://arxiv.org/abs/hep-ph/0012278}{{\tt hep-ph/0012278}}].

\bibitem{Delgado:2015kxa}
R.~L. Delgado, A.~Dobado, and F.~J. Llanes-Estrada, {\it {Unitarity,
  analyticity, dispersion relations, and resonances in strongly interacting
  $W_LW_L$, $Z_LZ_L$, and $hh$ scattering}},  {\em Phys.Rev.} {\bf D91} (2015),
  no.~7 075017, [\href{http://arxiv.org/abs/1502.04841}{{\tt
  arXiv:1502.04841}}].

\bibitem{Guo:2006br}
Z.-H. Guo, L.~Y. Xiao, and H.~Q. Zheng, {\it {Is the f0(600) meson a
  dynamically generated resonance? A Lesson learned from the O(N) model and
  beyond}},  {\em Int. J. Mod. Phys. A} {\bf 22} (2007) 4603--4616,
  [\href{http://arxiv.org/abs/hep-ph/0610434}{{\tt hep-ph/0610434}}].

\bibitem{Gulmez:2016scm}
D.~G\"ulmez, U.~G. Mei\ss{}ner, and J.~A. Oller, {\it {A chiral covariant
  approach to $\rho\rho$ scattering}},  {\em Eur. Phys. J. C} {\bf 77} (2017),
  no.~7 460, [\href{http://arxiv.org/abs/1611.00168}{{\tt arXiv:1611.00168}}].

\bibitem{Oller:2020guq}
J.~A. Oller, {\it {Unitarization Technics in Hadron Physics with Historical
  Remarks}},  {\em Symmetry} {\bf 12} (2020), no.~7 1114,
  [\href{http://arxiv.org/abs/2005.14417}{{\tt arXiv:2005.14417}}].

\bibitem{Amsler:1995td}
C.~Amsler and F.~E. Close, {\it {Is f0 (1500) a scalar glueball?}},  {\em Phys.
  Rev. D} {\bf 53} (1996) 295--311,
  [\href{http://arxiv.org/abs/hep-ph/9507326}{{\tt hep-ph/9507326}}].

\bibitem{Faessler:2003yf}
A.~Faessler, T.~Gutsche, M.~Ivanov, V.~E. Lyubovitskij, and P.~Wang, {\it {Pion
  and sigma meson properties in a relativistic quark model}},  {\em Phys. Rev.
  D} {\bf 68} (2003) 014011, [\href{http://arxiv.org/abs/hep-ph/0304031}{{\tt
  hep-ph/0304031}}].

\bibitem{Coito:2019cts}
S.~Coito and F.~Giacosa, {\it {On the Origin of the $Y(4260)$}},  {\em Acta
  Phys. Polon. B} {\bf 51} (2020), no.~8 1713--1737,
  [\href{http://arxiv.org/abs/1902.09268}{{\tt arXiv:1902.09268}}].

\bibitem{Mai:2022eur}
M.~Mai, U.-G. Mei\ss{}ner, and C.~Urbach, {\it {Towards a theory of hadron
  resonances}},  \href{http://arxiv.org/abs/2206.01477}{{\tt
  arXiv:2206.01477}}.

\bibitem{Hayashi:1967bjx}
K.~Hayashi, M.~Hirayama, T.~Muta, N.~Seto, and T.~Shirafuji, {\it
  {Compositeness criteria of particles in quantum field theory and S-matrix
  theory}},  {\em Fortsch. Phys.} {\bf 15} (1967), no.~10 625--660.

\bibitem{Cahn:1983vi}
R.~N. Cahn and M.~Suzuki, {\it {The Higgs - Higgs Bound State}},  {\em Phys.
  Lett. B} {\bf 134} (1984) 115--119.

\bibitem{Baru:2003qq}
V.~Baru, J.~Haidenbauer, C.~Hanhart, Y.~Kalashnikova, and A.~E. Kudryavtsev,
  {\it {Evidence that the a(0)(980) and f(0)(980) are not elementary
  particles}},  {\em Phys. Lett. B} {\bf 586} (2004) 53--61,
  [\href{http://arxiv.org/abs/hep-ph/0308129}{{\tt hep-ph/0308129}}].

\bibitem{Dong:2008mt}
Y.-b. Dong, A.~Faessler, T.~Gutsche, and V.~E. Lyubovitskij, {\it
  {Phenomenological Lagrangian approach to the electromagnetic deuteron form
  factors}},  {\em Phys. Rev. C} {\bf 78} (2008) 035205,
  [\href{http://arxiv.org/abs/0806.3679}{{\tt arXiv:0806.3679}}].

\bibitem{Boglione:2002vv}
M.~Boglione and M.~R. Pennington, {\it {Dynamical generation of scalar
  mesons}},  {\em Phys. Rev. D} {\bf 65} (2002) 114010,
  [\href{http://arxiv.org/abs/hep-ph/0203149}{{\tt hep-ph/0203149}}].

\bibitem{Kapusta:2006pm}
J.~I. Kapusta and C.~Gale, {\em {Finite-temperature field theory: Principles
  and applications}}.
\newblock Cambridge Monographs on Mathematical Physics. Cambridge University
  Press, 2011.

\bibitem{Rischke:2003mt}
D.~H. Rischke, {\it {The Quark gluon plasma in equilibrium}},  {\em Prog. Part.
  Nucl. Phys.} {\bf 52} (2004) 197--296,
  [\href{http://arxiv.org/abs/nucl-th/0305030}{{\tt nucl-th/0305030}}].

\bibitem{Brambilla:2014jmp}
N.~Brambilla et~al., {\it {QCD and Strongly Coupled Gauge Theories: Challenges
  and Perspectives}},  {\em Eur. Phys. J. C} {\bf 74} (2014), no.~10 2981,
  [\href{http://arxiv.org/abs/1404.3723}{{\tt arXiv:1404.3723}}].

\bibitem{Lenaghan:1999si}
J.~T. Lenaghan and D.~H. Rischke, {\it {The O(N) model at finite temperature:
  Renormalization of the gap equations in Hartree and large N approximation}},
  {\em J. Phys. G} {\bf 26} (2000) 431--450,
  [\href{http://arxiv.org/abs/nucl-th/9901049}{{\tt nucl-th/9901049}}].

\bibitem{Papazoglou:1996hf}
P.~Papazoglou, J.~Schaffner, S.~Schramm, D.~Zschiesche, H.~Stoecker, and
  W.~Greiner, {\it {Phase transition in the chiral sigma - omega model with
  dilatons}},  {\em Phys. Rev. C} {\bf 55} (1997) 1499--1508,
  [\href{http://arxiv.org/abs/nucl-th/9609035}{{\tt nucl-th/9609035}}].

\bibitem{Amelino-Camelia:1996sfy}
G.~Amelino-Camelia, {\it {On the CJT formalism in multifield theories}},  {\em
  Nucl. Phys. B} {\bf 476} (1996) 255--274,
  [\href{http://arxiv.org/abs/hep-th/9603135}{{\tt hep-th/9603135}}].

\bibitem{Tolos:2008di}
L.~Tolos, D.~Cabrera, and A.~Ramos, {\it {Strange mesons in nuclear matter at
  finite temperature}},  {\em Phys. Rev. C} {\bf 78} (2008) 045205,
  [\href{http://arxiv.org/abs/0807.2947}{{\tt arXiv:0807.2947}}].

\bibitem{GomezNicola:2002tn}
A.~Gomez~Nicola, F.~J. Llanes-Estrada, and J.~R. Pelaez, {\it {Finite
  temperature pion scattering to one loop in chiral perturbation theory}},
  {\em Phys. Lett. B} {\bf 550} (2002) 55--64,
  [\href{http://arxiv.org/abs/hep-ph/0203134}{{\tt hep-ph/0203134}}].

\bibitem{Carrington:1999bw}
M.~E. Carrington, D.-f. Hou, and R.~Kobes, {\it {Shear viscosity in phi**4
  theory from an extended ladder resummation}},  {\em Phys. Rev. D} {\bf 62}
  (2000) 025010, [\href{http://arxiv.org/abs/hep-ph/9910344}{{\tt
  hep-ph/9910344}}].

\bibitem{Luo:2004rj}
X.-Q. Luo, Y.-Y. Li, and H.~Kroger, {\it {Bound states and critical behavior of
  the Yukawa potential}},  {\em Sci. China G} {\bf 35} (2005) 631--642,
  [\href{http://arxiv.org/abs/hep-ph/0407258}{{\tt hep-ph/0407258}}].

\bibitem{Napsuciale:2020ehf}
M.~Napsuciale and S.~Rodriguez, {\it {Complete analytical solution to the
  quantum Yukawa potential}},  {\em Phys. Lett. B} {\bf 816} (2021) 136218,
  [\href{http://arxiv.org/abs/2012.12969}{{\tt arXiv:2012.12969}}].

\bibitem{Yen:1997rv}
G.~D. Yen, M.~I. Gorenstein, W.~Greiner, and S.-N. Yang, {\it {Excluded volume
  hadron gas model for particle number ratios in A+A collisions}},  {\em Phys.
  Rev. C} {\bf 56} (1997) 2210--2218,
  [\href{http://arxiv.org/abs/nucl-th/9711062}{{\tt nucl-th/9711062}}].

\bibitem{Note2}
When subtractions are considered, $\protect \operatorname {Im}\Sigma (z=s)$ is
  indeed exactly zero at the subtraction points. Moreover, $\protect
  \operatorname {Im}\Sigma (z=s)\propto \varepsilon $ for $s<m^{2},$
  $-\varepsilon $ for $m^{2}<s<3m^{2}$, and $\varepsilon $ for $3m^{2}<s\ll
  4m^{2}-\varepsilon $. This feature shows also why the bound state is expected
  above $3m^{2}$. Nevertheless, the qualitative discussion remains unchanged.

\end{thebibliography}\endgroup

\end{document}